\documentclass[submission, Phys]{SciPost}
\usepackage{graphicx}
\usepackage{dcolumn}
\usepackage{bm}
\usepackage{upgreek}
\usepackage{hyperref}
\usepackage{subcaption}
\usepackage{multirow}
\usepackage[utf8]{inputenc} 
\usepackage[T1]{fontenc}    
\usepackage{url}            
\usepackage{booktabs}       
\usepackage{nicefrac}       
\usepackage{microtype}      
\usepackage{lipsum}
\usepackage{xcolor}
\usepackage{amsmath}
\usepackage{empheq} 
\usepackage{ptdr-definitions}
\usepackage{bbm}
\usepackage{xspace}
\usepackage{relsize}
\usepackage{multirow}
\usepackage[countmax]{subfloat}
\usepackage{graphicx}
\usepackage{dcolumn}
\usepackage[most]{tcolorbox} 
\usepackage[export]{adjustbox} 
\graphicspath{ {./images/} }

\newcommand{\PZ}{\ensuremath{\cmsSymbolFace{Z}}\xspace}

\newcommand{\pTmiss}{\ensuremath{\vec{p}_\text{T}^{\,\text{miss}}}\xspace}
\newcommand{\mT}{\ensuremath{m_\text{T}}\xspace}

\newcommand{\VV}{\ensuremath{\cmsSymbolFace{VV}}\xspace}

\usepackage[utf8]{inputenc} 
\usepackage[T1]{fontenc} 	
\usepackage[english]{babel} 

\usepackage[bitstream-charter]{mathdesign}
\usepackage{geometry} 		
\usepackage{amsmath} 		
\usepackage{mathtools} 		
\usepackage{float} 			
\usepackage{graphicx} 		
\usepackage{tabularx} 		
\usepackage{booktabs} 		
\usepackage{color, xcolor} 	
\usepackage{pdfpages} 		
\usepackage{extarrows} 		
\usepackage{multirow} 		
\usepackage{multicol} 		
\usepackage{adjustbox}
\usepackage{enumitem} 		
\usepackage{xspace} 		
\usepackage{stackrel} 		
\usepackage{tikz} 			
\usepackage{braket} 		
\usepackage{bm} 			
\usepackage{tensor} 		
\usepackage{slashed} 		
\usepackage{siunitx} 		
\usepackage{lastpage} 		
\usepackage{cite} 			
\usepackage[normalem]{ulem} 
\usepackage{fontawesome} 	
\usepackage{tocloft} 		
\usepackage{titlesec} 		
\usepackage{doi} 			
\usepackage{hyperref} 		
\usepackage[most]{tcolorbox} 					
\usepackage[nameinlink, capitalize]{cleveref} 	
\usepackage[nottoc, notlot, notlof]{tocbibind} 	
\usepackage[ruled, vlined]{algorithm2e} 		
\usepackage{makecell}
\usepackage[makeroom]{cancel}

\DeclareSymbolFont{usualmathcal}{OMS}{cmsy}{m}{n}
\DeclareSymbolFontAlphabet{\mathcal}{usualmathcal}

\makeatletter
\def\BState{\State\hskip-\ALG@thistlm}
\makeatother

\makeatletter
\@ifundefined{pdfoutput}{}{\DeclareGraphicsRule{*}{mps}{*}{}}
\makeatother

\makeatletter
\DeclareRobustCommand*{\bfseries}{%
   \not@math@alphabet\bfseries\mathbf
   \fontseries\bfdefault\selectfont
   \boldmath
}
\makeatother

\definecolor{EmeraldGreen}{HTML}{1ea78d}
\definecolor{EnglishRed}{HTML}{b02427}
\hypersetup{
	pdftitle={Higgs Signal Strength Estimation with Machine Learning under Systematic Uncertainties},
	pdfauthor={He et al.},
	colorlinks=true, 			
	linkcolor=EnglishRed, 	
	citecolor=EmeraldGreen, 	
	urlcolor=EmeraldGreen 	
} 

\SetArgSty{textnormal}
\SetKwComment{Comment}{{\small\#}~}{}
\SetCommentSty{mycommfont}

\setitemize{itemsep=2pt,topsep=2pt,parsep=0pt,partopsep=0pt,leftmargin=*}
\setenumerate{itemsep=0pt,topsep=2pt,parsep=0pt,partopsep=0pt,labelindent=3pt,leftmargin=*}
\usepackage{amsmath}
 
\usepackage{amsthm} 		
\theoremstyle{definition}
 
\definecolor{red_cb}{HTML}{e41a1c}
\definecolor{blue_cb}{HTML}{377eb8}
\definecolor{green_cb}{HTML}{4daf4a}
\definecolor{purple_cb}{HTML}{984ea3}
\definecolor{orange_cb}{HTML}{ff7f00}

\newcommand{\arXiv}[2][]{%
	\ifthenelse{\equal{#1}{}}%
	{\href{http://arxiv.org/abs/#2}{arXiv:#2}}%
	{\href{http://arxiv.org/abs/#2}{arXiv:#2~[#1]}}}

\def\slashchar#1{\setbox0=\hbox{$#1$}           
   \dimen0=\wd0                                 
   \setbox1=\hbox{/} \dimen1=\wd1               
   \ifdim\dimen0>\dimen1                        
      \rlap{\hbox to \dimen0{\hfil/\hfil}}      
      #1                                        
   \else                                        
      \rlap{\hbox to \dimen1{\hfil$#1$\hfil}}   
      /                                         
   \fi}

\newcommand{\tikznode}[2]{%
\ifmmode%
\tikz[remember picture,baseline=(#1.base),inner sep=0pt] \node (#1) {$#2$};%
\else
\tikz[remember picture,baseline=(#1.base),inner sep=0pt] \node (#1) {#2};%
\fi}

\def\mathswitchr#1{\relax\ifmmode{\mathrm{#1}}\else$\mathrm{#1}$\xspace\fi}
\def\mathswitch#1{\relax\ifmmode#1\else$#1$\xspace\fi}

\newcommand{\Pep}{\mathswitchr {e^+}}
\newcommand{\Pem}{\mathswitchr {e^-}}

\newcommand{\Pp}{\mathswitchr p}

\graphicspath{{./figs/}}

\begin{document}
\begin{flushright}
    \vspace*{-1cm}
    HEPHY-ML-25-03
    \vspace*{0.5cm}
\end{flushright}

\begin{center}{\Large \textbf{
Higgs Signal Strength Estimation with Machine Learning under Systematic Uncertainties
}}\end{center}

\begin{center}\textbf{
Minxuan He\textsuperscript{1},
Claudius Krause\textsuperscript{2}, and
Daohan Wang\textsuperscript{2}
}\end{center}

\begin{center}

{\bf 1} School of Economics and Management, Shaanxi University of Science \& Technology, Xi'an 710021, PR China\\
{\bf 2} Marietta Blau Institute for Particle Physics (MBI), Austrian Academy of Sciences (OeAW), 1010 Vienna, Austria
\\
\end{center}

\begin{center}
\today
\end{center}

\section*{Abstract}
{\bf We present a dedicated graph neural network (GNN)-based methodology for the extraction of the Higgs boson signal strength $\mu$, incorporating systematic uncertainties. The architecture features two branches: a deterministic GNN that processes kinematic variables unaffected by nuisance parameters, and an uncertainty-aware GNN that handles inputs modulated by systematic effects through gated attention-based message passing. Their outputs are fused to produce classification scores for signal-background discrimination. During training we sample nuisance-parameter configurations and aggregate the loss across them, promoting stability of the classifier under systematic shifts and effectively decorrelating its outputs from nuisance variations. The resulting binned classifier outputs are used to construct a Poisson likelihood, which enables profile likelihood scans over signal strength, with nuisance parameters profiled out via numerical optimization. We validate this framework on the FAIR Universe Higgs Uncertainty Challenge dataset, yielding accurate estimation of signal strength $\mu$ and its 68.27\% confidence interval, achieving competitive coverage and interval widths in large-scale pseudo-experiments. Our code "Systematics-Aware Graph Estimator" (SAGE) is publicly available.}

\vspace{10pt}
\noindent\rule{\textwidth}{1pt}
\tableofcontents
\noindent\rule{\textwidth}{1pt}
\vspace{10pt}

\clearpage

\section{Introduction}
\label{sec:intro}

With the increasing volume of data collected at the Large Hadron Collider (LHC), precision measurements are no longer limited by statistical uncertainty but by systematic effects stemming from detector calibration, theoretical modeling, and event reconstruction~\cite{CMS:2022dwd,ATLAS:2022vkf}. This shift underscores the growing importance of accurately modeling multiple nuisance parameters and their combined impact on classifier outputs and likelihood inference when extracting physical observables, such as signal strengths, from experimental data. The challenge is particularly pronounced when employing machine learning classifiers for signal discrimination, as subtle systematic variations can affect the learned decision boundaries and ultimately propagate to the estimation of the signal strength and its confidence interval.

Machine learning, particularly deep neural networks, has demonstrated impressive performance in classification tasks within Higgs boson analyses~\cite{Aad:2019yxi,Aad:2020hzm,Sirunyan:2020hwz,CMS:2022fxs,ATLAS:2023dnm,ATLAS:2024auw,CMS:2024fkb,ATLAS:2024itc,CMS:2024ddc}. However, most standard architectures fail to propagate or incorporate systematic uncertainties in a manner that supports downstream statistical interpretation. As a result, estimations, such as signal strengths and confidence intervals derived from likelihood functions built on classifier scores, may be biased or under-covered. This disconnect between model prediction and statistical inference undermines the robustness of likelihood-based signal strength estimation. To achieve robust and trustworthy measurements, it is therefore essential to incorporate systematic variations directly into the learning workflow, allowing the classifier to learn a smooth and statistically consistent response to these uncertainties.

Traditionally, LHC analyses simplify complex measurements by projecting high-dimensional data into low-dimensional summary statistics through binning and physically motivated observables~\cite{Cowan:2010js,ATLAS:2011tau,Cranmer:2014lly,CMS:2024onh}. This dimensional reduction is essential to make the likelihood function analytically tractable, particularly when incorporating systematic uncertainties via shape variations in histograms induced by nuisance parameters. Such binned methods remain standard in precision Higgs measurements and provide a robust framework for statistical inference. In parallel, modern machine-learning methods enable unbinned analyses by learning surrogate likelihoods or posteriors directly from high-dimensional features and nuisance-aware simulations, thereby avoiding histogramming altogether. As the complexity of datasets and the number of systematic sources grow, recent developments have explored machine-learning approaches that extend beyond traditional binning. These strategies fall into several categories:

\begin{itemize}
    \item[(i)] \textbf{Simulation-based inference (SBI)}~\cite{Andreassen:2019nnm,Stoye:2018ovl,Hollingsworth:2020kjg,Brehmer:2018kdj,Brehmer:2018eca,Brehmer:2019xox,Brehmer:2018hga,Cranmer:2015bka,Andreassen:2020gtw,Coogan:2020yux,Flesher:2020kuy,Bieringer:2020tnw,Nachman:2021yvi,Chatterjee:2021nms,NEURIPS2020_a878dbeb,Mishra-Sharma:2021oxe,Barman:2021yfh,Bahl:2021dnc,Arganda:2022qzy,Kong:2022rnd,Arganda:2022zbs,Butter:2022vkj,Neubauer:2022gbu,Rizvi:2023mws,Heinrich:2023bmt,Morandini:2023pwj,Barrue:2023ysk,Chen:2023ind,Heimel:2023mvw,Chai:2024zyl,Chatterjee:2024pbp,Alvarez:2024owq,Diaz:2024yfu,Mastandrea:2024irf,JETSCAPE:2024cqe,Bahl:2024meb,Maitre:2024hzp,Heimel:2024drk,ATLAS:2024ynn,Benato:2025rgo,Ghosh:2025fma,Ghosh:2021roe,Chen:2022pzc,Farrell:2022lfd,Brandes:2024vhw} directly learns an unbinned surrogate likelihood or posterior from high-dimensional features and nuisance-aware simulations, eliminating the need for binning or simplified observables. This enables precision analyses without sacrificing information from detector or modeling effects.
    
    \item[(ii)] \textbf{Decorrelation techniques}~\cite{Blance:2019ibf,Englert:2018cfo,Dolen:2016kst,Moult:2017okx,Stevens:2013dya,Shimmin:2017mfk,Bradshaw:2019ipy,ATLAS:2018ibz,DiscoFever,Wunsch:2019qbo,Rogozhnikov:2014zea,10.1088/2632-2153/ab9023,clavijo2020adversarial,Kasieczka:2020pil,Kitouni:2020xgb,Estrade:2019gzk,Ghosh:2021hrh,Nachman:2019dol}, such as adversarial training and penalty-based regularization, seek to reduce the sensitivity of classifier outputs to nuisance parameters, enhancing robustness but without yielding an explicit likelihood function.
    \item[(iii)]\textbf{Training with downstream test statistics}~\cite{DAgnolo:2018cun,Grosso:2023scl,Letizia:2022xbe,dAgnolo:2021aun,Chen:2020mev,Wunsch:2020iuh,Elwood:2020pik,Xia:2018kgd,deCastro:2018mgh,Charnock_2018,Alsing:2019dvb,Simpson:2022suz,Feichtinger:2021uff,Layer:2023lwi}, integrates statistical inference goals directly into the training objective, using approximations of the profile likelihood or related test statistics to optimize model parameters in a statistically meaningful way.

    \item[(iv)]\textbf{Bayesian neural networks}~\cite{Kasieczka:2020vlh,Bollweg:2019skg,Araz:2021wqm,Bellagente:2021yyh} aim to model uncertainties from the model and from the data by placing distributions over network weights or outputs, allowing for uncertainty estimation in predictions that can be propagated into inference procedures.
\end{itemize}
An alternative method to obtain uncertainties on neural network outputs involves ensembling together with special training objectives~\cite{Bahl:2024gyt,ATLAS:2024rpl,Benevedes:2025nzr}. In addition to these categories, recent work~\cite{Elsharkawy:2025six} proposes a generative–discriminative hybrid: a contrastive normalizing-flow + DNN pipeline trained across multiple nuisance settings to learn nuisance-aware features, with inference performed via a binned likelihood in score space.

Beyond these training paradigms, the event representation and inductive bias are equally crucial. Graph neural networks (GNNs) naturally encode collider events as variable-size sets of physics objects (nodes) linked by kinematic or geometric relations (edges); permutation-invariant message passing aggregates local neighborhoods and captures non-Euclidean, long-range correlations. Variants with dynamic edge construction and attention achieve state-of-the-art results in tracking, jet tagging, particle-flow, and event-level classification, making GNNs strong baselines for LHC analyses~\cite{Thais:2022iok}. In this work, we present a dedicated GNN-based method for estimating the Higgs boson signal strength $\mu$ in the presence of systematic uncertainties. Our model architecture features two branches: a deterministic GNN that processes features unaffected by nuisance shifts, and an uncertainty-aware GNN that handles systematics-perturbed features. By randomly sampling 100 nuisance parameter configurations per training epoch and injecting the corresponding perturbed datasets, the classifier learns to produce outputs that are robust to systematic effects, without explicitly taking nuisance parameters as input. This framework falls into the category of decorrelation techniques, where the goal is to suppress the sensitivity of classifier outputs to nuisance variations while preserving discriminative power between signal and background. Our code ``Systematics-Aware Graph Estimator'' (SAGE) is publicly available~\cite{SAGE}.

We validate our method using the FAIR Universe Higgs Uncertainty Challenge~\cite{Bhimji:2024bcd}~(FAIR-HUC) dataset, which simulates $\PH \rightarrow \tau \tau$ events along with major Standard Model backgrounds and multiple sources of systematic uncertainty, including six nuisance parameters: three related to calibration and three related to background normalization. Using template morphing and profile likelihood techniques, we extract $\mu$ and its confidence interval in a statistically consistent manner. The results demonstrate competitive performance in coverage and precision, while maintaining a physics-motivated interpretation of systematic effects.

The structure of this paper is as follows. In Sec.~\ref{sec:dataset}, we describe the simulated datasets provided in the FAIR-HUC, including available feature definitions, and the modeling of systematic uncertainties through six nuisance parameters. In Sec.~\ref{sec:model}, we present the dual-branch graph network architecture, which separates the treatment of deterministic and uncertainty-sensitive features and integrates attention and gating mechanisms in the message passing. Sec.~\ref{sec:scan} details the procedure for estimating the Higgs signal strength $\mu$ and its confidence interval using classifier outputs under systematic variations, including interpolation and numerical profiling. In Sec.~\ref{sec:toy}, we validate the method using an Asimov dataset~\cite{Cowan:2010js} and large-scale pseudo-experiments, assessing the coverage, interval width, and robustness of the inferred signal strength, as well as the accuracy and stability of the nuisance parameter estimates. We conclude in Sec.~\ref{sec:conclusions}. In the appendices we collect further studies for architectural choices. 

\section{Event Samples and Nuisance Modeling}
\label{sec:dataset}
\subsection{The \texorpdfstring{$\PH\to\tau\tau$}{H→tau tau} process}

The decay of the Higgs boson into a pair of $\tau$ leptons ($\PH\to\tau\tau$) plays a pivotal role in probing the Yukawa sector of the Standard Model (SM), offering direct access to the coupling of the Higgs field to third-generation leptons. Over the past decade, both the ATLAS and CMS collaborations~\cite{CMS:2014wdm,ATLAS:2015xst,CMS:2017zyp,ATLAS:2018ynr} have accumulated extensive evidence for this process across multiple data-taking periods and center-of-mass energies, ultimately culminating in its observation at more than 5$\sigma$ significance. In the era of precision Higgs physics, measurements of the $\PH\to\tau\tau$ rate have transitioned from inclusive cross-section determinations~\cite{CMS:2022kdi,ATLAS:2022yrq} to increasingly differential studies targeting distinct production modes and kinematic regions~\cite{CMS:2021gxc,CMS:2024jbe,ATLAS:2024wfv}. The $\tau\tau$ final state thus serves as a key benchmark channel for testing SM predictions and exploring Higgs boson properties at the LHC. Beyond the SM, $\PH\to\tau\tau$ also provides clean sensitivity to SMEFT effects: dimension-six operators can rescale or rephase the $\tau$ Yukawa and introduce CP-odd admixtures. Consequently, precision measurements of kinematic shapes in this channel place competitive constraints on the corresponding Wilson coefficients and complement global SMEFT fits~\cite{Ethier:2021bye}.

Tau leptons decay promptly either leptonically ($\tau_\text{lep}$) into electrons or muons plus neutrinos, or hadronically ($\tau_\text{had}$) into narrow jets of hadrons and a neutrino. These decay channels give rise to several final-state configurations, each with distinct experimental characteristics. Fully leptonic decays yield clean signatures with reduced backgrounds but limited mass resolution due to multiple invisible neutrinos. Fully hadronic decays benefit from larger branching fractions yet face challenges from QCD jet backgrounds. The semi-leptonic topology, where one $\tau$ decays hadronically and the other leptonically, strikes a balance between signal yield and background rejection, and is the configuration adopted in FAIR-HUC.

\subsection{Primary and derived features}
This study is based on the full training dataset provided by the FAIR-HUC~\cite{Bhimji:2024bcd}, which simulates proton-proton collisions at the LHC. The $\PH \to \tau_\text{lep} \tau_\text{had}$ process serves as the signal, accompanied by three dominant SM backgrounds: $\PZ\to\tau\tau$, top quark pair production~($\ttbar$), and diboson processes (primarily $WW$ with minor contributions from $WZ$ and $ZZ$). Both the signal process and the three SM backgrounds are generated using \textsc{Pythia8.2}\cite{Sjostrand:2014zea} for parton-level event generation, followed by fast detector simulation with \textsc{Delphes3.5.0}~\cite{deFavereau:2013fsa}. As documented in Ref.~\cite{Bhimji:2024bcd}, the FAIR-HUC samples were generated with a leading-order PYTHIA setup. PYTHIA includes parton-shower modelling, but Ref.~\cite{Bhimji:2024bcd} does not specify a matching or merging configuration or a formal resummation accuracy for these samples. We therefore do not assign NLO/NNLO or a specified logarithmic accuracy to the differential predictions used in this study. The dataset includes only events with exactly one electron or muon, exactly one hadronically decaying $\tau$, up to two reconstructed jets, and missing transverse energy. Basic kinematic selections are applied at the generator level: the lepton is required to have transverse momentum $p_\text{T}^\ell > 20~\GeV$ and pseudorapidity $|\eta^\ell| < 2.5$, while the hadronic $\tau$ must satisfy $p_\text{T}^{\tau_\text{had}} > 26~\GeV$ and $|\eta^{\tau_\text{had}}| < 2.69$. In order to maximize statistical power and avoid selection-induced biases in classifier training, we do not perform event-level selections and retain all available training events.

The original FAIR-HUC dataset provides 28 features for each event, designed to capture the essential kinematic information of the $\tau_\text{had}$, the lepton, reconstructed jets, and missing transverse momentum (MET). These include primary observables such as transverse momentum ($p_\text{T}$), pseudorapidity ($\eta$), and azimuthal angle ($\phi$) of the visible final-state particles, as well as the missing transverse momentum and its azimuthal angle. Additional variables summarize global properties of the event, such as the number of reconstructed jets, the scalar sum of all the jets' $p_T$ of the events. The derived features encode correlations between particles and characterize the overall event topology. These include the transverse mass of the lepton and MET ($m_\text{T}$), the visible invariant mass of the lepton and $\tau_\text{had}$ system ($m_\text{vis}$), and the modulus of the vector sum of the lepton, $\tau_\text{had}$, and MET momenta ($p_\text{T}^{\text{H}}$). Additional features include the invariant mass of the two leading jets, the ratio of lepton to $\tau_\text{had}$ transverse momentum, and the angular separation $\Delta R$ between the lepton and $\tau_\text{had}$. The global event kinematics is characterized by the scalar and vector sums of all reconstructed transverse momenta ($\sum p_\text{T}$ and $p_\text{T}^\text{tot}$), while centrality and topology are captured by the azimuthal centrality of MET with respect to the lepton and $\tau_\text{had}$ ($C^\text{miss}_\phi$), and the pseudorapidity centrality of the lepton with respect to the jets ($C^\ell_\eta$). Finally, the pseudorapidity separation and the product of pseudorapidities of the two leading jets are included as jet-correlated shape observables. These engineered features are designed to enhance sensitivity to the Higgs boson decay topology and improve background rejection. Participants in the FAIR-HUC are free to utilize, modify, or omit these features depending on their analysis strategy. In our implementation, we exclude two variables—pseudorapidity separation and product of pseudorapidities of the leading jets—and extend the original feature set with nine additional angular distance variables between all pairs of final-state particles. This modification is tailored to the architecture of our graph neural network, which benefits from explicitly encoding geometric relations between nodes. A complete list of the 35 features used in our model is summarized in Table~\ref{tab:event_features}, while the definitions of the original 28 FAIR-HUC observables are available in Ref.~\cite{Bhimji:2024bcd}.

\begin{table*}
  \caption{List of event features used in the analysis. Primary (PRI) variables are direct kinematic observables, while derived (DER) variables encode higher–level correlations and global event properties. Features shown in blue denote the additional nine angular distance features introduced in this study.}
  \label{tab:event_features}
  \centering
  {\footnotesize  
  \renewcommand{\arraystretch}{1.3}
  \begin{tabular}{l l l l}
    \hline\hline
    \multicolumn{4}{c}{\textbf{Primary Features (PRI)}} \\
    \hline
    \textbf{Symbol} & \textbf{Description} & \textbf{Symbol} & \textbf{Description} \\
    \hline
    $\pt^{\ell}$                & Transverse momentum of the lepton            &
    $\eta^{\ell}$               & Pseudorapidity of the lepton                 \\
    $\phi^{\ell}$               & Azimuthal angle of the lepton                &
    $\pt^{\tau_\text{had}}$     & Transverse momentum of the hadronic $\tau$   \\
    $\eta^{\tau_\text{had}}$    & Pseudorapidity of the hadronic $\tau$        &
    $\phi^{\tau_\text{had}}$    & Azimuthal angle of the hadronic $\tau$       \\
    $\pt^{j_1}$                 & Transverse momentum of the leading jet       &
    $\eta^{j_1}$                & Pseudorapidity of the leading jet            \\
    $\phi^{j_1}$                & Azimuthal angle of the leading jet           &
    $\pt^{j_2}$                 & Transverse momentum of the subleading jet    \\
    $\eta^{j_2}$                & Pseudorapidity of the subleading jet         &
    $\phi^{j_2}$                & Azimuthal angle of the subleading jet        \\
    $N_{j}$                     & Number of reconstructed jets                 &
    $\sum_{\text{jets}}\!p_\text{T}$ & Scalar sum of transverse momenta of all jets \\
    $\pTmiss$                   & Missing transverse momentum                  &
    $\phi^{\text{miss}}$        & Azimuthal angle of missing transverse momentum \\
    \hline
    \multicolumn{4}{c}{\textbf{Derived Features (DER)}} \\
    \hline
    \textbf{Symbol} & \textbf{Description} & \textbf{Symbol} & \textbf{Description} \\
    \hline
    $m_\text{T}(\ell,\pTmiss)$  & Transverse mass of lepton and $\pTmiss$      &
    $m_{\text{vis}}$            & Visible invariant mass of $\tau_\text{had}$ and $\ell$ \\
    $\pt^{H}$                   & Vector sum of $\pt^{\tau_\text{had}},\pt^{\ell},\pTmiss$ &
    $m^{j_1j_2}$                & Invariant mass of the two leading jets\\
    \textcolor{blue}{$\Delta R(\ell,miss)$} & \textcolor{blue}{Angular distance between $\ell$ and MET}
           &
    \textcolor{blue}{$\Delta R(\tau_\text{had},miss)$} & \textcolor{blue}{Angular distance between $\tau_\text{had}$ and MET} \\
    \textcolor{blue}{$\Delta R(j_1,miss)$} & \textcolor{blue}{Angular distance between $j_1$ and MET}       &
    \textcolor{blue}{$\Delta R(j_2,miss)$} & \textcolor{blue}{Angular distance between $j_2$ and MET} \\
    \textcolor{blue}{$\Delta R(j_1,\tau_\text{had})$} & \textcolor{blue}{Angular distance between $j_1$ and $\tau_\text{had}$}       &
    \textcolor{blue}{$\Delta R(j_2,\tau_\text{had})$} & \textcolor{blue}{Angular distance between $j_2$ and $\tau_\text{had}$} \\
    \textcolor{blue}{$\Delta R(j_1,\ell)$} & \textcolor{blue}{Angular distance between $j_1$ and $\ell$}       &
    \textcolor{blue}{$\Delta R(j_2,\ell)$} & \textcolor{blue}{Angular distance between $j_2$ and $\ell$} \\
    \textcolor{blue}{$\Delta R(j_1,j_2)$} & \textcolor{blue}{Angular distance between $j_1$ and $j_2$}  &
    $\Delta R(\tau_\text{had},\ell)$ & Angular distance between $\tau_\text{had}$ and $\ell$ \\
    $\pt^\text{tot}$            & Vector sum of all visible momenta and $\pTmiss$ &
    $\sum p_\text{T}$           & Scalar sum of all visible momenta and $\pTmiss$ \\
    $C^{\text{miss}}_{\phi}$    & Azimuthal centrality of $\pTmiss$ w.r.t.\ $\ell,\tau_\text{had}$ &
    $C^{\ell}_{\eta}$          & Pseudorapidity centrality of the lepton w.r.t.\ the jets \\
    $\pt^{\ell}/\pt^{\tau_\text{had}}$ & Transverse–momentum ratio of lepton to $\tau_\text{had}$ \\
    \hline\hline
  \end{tabular}
  } 
\end{table*}
\subsection{Systematic uncertainties}
\label{sec:sysunc}
To reflect realistic sources of experimental uncertainty, each event in the FAIR-HUC dataset is subject to systematic variations governed by six predefined nuisance parameters, grouped into two categories:
\begin{itemize}
    \item \textbf{Calibration-type uncertainties} affect the primary features of final-state objects directly and propagate to derived features. These include the jet energy scale (JES), $\tau$ energy scale (TES), and the soft missing transverse energy (soft MET) scale. Concretely, the corresponding variations are applied to the reconstructed objects as
    \begin{align}
        P^{\text{biased}}_{\text{had}}&=\alpha_{\text{tes}} P_{\text{had}},\nonumber \\ 
        P^{\text{biased}}_{\text{jet}}&=\alpha_{\text{jes}} P_{\text{jet}}, \nonumber \\
        P^{\text{biased}}_{\text{MET}}&=P_{\text{MET}}+\left( \begin{array}{c}
\text{Gauss} (0,\alpha_{\text{met}}) \\
\text{Gauss} (0,\alpha_{\text{met}}) \\
\end{array} \right) 
    \end{align}

    The JES and TES variations are parameterized by standard normal nuisance parameters, $\alpha_\text{jes}, \alpha_\text{tes} \sim \mathcal{N}(1,0.01)$, subject to the constraint $0.9 \leq \alpha_\text{tes}, \alpha_\text{jes} \leq 1.1$, as prescribed by the FAIR-HUC benchmark. The MET scale variation is controlled by a log-normal nuisance parameter $\alpha_\text{met}$ with mean 0 and unit width in log-space, constrained to the range $[0,5]$. For all three sources, the FAIR-HUC organizers provide code to modify both primary and derived features as a function of these nuisance parameters.

    \item \textbf{Normalization-type uncertainties} affect only the event weights of background processes. These include a global background normalization factor (\(\alpha_\text{bkg}\)) and process-specific modifiers for the \ttbar (\(\alpha_\text{$t\overline{t}$}\)) and diboson (\(\alpha_\text{VV}\)) processes. All three nuisance parameters follow Gaussian distributions centered at 1, with standard deviations \(\sigma_\text{bkg} = 0.001\), \(\sigma_\text{$t\overline{t}$} = 0.02\), and \(\sigma_\text{VV} = 0.25\), and are clipped to the ranges specified by FAIR-HUC $[0.99, 1.01]$, $[0.8, 1.2]$ and $[0, 2]$, respectively. Their effect is implemented via rescaling of the event weights:
    \[
    w_\text{$Z\rightarrow{\tau\tau}$}^{\text{biased}} = \alpha_\text{bkg} \times w_\text{$Z\rightarrow{\tau\tau}$}, \quad 
    w_\text{$t\overline{t}$}^{\text{biased}} = \alpha_\text{bkg} \times \alpha_\text{$t\overline{t}$} \times w_\text{$t\overline{t}$}, \quad 
    w_\text{VV}^{\text{biased}} = \alpha_\text{bkg} \times \alpha_\text{VV} \times w_\text{VV}.
    \]
These transformations preserve the kinematic shapes of the background processes while altering their relative and total contributions, following the parameter definitions from the FAIR-HUC~\cite{Bhimji:2024bcd}.
\end{itemize}

The central values, prior widths, and clipping intervals quoted above are adopted from the FAIR-HUC benchmark~\cite{Bhimji:2024bcd}. The prior widths specify the assumed typical variations, while the clipping intervals bound the allowed parameter values; the latter should not be interpreted as one-standard-deviation ranges. For example, the JES and TES prior widths are 1\%, whereas their allowed interval, \([0.9,1.1]\), extends to ten prior standard deviations on either side of the nominal value. We retain these prescribed settings to evaluate the method under the FAIR-HUC uncertainty model.

Together, these six nuisance parameters describe the calibration and background-normalization variations prescribed by FAIR-HUC. The former affect primary and derived kinematic features, while the latter rescale background event weights. Differential theoretical uncertainties are not included in this benchmark. Within this prescribed uncertainty model, the impact of nuisance variations on the signal-strength fit is quantified in Sec.~\ref{sec:scan} and validated via Monte Carlo studies in Sec.~\ref{sec:toy}.

\section{Dual-Branch Graph Network Architecture}
\label{sec:model}

\subsection{Overall Model Architecture and Workflow}

To enable robust signal strength estimation under systematic uncertainties, we develop a graph-based classification model that processes each event using a dual-branch architecture. The model takes the full set of 35 features as input. These features capture the essential information of final-state particles and event topology, and are structured into a physics-motivated graph representation. We adopt a GNN for three practical reasons. First, collider events are intrinsically relational: discrimination depends on both pairwise correlations (e.g., $\Delta R$, invariant masses) and higher-order (three-object or more) structures---e.g., the azimuthal centrality of $p_T^{\text{miss}}$ with respect to $(\ell,\tau_{\text{had}})$---which message passing captures naturally. Second, the graph representation is permutation-invariant—message passing with symmetric (sum/mean) aggregation makes the prediction invariant to the arbitrary indexing of nodes within the event graph—and tolerant to missing objects via masking, providing robustness to varying jet multiplicities and threshold effects without ad-hoc ordering. Third, the node/edge/global split offers clean attachment points for systematics: calibration shifts on single objects or object pairs map naturally to nodes/edges, whereas calibration effects that couple three or more objects—handled here as ``global'' uncertainties—are injected as event-level context, and their influence propagates in a controlled, physically interpretable way.

By decoupling the processing of deterministic and uncertainty-sensitive features, the dual-branch design allows the model to leverage both sources of information for signal-background discrimination, while also ensuring a smooth and robust dependence of its predictions on nuisance parameter variations. An architectural ablation study presented in Appendix~\ref{Robustness-Architecture} further shows that this design provides a balanced solution, achieving robustness without sacrificing classification performance. A dedicated fusion module combines the learned representations from both branches, and the model is trained on systematically perturbed samples to ensure stable and controlled classifier responses in the downstream likelihood profiling. In the following, we describe the input representation in Section~\ref{sec:input}, the deterministic branch in Section~\ref{sec:det}, the uncertainty-aware branch in Section~\ref{sec:unc}, and the fusion module in Section~\ref{sec:fusion}.

\begin{figure}[bht!]
    \centering
    \includegraphics[width=1.1\textwidth]{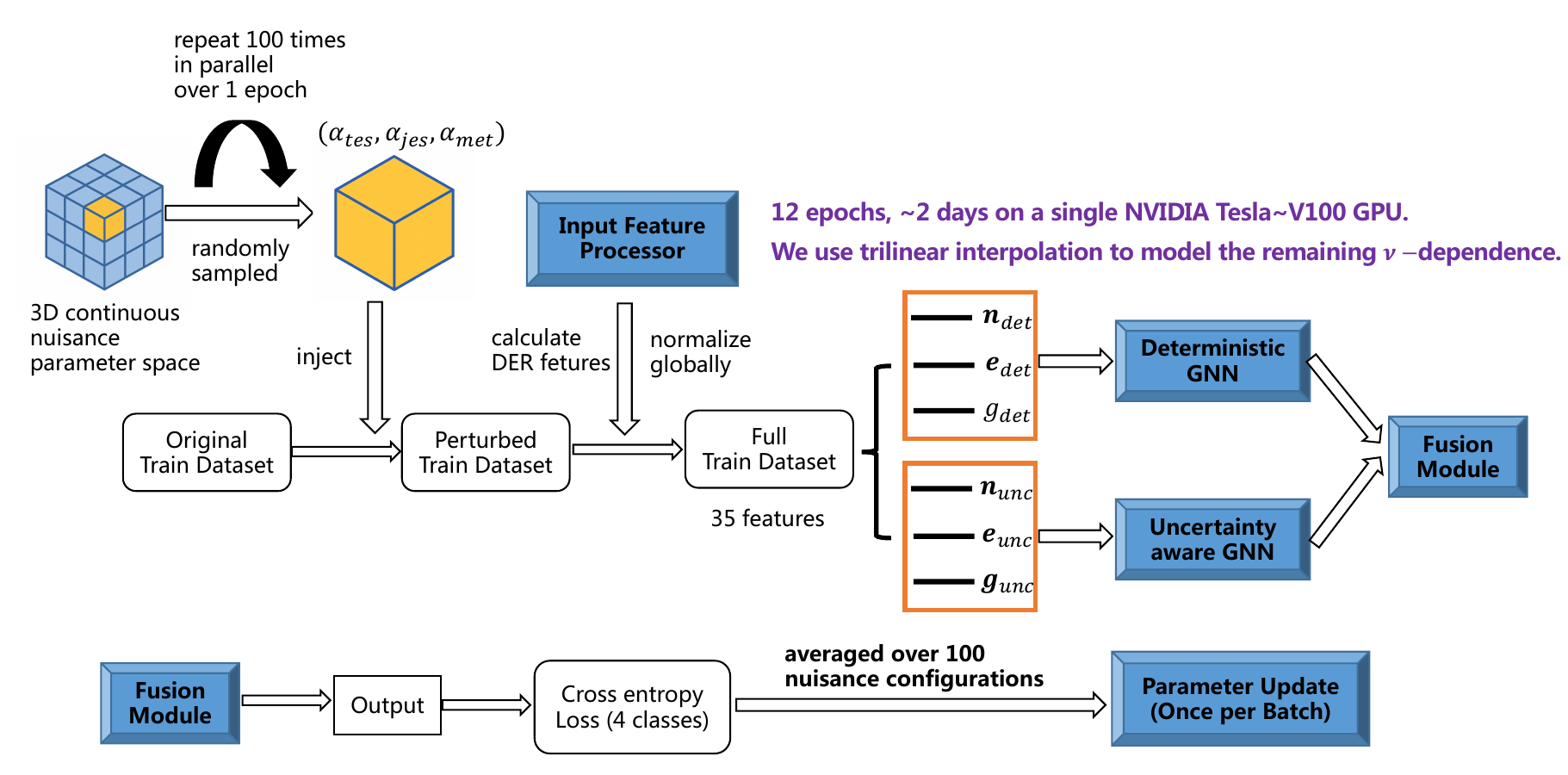}
    \caption{Illustration of the overall model architecture and training strategy.}
    \label{fig:overall}
\end{figure}

\begin{table}[t]
  \centering
  \setlength{\tabcolsep}{12pt}
  \caption{Event statistics of the FAIR-HUC training dataset. The raw event counts are the total numbers of simulated events provided in the dataset, which are subsequently split 90/10 into training and validation subsets. The event yields are the sums of per-event physical weights, \textit{i.e.} their expected number based on process cross section and luminosity.}
  \label{tab:event-stats}
  \begin{tabular}{lrr}
    \hline
    \textbf{Process} & \textbf{Raw event count} & \textbf{Event yield} \\
    \hline
    $H\!\to\!\tau\tau$ & 49450685 & 966.0 \\
    $Z\!\to\!\tau\tau$ & 90113695 & 901137.5 \\
    $t\overline t$      & 4128325  & 41283.4 \\
    VV                  & 343294      & 3433.5 \\
    \hline
  \end{tabular}
\end{table}

The training procedure, illustrated in Figure~\ref{fig:overall}, is designed to incorporate systematic uncertainties in a physically meaningful and computationally efficient manner. The full training set provided by FAIR-HUC is used without any event selection, and a randomly selected 10$\%$ subset is reserved for validation. Classification is performed with uniform event weighting, i.e., every event is assigned the same weight irrespective of its class. The raw and weighted event counts for all processes are summarized in Table~\ref{tab:event-stats}, where it is noted that the event counts are not based on luminosity times cross section and are not balanced across classes. Importantly, although the total dataset includes six nuisance parameters (three calibration-based and three normalization-based), only the three calibration-type parameters---the $\tau$ energy scale (TES), the jet energy scale (JES) and the soft MET scale---are used in the perturbation procedures. This choice reflects the fact that the training data itself does not contain normalization weights. The training dataset has a shape of $(N, 17)$, consisting of 16 primary features and one label. All 19 derived features (DER) used in the graph model are computed entirely on the fly, immediately after applying nuisance perturbations to the relevant primary (PRI) features, rather than being precomputed.

The training strategy incorporates stochastic sampling over nuisance parameters to marginalize systematic effects during training. At the beginning of each epoch, 100 distinct nuisance parameter vectors are sampled from their respective prior distributions (see Section~\ref{sec:unc} for details). For each training batch, the nominal events are duplicated $R=100$ times, called replicas, with each replica perturbed by one of the sampled nuisance configurations. The deterministic GNN branch processes the nominal features only once, while the uncertainty-aware branch simultaneously and independently processes all 100 perturbed variants in parallel. The two branches are merged via a gated fusion module, and the final prediction is obtained using a four-class cross-entropy loss, enabling the model to separate the signal from each of the three major SM background classes and to discriminate among the background classes. Denoting the model output for replica $r$ of event $i$ as $p^{(i,r)}$ and the one-hot target as $y^{(i)}$, the training objective is
\begin{equation}
\label{eq:loss}
\mathcal{L}=\frac{1}{BR}\sum_{i=1}^{B}\sum_{r=1}^{R}\left(-\sum_{c=1}^{4} y^{(i)}_{c}\,\log p^{(i,r)}_{c}\right),
\end{equation}
where $B$ is the batch size and $c$ indicates the signal/background class. For each training batch, the loss is computed for all 100 perturbed replicas of every event, and a single scalar loss value is obtained by averaging over all replicas and events in the batch. A single backward pass and parameter update is then performed using this averaged loss. This strategy encourages the model to produce predictions that are insensitive to nuisance-induced variations, while preserving discriminative power between signal and background across the relevant phase space. 

The model is optimized using the AdamW algorithm~\cite{loshchilov2019decoupled}, with a learning rate governed by a cosine annealing schedule characterized by a maximum iteration count of $T_{\text{max}} = 20$, a minimum learning rate of $\eta_{\text{min}} = 10^{-5}$, and a weight decay factor of $10^{-4}$. Training is carried out for 12 epochs, requiring approximately two days on a single NVIDIA Tesla V100 GPU, largely driven by the 100 times $3\cdot 10^{8}$ events dataset size. For validation, at the start of each epoch we draw a single nuisance-parameter vector at random and evaluate the loss once at that fixed configuration. Overall, the training strategy emphasizes robustness by integrating out systematic effects through stochastic averaging, supporting the construction of a robust model with \emph{reduced and controlled sensitivity} to nuisance fluctuations---rather than full decorrelation---and suitable for downstream statistical inference tasks such as maximum likelihood estimation and coverage evaluation.

To connect the trained classifier with the statistical inference pipeline, we introduce an inclusive region for extracting the signal strength and three control regions for constraining background normalization nuisance parameters, with precise definitions given in Sec.~\ref{sec:scan}. The downstream goal of the GNN is to turn the signal-class output probability $p_{H\rightarrow{\tau\tau}}$ into a binned observable per analysis region and then perform likelihood inference. Concretely, for each region we histogram $p_{H\rightarrow{\tau\tau}}$ using region-specific adaptive binning. Using simulated predictions at fixed calibration shifts, we precompute a region-wise interpolation table $T_R(\alpha_{\mathrm{tes}},\alpha_{\mathrm{jes}},\alpha_{\mathrm{met}})$ that maps any nuisance configuration to process-wise binned yields $\{s_{b,R}, z_{b,R}, t_{b,R}, v_{b,R}\}$. The surrogate likelihood is then formed as a product of Poisson terms over all bins: in the inclusive region the expectation includes the signal scaled by $\mu$ plus all background components, whereas in the control regions only backgrounds enter. We profile the nuisances to obtain $\hat\mu$ and extract confidence intervals from the profiled NLL curve. Full details of the downstream pipeline (adaptive binning, interpolation tables, likelihood construction, and profiling) are given in Section~\ref{sec:scan}.

\subsection{Input Representation}
\label{sec:input}

To enable efficient learning from collider events, we transform each event into a graph-structured representation that captures both the properties of individual particles and their kinematic correlations. The input representation consists of three components: node features $\bf{n}$, edge features $\bf{e}$, and global features $\bf{g}$, encompassing a total of 35 features. These features comprise both deterministic quantities and those affected by systematic variations. To account for their distinct characteristics, the two types are processed separately within the architecture. While the present study focuses on $H\!\to\!\tau\tau$ final states, the graph-based representation is particularly advantageous for more complex topologies and higher object multiplicities, where permutations of identical particles and correlations among them become more relevant.

Each event is encoded as a graph with five nodes, corresponding to the final-state objects: lepton, tau-jet ($\tau_{\text{had}}$), leading jet ($j_1$), subleading jet ($j_2$), and missing transverse momentum (MET). For each node, we construct two types of features: deterministic and uncertainty-aware. The deterministic node features include the lepton transverse momentum $p_T$ together with pseudorapidity $\eta$ and azimuthal angle $\phi$ of all the visible final objects, while uncertainty-aware node features consist of quantities sensitive to the three calibration-type nuisance parameters. These include the transverse momentum of the $\tau$ jet (affected by $\alpha_\text{tes}$), the transverse momentum of the two QCD jets (affected by $\alpha_\text{jes}$), and both the transverse momentum and azimuthal angle of the MET (affected by $\alpha_\text{met}$). For each node, the coordinates that do not belong to its current branch are set to zero. For example, the deterministic GNN branch zero-pads the tau-jet $p_T$ slot, whereas the uncertainty-aware branch zero-pads the tau-jet $\eta$ and $\phi$ slots. And each node is additionally tagged with a five-dimensional one-hot vector to indicate its identity. Table~\ref{tab:node_features} summarizes the structure and assignments of node features $\bf{n}$.

To characterize the spatial and kinematic relationships between final-state objects, we define edge features for each valid pair of nodes. These include angular separations as well as physics-motivated variables such as the invariant mass $m_{j_1j_2}$ (when both jets are present) and the transverse mass $m_\text{T}(\ell,\pTmiss)$. Edge features are classified as deterministic or uncertainty aware depending on whether they are affected by systematic variations; for example, $m_{j_1j_2}$ depends on the jet energy scale, while $\Delta R(j_1,j_2)$ is deterministic. If one or both particles in a pair are missing, the corresponding edge is simply omitted. Table~\ref{tab:edge_features} provides a complete list of all edge features  $\bf{e}$ and their assignments.

In addition to node and edge features, we extract seven global features $\bf{g}$ constructed from three or more final-state objects, such as the vector sum of all visible momenta and $\pTmiss$ or the pseudorapidity centrality of the lepton with respect to the jets. These features are partitioned into deterministic and uncertainty-aware subsets according to their dependence on systematically affected quantities. The definitions of global features $\bf{g}$ are given in Table~\ref{tab:global_features}.

To ensure numerical stability and consistency across nominal and systematically perturbed inputs, all features are standardized using global mean and standard deviation statistics computed once on the training set. These statistics are then frozen and reused throughout: for both the nominal and systematics-perturbed datasets, during both training and validation. Each feature dimension is normalized independently (channel-wise) using its mean and standard deviation of the nominal training dataset. For derived features that depend on jet availability, we construct three variants of the feature matrix (both jets present, leading-only, or neither). Each variant includes only the computable quantities, which are then standardized using the same global training-set statistics (not variant-specific ones). Based on these standardized tensors, we form the deterministic and uncertainty-sensitive node/edge/global inputs that feed the two GNN branches.

\begin{table}[htbp]
\centering
\caption{Node feature assignment for final-state objects. Each node is represented by a one-hot encoding indicating its type. Deterministic and uncertainty-aware features are assigned separately according to their sensitivity to systematic variations.}
\label{tab:node_features}
\footnotesize
\renewcommand{\arraystretch}{1.6} 
\setlength{\tabcolsep}{8pt}
\begin{tabular}{|l|l|l|l|l|}
\hline
\textbf{Node Type} & \textbf{Feature Type} & \textbf{Variables Used} & \textbf{Feature Role} & \textbf{Node Encoding} \\
\hline
\textbf{Lepton}         & Deterministic     & $p_T$, $\eta$, $\phi$         & Fully specified         & [1, 0, 0, 0, 0] \\
\hline
\multirow{2}{*}{\textbf{Tau-jet}}        & Deterministic     & $\eta$, $\phi$                & Spatial info          & [0, 1, 0, 0, 0] \\
                        & Uncertainty-aware & $p_T$                         & Affected by $\alpha_\text{tes}$         & [0, 1, 0, 0, 0] \\
\hline
\multirow{2}{*}{\textbf{Leading Jet}}    & Deterministic     & $\eta$, $\phi$                & Spatial info            & [0, 0, 1, 0, 0] \\
                        & Uncertainty-aware & $p_T$                         & Affected by $\alpha_\text{jes}$           & [0, 0, 1, 0, 0] \\
\hline
\multirow{2}{*}{\textbf{Subleading Jet}} & Deterministic     & $\eta$, $\phi$                & Spatial info            & [0, 0, 0, 1, 0] \\
                        & Uncertainty-aware & $p_T$                         & Affected by $\alpha_\text{jes}$           & [0, 0, 0, 1, 0] \\
\hline
\textbf{MET}            & Uncertainty-aware & $p_T$, $\phi$                 & Affected by $\alpha_\text{met}$      & [0, 0, 0, 0, 1] \\
\hline
\end{tabular}
\end{table}

\begin{table}[htbp]
\centering
\caption{Edge features constructed between pairs of final-state objects. Features are categorized as deterministic or uncertainty-aware based on their sensitivity to systematic variations.}
\label{tab:edge_features}
\footnotesize
\renewcommand{\arraystretch}{1.6} 
\setlength{\tabcolsep}{8pt}
\begin{tabular}{|l|l|l|}
\hline
\textbf{Edge Type}        & \textbf{Feature Name}                         & \textbf{Feature Type} \\
\hline
Tau-jet – MET             & $\Delta R(\tau_\text{had}, \text{miss})$     & Uncertainty-aware     \\
\hline
\multirow{2}{*}{Lepton – MET}              & $\Delta R(\ell, \text{miss})$                & Uncertainty-aware     \\
                          & $m_\text{T}(\ell, \pt^\text{miss})$          & Uncertainty-aware     \\
\hline
Jet1 – MET                & $\Delta R(j_1, \text{miss})$                 & Uncertainty-aware     \\
\hline
Jet2 – MET                & $\Delta R(j_2, \text{miss})$                 & Uncertainty-aware     \\
\hline
Tau-jet – Jet1            & $\Delta R(\tau_\text{had}, j_1)$             & Deterministic         \\
Tau-jet – Jet2            & $\Delta R(\tau_\text{had}, j_2)$             & Deterministic         \\
\hline
Lepton – Jet1             & $\Delta R(\ell, j_1)$                         & Deterministic         \\
Lepton – Jet2             & $\Delta R(\ell, j_2)$                         & Deterministic         \\
\hline
\multirow{2}{*}{Jet1 – Jet2}                & $\Delta R(j_1, j_2)$                          & Deterministic         \\
                          & $m^{j_1j_2}$                                  & Uncertainty-aware     \\
\hline
\multirow{3}{*}{Tau-jet – Lepton}          & $\Delta R(\tau_\text{had}, \ell)$            & Deterministic         \\
                          & $\pt^{\ell}/\pt^{\tau_\text{had}}$           & Uncertainty-aware     \\
                          & $m_\text{vis}(\tau_\text{had}, \ell)$        & Uncertainty-aware     \\
\hline
\end{tabular}
\end{table}

\begin{table}[htbp]
\centering
\caption{Global features constructed from three or more final-state objects. Each feature is categorized by its defining particles and its sensitivity to systematic uncertainties.}
\label{tab:global_features}
\footnotesize
\renewcommand{\arraystretch}{1.5}
\setlength{\tabcolsep}{8pt}
\begin{tabular}{|l|l|l|}
\hline
\textbf{Particles Involved} & \textbf{Feature Name} & \textbf{Feature Type} \\
\hline
All QCD jets & $N_{j}$ & Uncertainty-aware \\
\hline
Tau-jet, Lepton, Jet1, Jet2, MET & $\pt^\text{tot}$ & Uncertainty-aware \\
\hline
Tau-jet, Lepton, MET & $\pt^{H}$  & Uncertainty-aware \\
\hline
Tau-jet, Lepton, Jet1, Jet2, MET & $\sum p_\text{T}$  & Uncertainty-aware \\
\hline
Tau-jet, Lepton, MET & $C^{\text{miss}}_{\phi}$  & Uncertainty-aware \\
\hline
All QCD jets & $\sum_{\text{jets}}\!p_\text{T}$ & Uncertainty-aware \\
\hline
Lepton, Jet1, Jet2 & $C^{\ell}_{\eta}$  & Deterministic \\
\hline
\end{tabular}
\end{table}

\subsection{Deterministic GNN Branch}
\label{sec:det}

\begin{figure}[bht!]
    \centering
    \includegraphics[width=1.1\textwidth]{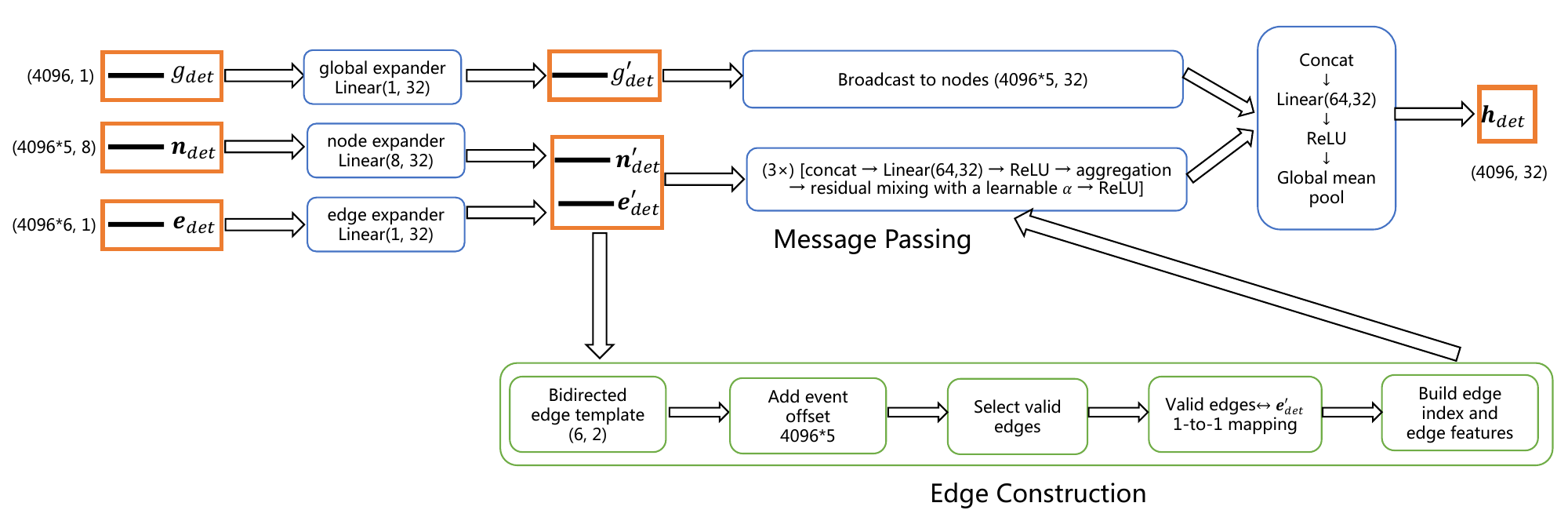}
    \caption{Illustration of the deterministic GNN architecture.}
    \label{fig:dgnn}
\end{figure}

The deterministic branch is designed to process features that are unaffected by systematic variations. It follows a graph-based message passing architecture, where encoded node features interact via predefined edge connections, and are further modulated by global event-level information. The architecture of the deterministic GNN branch is illustrated in Figure~\ref{fig:dgnn}. The input to this branch comprises three components: deterministic node features $\bf{n}_{det}$, deterministic edge features $\bf{e}_{det}$, and a single deterministic global feature $g_{det}$. Each event graph contains five nodes, and the node features are structured as vectors of dimension 8, consisting of three physical quantities ($p_T$, $\eta$, $\phi$) and a five-dimensional one-hot encoding that identifies the particle type. When flattened across the batch, the input node tensor has shape $(20480, 8)$, corresponding to 4096 events with five nodes each. Among the 15 physical quantities, only nine deterministic features are assigned non-zero values, while all other entries are intentionally zero-padded to reserve fixed positions for uncertainty-aware features. In addition to the node features, the input includes six deterministic edge features and one deterministic global feature. 

The 8-dimensional node vectors are passed through a linear node encoder to obtain per-node hidden representations. In parallel, each of the six deterministic edge features is expanded via a linear projection to a shared embedding space and assigned to the corresponding bidirected edge, as defined by a fixed edge template $\mathcal{T} \in \mathbb{Z}^{6 \times 2}$ that is consistently applied across all events. To accommodate batched processing, the same template is applied across all events, and event-wise node offsets, multiples of five corresponding to the number of nodes per graph, are added to correctly index nodes in the flattened batch. The graph construction module builds the edge index and attributes for each event based on a set of valid bidirected edges, conditioned on the presence or absence of the leading and subleading jets. Invalid edges involving missing nodes are automatically excluded.

The encoded node and edge features are processed through three message passing layers. In each layer $\ell = 1,2,3$, messages from sender node $i$ to receiver node $j$ are computed as:
\begin{equation}
    \mathbf{m}_{i \to j}^{(\ell)} = \text{ReLU} \left( \mathbf{W}_\ell^{\text{msg}} \left[ \mathbf{h}_i^{(\ell-1)} \, \| \, \mathbf{e}_{ij} \right] + \mathbf{b}_\ell^{\text{msg}} \right),
\end{equation}
where $\|\,$ denotes vector concatenation; $\mathbf{h}_i^{(\ell-1)}$ is the node-$i$ embedding from the previous layer, $\mathbf{e}_{ij}$ is the encoded edge feature for node pair $(i,j)$, $\mathbf{W}_\ell^{\text{msg}}$ and $\mathbf{b}_\ell^{\text{msg}}$ are learnable parameters at layer $\ell$. At each node, the messages from its connected nodes are summed together:
\begin{equation}
    \mathbf{a}_j^{(\ell)} = \sum_{i \in \mathcal{N}(j)} \mathbf{m}_{i \to j}^{(\ell)},
\end{equation}
and combined with the previous node states via residual mixing
\begin{equation}
    \mathbf{h}_j^{(\ell)} = \text{ReLU} \left( \alpha \cdot \mathbf{h}_j^{(\ell-1)} + (1 - \alpha) \cdot \mathbf{a}_j^{(\ell)} \right),
\end{equation}
where $\alpha$ is a learnable scalar gate, shared across nodes and layers, and hard-projected to $[0,1]$ at each forward pass.

To incorporate global context, the global deterministic feature is passed through a separate linear encoder and broadcasted to all five nodes per graph. The node states are updated by concatenating them with the broadcasted global embedding and applying another linear transformation followed by ReLU activation. Finally, the updated node features are aggregated across all nodes within each event by taking their element-wise mean, a procedure known as global mean pooling. This operation converts a variable-size set of node representations into a single fixed-size graph-level embedding, resulting in a tensor of shape (4096, 32) per batch. This vector, $\bf{h}_{det}$, serves as the deterministic summary of each event, capturing spatial, kinematic, and topological correlations among visible particles through message passing.

\subsection{Uncertainty-aware GNN Branch}
\label{sec:unc}

\begin{figure}[bht!]
    \centering
    \includegraphics[width=1.1\textwidth]{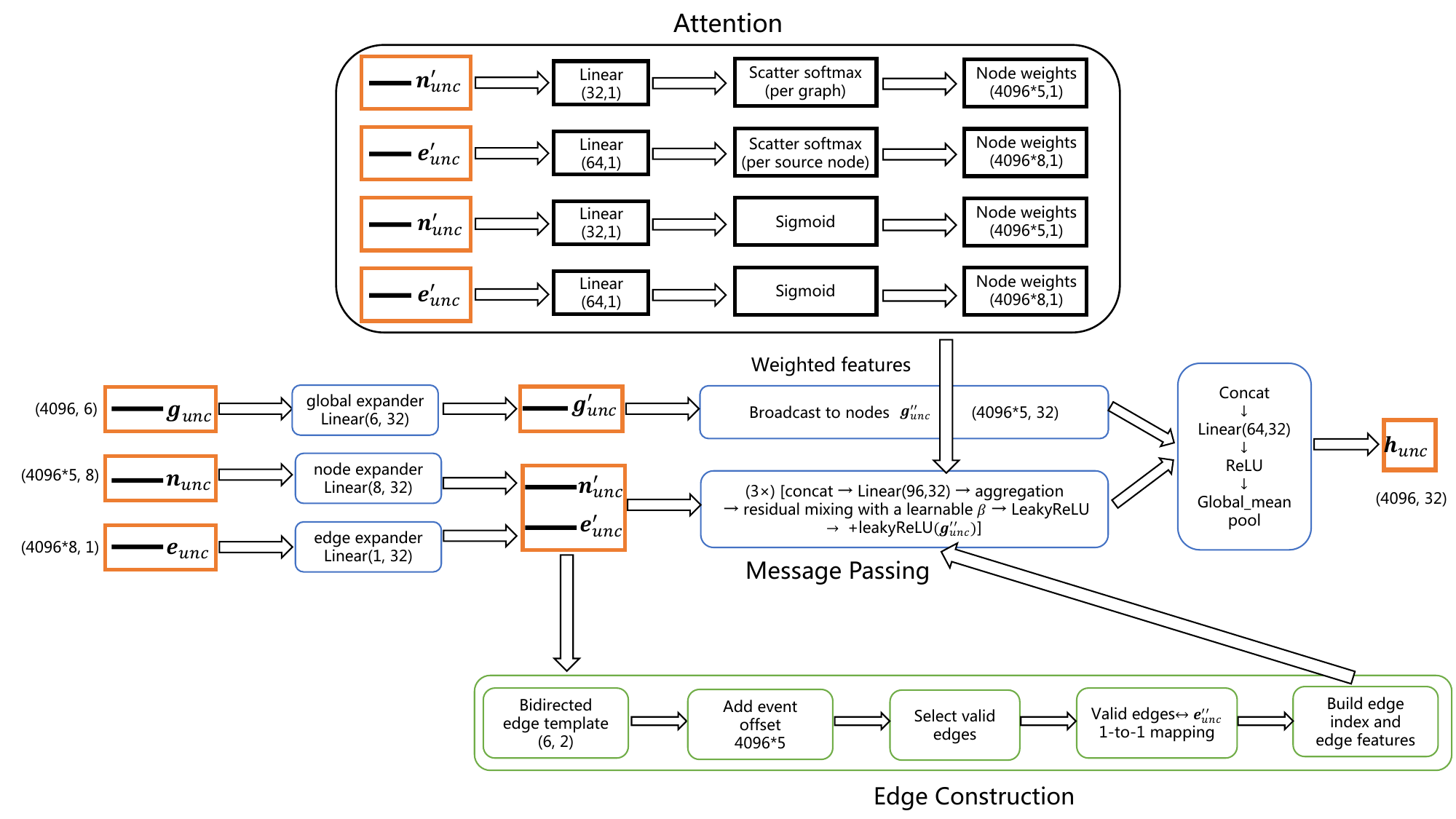}
    \caption{Illustration of the uncertainty-aware GNN architecture.}
    \label{fig:ugnn}
\end{figure}

The uncertainty-aware branch is designed to process node features $\bf{n}_{unc}$, edge features $\bf{e}_{unc}$, and global features $\bf{g}_{unc}$ that are sensitive to systematic variations. Conceptually, this branch mirrors the deterministic one in its graph-based message passing structure, but incorporates additional attention and gating mechanisms that enable adaptive modulation of the information flow. This design is introduced only in the uncertainty-aware branch, because the deterministic inputs are fixed quantities, whereas the uncertainty-aware inputs explicitly encode features that vary under systematic perturbations and therefore require adaptive weighting. This design is consistent with our training objective, which averages the cross-entropy over $R{=}100$ nuisance replicas per event and thus averages out fluctuations in the predictions under systematic shifts. The architecture of the uncertainty-aware GNN branch is illustrated in Figure~\ref{fig:ugnn}. The input to this branch consists of three components: uncertainty-aware node features of shape $(20480, 8)$, uncertainty-aware edge features of shape $(32768, 1)$, corresponding to eight bidirected edges per event across a batch of 4096 events, and global uncertainty-aware features of shape $(4096, 6)$, representing six global features per event. Each event graph contains five nodes, and among the 15 physical quantities assigned to nodes, only five uncertainty-aware variables are given non-zero values while all other entries are zero-padded to maintain alignment with the deterministic input structure. Edge construction reuses the same logic as the deterministic branch, with masking based on jet availability to select valid edges and index offsets applied to correctly locate nodes in the flattened batch. 

At the beginning of forward propagation, the 8-dimensional node feature vectors are passed through a linear encoder followed by LeakyReLU activation, producing the initial node embeddings. These embeddings are then used to compute scalar attention scores via a linear layer of shape $(32, 1)$, followed by softmax normalization within each event. In parallel, a separate linear layer of shape $(32, 1)$ with sigmoid activation is applied to each node embedding to produce node-specific gates that modulate the contribution of self-information during aggregation. For edges, each of the eight scalar edge features is first projected into a shared 32-dimensional embedding space via a linear transformation with LeakyReLU activation. Meanwhile, the 6-dimensional global feature vector for each event is passed through a linear encoder into a 32-dimensional representation, which is then broadcasted to all five nodes in the corresponding event graph. For each edge, its embedding is concatenated with the corresponding global context vector associated with the sender node, forming a 64-dimensional input. This is fed into two separate linear layers: one of shape $(64, 1)$ and softmax normalization to compute edge attention weights, and another with shape of $(64, 1)$ and sigmoid activation to compute learnable edge gates that modulate the information flow during message passing.

To perform uncertainty-aware message passing, the model uses the previously computed attention and gating weights to construct and propagate messages across the event graph. For each edge $(i \to j)$, the modulated sender node representation and edge embedding are computed as:
\begin{align}
h_i^{\text{mod}} &= a_i \cdot g_i \cdot h_i \\
e_{ij}^{\text{mod}} &= a_{ij} \cdot g_{ij} \cdot e_{ij}
\end{align}
where $a_i$ and $a_{ij}$ denote the attention weights for node $i$ and edge $(i,j)$, respectively, and $g_i$ and $g_{ij}$ represent the corresponding gating weights. The message is then formed by concatenating $h_i^{\text{mod}}$, $e_{ij}^{\text{mod}}$, and the global context $g_e$ of the corresponding event (broadcasted to all nodes):
\begin{equation}
m_{ij} = \phi\left( \text{concat}(h_i^{\text{mod}}, e_{ij}^{\text{mod}}, g_e) \right)
\end{equation}
where $\phi$ denotes a linear transformation of shape $(96, 32)$ followed by LeakyReLU activation, which outputs learned message vectors. These modulated messages are then aggregated at the receiver nodes using a scatter-based summation operation over all incoming edges:
\begin{equation}
\hat{h}_j = \sum_{i \in \mathcal{N}(j)} m_{ij}
\end{equation}
To preserve existing node representations and stabilize training, the aggregated messages are combined with the previous node embeddings via residual mixing:
\begin{equation}
h_j^{\text{new}} = \text{LeakyReLU} \left( \beta h_j + (1 - \beta) \hat{h}_j \right)
\end{equation}
where $\beta \in [0, 1]$ is a learnable scalar parameter shared across the model and governs the balance between the new incoming messages and the previous hidden states. After the residual update, a LeakyReLU activation is applied. Additionally, the global feature embedding is added to the node representation at every layer to reinforce global event-level information.

The entire message passing process is repeated three times with independent edge encoders but a shared attention matrix, which enforces a consistent notion of importance across layers while still allowing the model to capture higher-order interactions among the final-state particles. After the final propagation step, the updated node representations are concatenated with their corresponding broadcasted global feature vectors and passed through a final linear transformation $(64, 32)$ to produce the final node states. To generate a per-event summary representation, global mean pooling is applied across the five nodes within each event. This yields a final uncertainty-aware embedding, $\bf{h}_{unc}$, of shape $(4096, 32)$, which encapsulates the structural, kinematic, and topological characteristics of each event, modulated by the underlying systematic variations.

\begin{figure}[bht!]
    \centering
    \includegraphics[width=\textwidth]{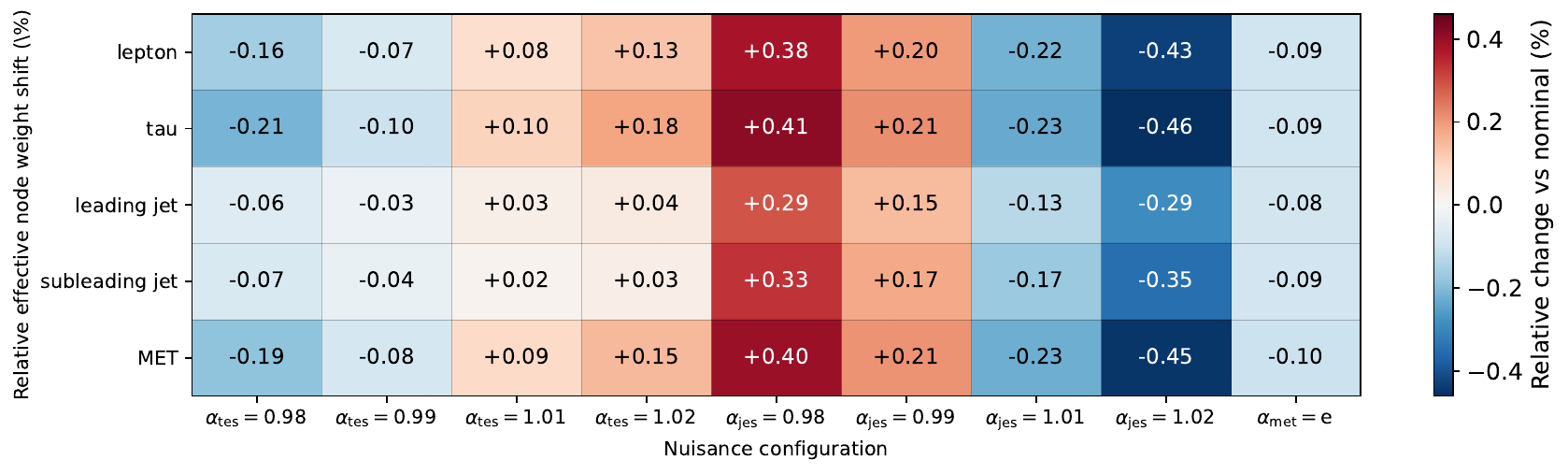}
    \includegraphics[width=\textwidth]{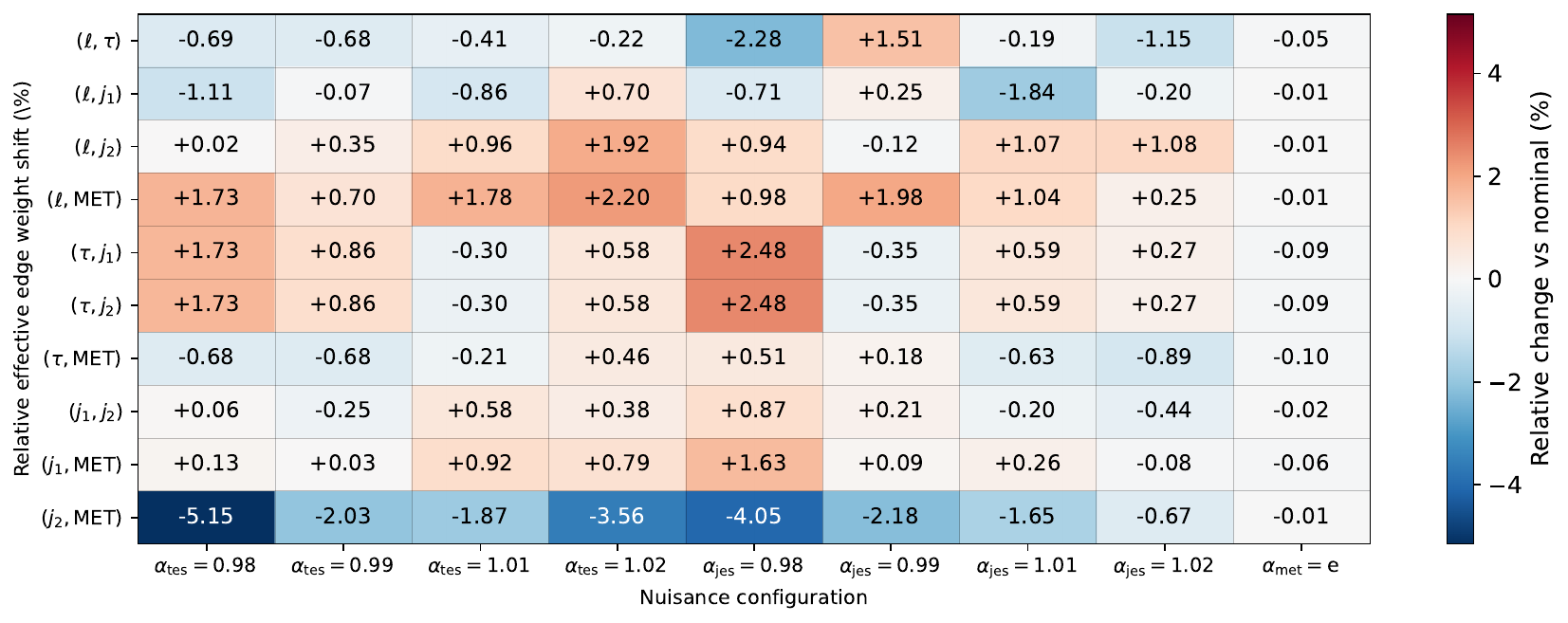}
    \caption{Relative changes of effective node- and edge-level weights in the uncertainty-aware GNN branch with respect to the nominal nuisance configuration. The upper (lower) panel shows node-level (edge-level) shifts under individual nuisance variations. The color scale indicates the relative change in percent.}
    \label{fig:heatmap}
\end{figure}

Moreover, to further illustrate how the uncertainty-aware attention mechanism responds to systematic variations, we visualize the relative changes of the effective node- and edge-level weights under representative nuisance configurations, with respect to the nominal setting. To this end, a nominal toy dataset and nine systematically perturbed toy datasets are constructed from the same nominal configuration, each containing events from one signal and three background processes. The effective node and edge weights are defined as
\(a_i \cdot g_i\) and \(a_{ij} \cdot g_{ij}\), respectively, corresponding to the attention and gating weights entering the modulated message construction. The heatmaps shown in Fig.~\ref{fig:heatmap} display the relative changes of these effective weights, averaged over all events in each dataset, and provide a diagnostic view of how the Uncertainty-aware GNN redistributes its focus across final-state objects and their pairwise relations under nuisance perturbations. At the node level, the variations remain moderate in magnitude but exhibit a coherent and systematic dependence on the nuisance parameters. In particular, the effective node weights show opposite responses under $\tau$ and jet energy scale variations, indicating a non-monotonic, nuisance-dependent internal response. In contrast, MET-related perturbations lead to a uniform suppression of node effective weights, attributable to the stochastic nature of the MET variation. At the edge level, the effective weights display a more pronounced and structured nuisance dependence than at the node level. Edges involving QCD jets, the hadronic $\tau$ jet, and MET exhibit the largest variations under calibration nuisances, while lepton-related edges remain comparatively stable. As observed at the node level, MET-related perturbations induce an overall mild suppression of edge effective weights, consistent with their stochastic treatment. Overall, these observations show that the uncertainty-aware attention and gating mechanisms induce a structured, nuisance-dependent redistribution of importance across both nodes and edges of the event graph, rather than acting as uniform reweighting factors. This behavior is consistent with the physical origin of the considered systematic uncertainties and supports the interpretability and robustness of the Uncertainty-aware GNN branch.
\subsection{Fusion Module}
\label{sec:fusion}

\begin{figure}[bht!]
    \centering
    \includegraphics[width=\textwidth]{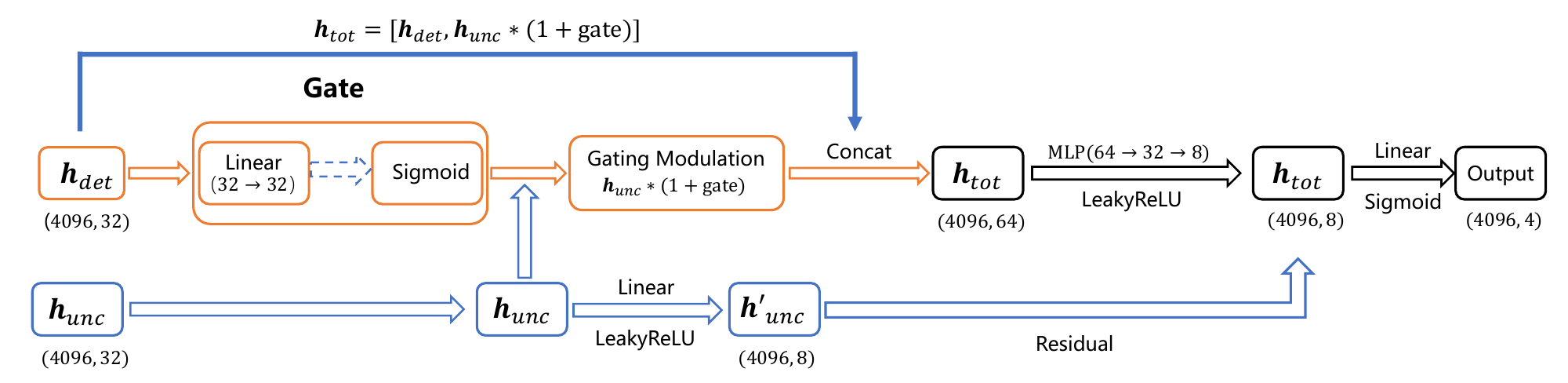}
    \caption{Illustration of the final fusion module.}
    \label{fig:fusion}
\end{figure}

The fusion module combines the output embeddings produced by the deterministic $\bf{h}_{det}$ and uncertainty-aware $\bf{h}_{unc}$ branches and transforms the joint representation into the final prediction, as illustrated in Figure~\ref{fig:fusion}. Each branch outputs a 32-dimensional feature vector for every event in the batch. These two vectors are first processed by a gating mechanism that adaptively modulates the uncertainty-aware embedding using information from the deterministic representation. This design is motivated by the observation that deterministic features are typically more stable and reliable, making them well-suited to guide the modulation of uncertainty-sensitive information in a data-driven and physically informed manner. Specifically, the deterministic feature vector is passed through a linear layer of shape (32, 32), followed by a sigmoid activation, producing a gate vector with shape (4096, 32) in the range of [0, 1]. This gate acts as a set of soft, learnable scaling factors applied element-wise to the uncertainty-aware embedding, which is then shifted by adding 1 to each component—thereby enhancing it according to the deterministic context. 
\begin{equation}
\bf{h}_{\text{unc\_enhanced}} \Rightarrow \bf{h}_{\text{unc}} \times (1+\text{gate})
\end{equation}
This operation enhances each component of the uncertainty-aware vector in a way that depends on the corresponding deterministic information.

These transformations allow the model to capture nonlinear interactions between the two branches and extract higher-level features for the final prediction. To preserve direct information from the uncertainty-aware branch, a residual connection is introduced. The original 32-dimensional uncertainty-aware vector is projected to 8 dimensions through a separate linear layer with LeakyReLU, and added element-wise to the 8-dimensional output of the main path, yielding the final hidden representation. Finally, the resulting 8-dimensional vector is passed through a final linear layer to produce the model output, a 4-dimensional vector for multi-class classification. The output corresponds to raw logits; during training the softmax operation is implicitly applied within the cross-entropy loss, while during inference an explicit softmax function can be added to obtain class probabilities.

\subsection{Signal-Class Output Distributions under Nuisance Variations}

\begin{figure}[ht]
    \includegraphics[width=8cm,height=6cm]{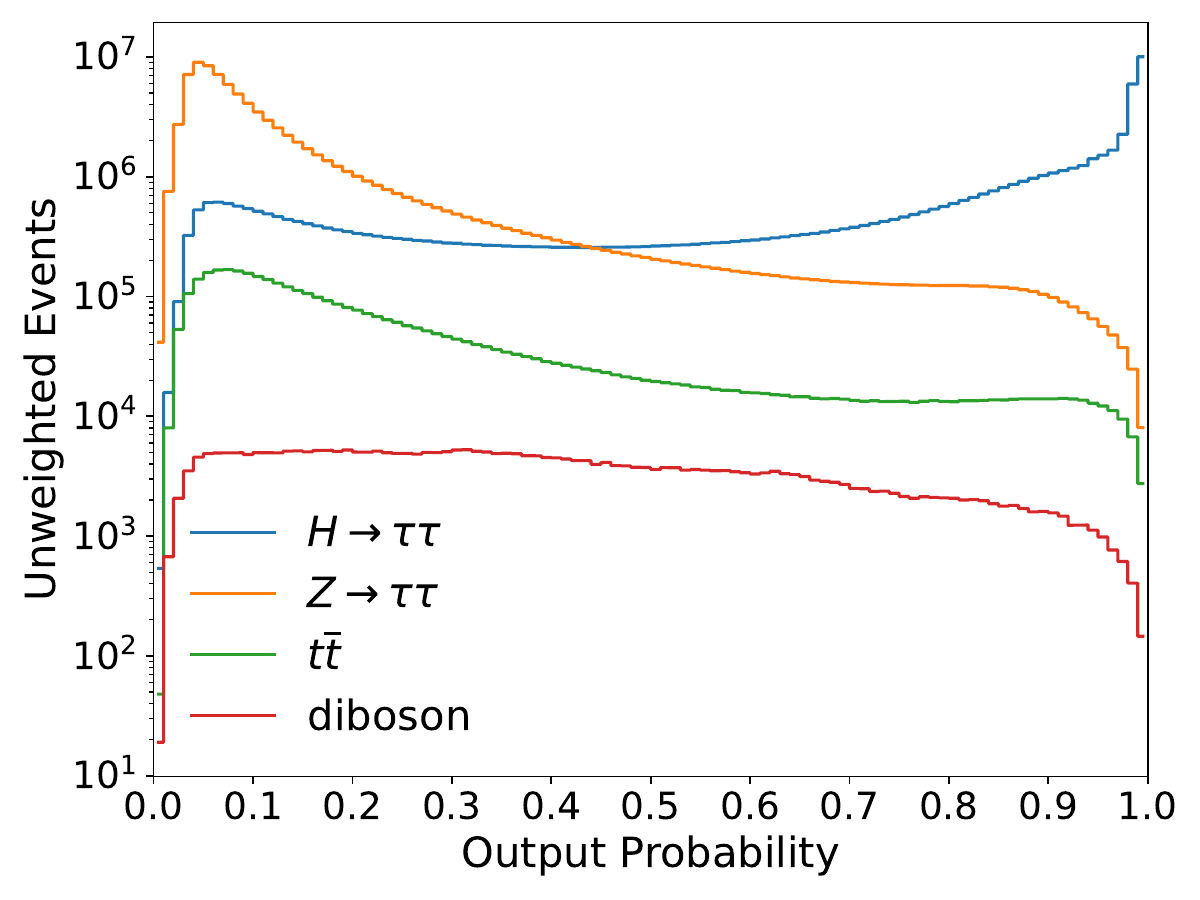}
    \includegraphics[width=8cm,height=6cm]{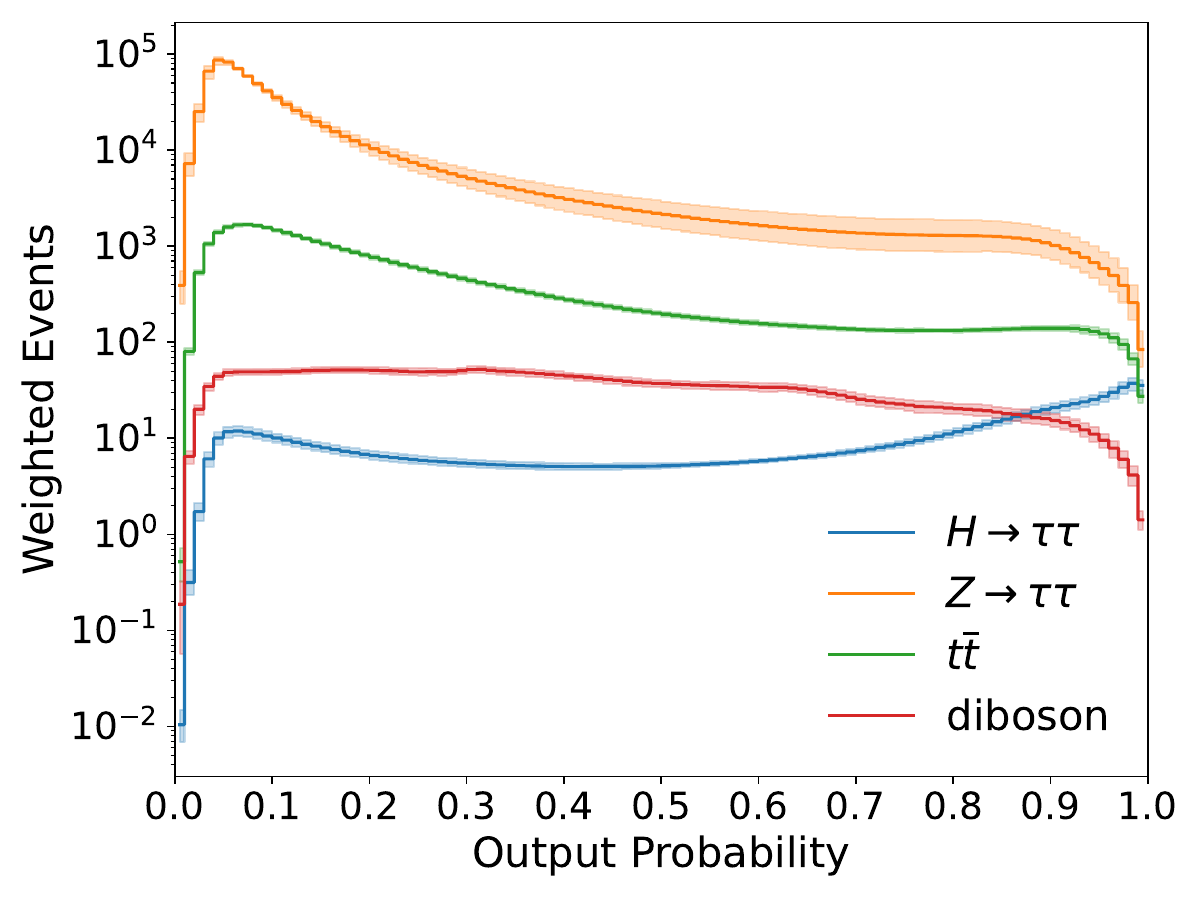}
    \caption{Distributions of the predicted signal-class probabilities for the four processes based on the output of the four-class classifier. The left panel shows the unweighted event distributions, while the right panel includes the expected event yields after applying cross-section-based event weights. In the weighted plot, the solid lines indicate the average count of events per bin evaluated across a grid of 11849 fixed nuisance-parameter settings, uniformly spanning $\alpha_{\text{tes}}\in[0.96,1.04]$, $\alpha_{\text{jes}}\in[0.96,1.04]$, and $\alpha_{\text{met}}\in[0,5]$. The shaded bands correspond to the envelope formed by the minimum and maximum bin counts over these variations.}
    \label{fig:signal_dist}
\end{figure}

To evaluate the four-class classifier's behavior under systematic uncertainties, we examine the distribution of predicted signal-class probabilities for all four physical processes. Figure~\ref{fig:signal_dist} shows these distributions both without and with yield weighting. In the left panel, each simulated event counts once (unweighted), directly representing the classifier output for each event. It highlights the intrinsic separation between signal and background classes achieved by the model: signal events ($H\rightarrow{\tau\tau}$) tend to cluster near high probability values, while background processes including $t\overline{t}$, $Z\rightarrow{\tau\tau}$, and diboson production dominate the low-probability region.

In the right panel, each simulated event is assigned a per-event weight such that the total of each process equals its expected yield (cross section $\times$ luminosity); thus the curves represent expected event counts rather than raw sample frequencies. The horizontal axis is the classifier's predicted probability for the signal class, with larger values corresponding to more signal-like events. Systematic variations are evaluated on a three-dimensional grid in $(\alpha_{\text{tes}},\alpha_{\text{jes}},\alpha_{\text{met}})$ with size $17\times17\times41$, corresponding to $17\cdot17\cdot41=11849$ fixed nuisance-parameter configurations, uniformly covering $\alpha_{\text{tes}},\alpha_{\text{jes}}\in[0.96,1.04]$ and $\alpha_{\text{met}}\in[0,5]$; see Section~\ref{parametrisation} for construction details. For each bin of the output probability, the solid lines denote the average expected yield across all nuisance points, while the shaded bands show the full variation envelope bounded by the minimum and maximum bin counts over the parameter grid. The relatively narrow envelopes indicate that the classifier output distributions remain stable over the considered nuisance-parameter range. A direct comparison with models trained exclusively on nominal samples is presented below to assess more explicitly the effects of nuisance-aware training and of the dual-branch architecture.

To further investigate whether the separation of deterministic and uncertainty-sensitive information into two branches introduces an information bottleneck, and to isolate the role of nuisance-aware training, we consider two additional reference models. The first, denoted as ``Baseline + nominal training'', retains exactly the same dual-branch architecture as the Baseline model but is trained exclusively on nominal samples, without injecting nuisance variations during training. The second, denoted as ``Single GNN + nominal training'', removes the explicit deterministic/uncertainty-aware branch separation. Instead, all 35 physical input features are organized as node, edge, and global features of a single event graph and are processed jointly from the beginning of the message-passing stage. The graph contains the same five object nodes as the Baseline representation, while pairwise information is encoded on the valid object--object edges. The node representations are propagated through three message-passing layers with residual mixing, after which the global event features are combined with the node embeddings and a global mean pooling operation produces the event-level representation. This single-branch model contains no uncertainty-specific attention or inter-branch fusion module and, like the nominally trained dual-branch reference, is trained only on nominal samples.

\begin{figure}[ht]
    \centering
    \includegraphics[width=\textwidth]{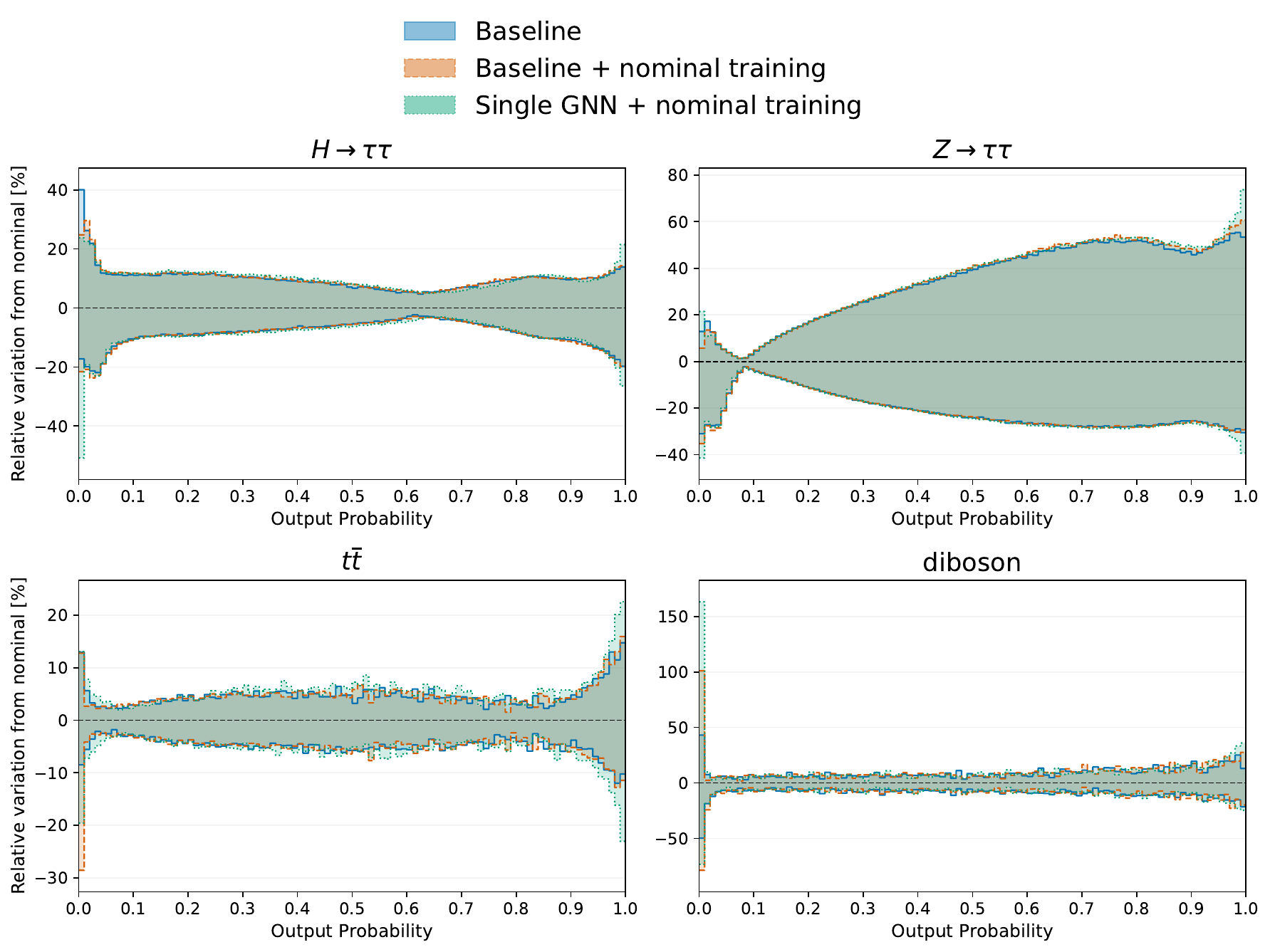}
    \caption{Relative variation of the weighted signal-class output distributions with respect to the nominal prediction for $H\rightarrow\tau\tau$, $Z\rightarrow\tau\tau$, $t\overline{t}$, and diboson events. The shaded envelopes show the minimum and maximum relative bin-yield variations over the same 11849 fixed configurations in $(\alpha_{\text{tes}},\alpha_{\text{jes}},\alpha_{\text{met}})$ for three models: the nuisance-aware dual-branch Baseline, the same dual-branch architecture trained only on nominal samples (Baseline + nominal training), and a single-branch GNN in which all 35 input features are processed jointly and which is also trained only on nominal samples (Single GNN + nominal training).}
    \label{fig:ratio_model_comparison}
\end{figure}

To compare the sensitivity of the three models to calibration uncertainties more directly, Figure~\ref{fig:ratio_model_comparison} shows the relative variation of the weighted event yield with respect to the nominal prediction for each physical process. For a given model, process, and classifier-output bin $b$, we define the upper and lower variation envelopes as
\begin{equation}
\delta^{\text{max/min}}_b
=
\frac{
N^{\text{max/min}}_b-N^{\text{nom}}_b
}{
N^{\text{nom}}_b
},
\end{equation}
where $N^{\text{nom}}_b$ is the bin yield at the nominal nuisance configuration $(\alpha_{\text{tes}},\alpha_{\text{jes}},\alpha_{\text{met}})=(1,1,0)$, and $N^{\text{max}}_b$ and $N^{\text{min}}_b$ are the maximum and minimum yields obtained over the same 11849 nuisance configurations used in Figure~\ref{fig:signal_dist}. The comparison is performed independently for the Baseline, Baseline + nominal training, and Single GNN + nominal training models.

As shown in Figure~\ref{fig:ratio_model_comparison}, the three configurations exhibit very similar responses to the nuisance variations over most of the classifier-output range, with their relative variation envelopes largely overlapping for all four processes. Nevertheless, a mild overall hierarchy can be observed in the systematic response: the nuisance-aware Baseline generally exhibits the narrowest variation envelope, followed closely by Baseline + nominal training, while Single GNN + nominal training tends to show slightly larger variations. The differences are small and occur mainly in the extreme score bins, where the event populations are comparatively small. This behavior is consistent with the corresponding Wasserstein-distance comparison presented in Appendix~\ref{Robustness-Architecture}, which likewise shows only modest differences among the three configurations, with the Baseline exhibiting the best overall stability under nuisance variations.

The comparison also shows that allowing deterministic and nuisance-sensitive information to interact jointly from the beginning of the graph message passing does not lead to an appreciable improvement in classifier performance. The nominal-test AUC values reported in Appendix~\ref{Robustness-Architecture} are 0.8883 for Baseline + nominal training and 0.8911 for Single GNN + nominal training, demonstrating that the two architectures provide essentially identical nominal discrimination power. Together with their closely comparable nuisance-response envelopes, these results indicate that processing all input information jointly within a single GNN does not provide a significant performance advantage over the dual-branch representation considered in this work.

\section{Signal Strength Profiling via Surrogate Likelihood}
\label{sec:scan}

\subsection{Region-wise Interpolation and Adaptive Binning}
\label{binning}

To extract the signal strength $\mu$, we construct a surrogate likelihood from binned event counts of the signal-class probability distributions across multiple analysis regions. For each region, we build a separate interpolation table that maps nuisance-parameter configurations to the process-wise binned yields. As summarized in Table~\ref{table:interpolation}, we consider four regions whose selection criteria are taken from Ref.~\cite{Benato:2025rgo}:

The \textbf{inclusive region} comprises all events that pass the baseline selection, without imposing any further cuts. It contains approximately 966 expected $H\rightarrow{\tau\tau}$ events, with a signal-to-background ratio around $10^{-3}$.

The \textbf{highMT-VBFJet region} (denoted as CR1 in Table~\ref{table:interpolation}) aims to select vector boson fusion (VBF) Higgs production but is depleted of signal due to the high transverse mass requirement. Events are required to have at least two jets with transverse momenta $\pt^{j_1} > 50\,\mathrm{GeV}$ and $\pt^{j_2} > 30\,\mathrm{GeV}$, mimicking the VBF topology. In addition, the transverse mass between the lepton and the missing transverse momentum must satisfy $\mT > 70\,\mathrm{GeV}$. This selection ensures that most signal-like events are removed, while retaining a substantial number of $t\overline{t}$ events, resulting in a signal-to-background ratio of $8.3 \times 10^{-4}$. The region is therefore strongly $t\overline{t}$ dominated and provides an effective constraint on the $\alpha_\text{$t\overline{t}$}$ nuisance parameter.

The \textbf{highMT-noVBFJet-tt} (denoted as CR2 in Table~\ref{table:interpolation}) is designed to enhance $t\overline{t}$-like events that are not already captured by the VBF selection. Events are selected to have high transverse mass ($\mT > 70\,\mathrm{GeV}$) and must veto the VBF topology. Furthermore, it requires that the class output probability of the four-class classifier $\hat p_{t\overline t}$ exceed 0.4. This requirement ensures that the selected events are strongly $t\overline{t}$-like in nature. In this region, the signal contribution is negligible (only 2.7 events), while the expected background yield from $t\overline{t}$ exceeds 3600, making it highly effective in constraining the $\alpha_\text{$t\overline{t}$}$ nuisance parameter.

The \textbf{highMT-noVBFJet-VV} (denoted as CR3 in Table~\ref{table:interpolation}) is constructed analogously to CR2, but instead focuses on isolating the diboson background. It also applies the high-$\mT$ cut and vetoes the VBF topology. In this region, the event selection requires that the diboson class output of the four-class classifier $\hat p_{VV}$ exceed 0.2. This yields a sample enriched in diboson events, with relatively suppressed contributions from signal and other backgrounds. The expected diboson yield is 597.4, compared to only 1.8 from the signal process and 207.6 from $t\overline{t}$. Although CR3 has the smallest total event count among all regions, its relatively high purity makes it useful for constraining the $\alpha_\text{VV}$ nuisance parameter.

\begin{table}[htbp]
  \footnotesize
  \centering
  \renewcommand{\arraystretch}{1.25}
  \setlength{\tabcolsep}{5pt}
  \begin{tabular}{|l|l|c|c|c|c|c|c|}
    \hline
    \multirow{2}{*}{\textbf{Region}} &
    \multirow{2}{*}{\textbf{Requirements}} &
    \multirow{2}{*}{\textbf{Type}} &
    \multicolumn{4}{c|}{\textbf{Poisson yield} $\mathcal{L}\sigma$} &
    \multirow{2}{*}{$S/B$} \\
    \cline{4-7}
    & & & $H\!\to\!\tau\tau$ & $Z\!\to\!\tau\tau$ & $t\overline t$ & VV & \\ \hline
    \texttt{inclusive} & -- & -- &
    966.0 & 901\,137.5 & 41\,283.4 & 3\,433.5 & $1.02\times10^{-3}$ \\ \hline
    \multirow{3}{*}{\texttt{highMT-VBFJet}} &
      $\pt^{j_1}\!>\!50$ GeV & \multirow{3}{*}{CR1} &
      \multirow{3}{*}{14.7} & \multirow{3}{*}{721.7} & \multirow{3}{*}{16\,768.6} & \multirow{3}{*}{193.2} & \multirow{3}{*}{$8.30\times10^{-4}$} \\ \cline{2-2}
    & $\pt^{j_2}\!>\!30$ GeV & & & & & & \\ \cline{2-2}
    & $\mT\!>\!70$ GeV      & & & & & & \\ \hline
    \multirow{3}{*}{\texttt{highMT-noVBFJet-tt}} &
      $\mT\!>\!70$ GeV & \multirow{3}{*}{CR2} &
      \multirow{3}{*}{2.7} & \multirow{3}{*}{202.7} &
      \multirow{3}{*}{3\,607.1} & \multirow{3}{*}{268.8} &
      \multirow{3}{*}{$6.62\times10^{-4}$} \\ \cline{2-2}
    & veto on \texttt{VBFJet} & & & & & & \\ \cline{2-2}
    & $\hat p_{t\overline t}\!>\!0.4$ & & & & & & \\ \hline
    \multirow{4}{*}{\texttt{highMT-noVBFJet-tt*}} &
      $\mT\!>\!70$ GeV & \multirow{4}{*}{CR2} &
      \multirow{4}{*}{4.3} & \multirow{4}{*}{384.7} &
      \multirow{4}{*}{2306.5} & \multirow{4}{*}{172.9} &
      \multirow{4}{*}{$1.50\times10^{-3}$} \\ \cline{2-2}
    & veto on \texttt{VBFJet} & & & & & & \\ \cline{2-2}
    & $\sum p_T\!>\!200$ GeV & & & & & & \\ \cline{2-2}
    & $p_T^H\!>\!29$ GeV & & & & & & \\ \hline
    
    \multirow{3}{*}{\texttt{highMT-noVBFJet-VV}} &
      $\mT\!>\!70$ GeV & \multirow{3}{*}{CR3} &
      \multirow{3}{*}{1.8} & \multirow{3}{*}{189.5} &
      \multirow{3}{*}{207.6} & \multirow{3}{*}{597.4} &
      \multirow{3}{*}{$1.8\times10^{-3}$} \\ \cline{2-2}
    & veto on \texttt{VBFJet} & & & & & & \\ \cline{2-2}
    & $\hat p_{VV}\!>\!0.2$ & & & & & & \\ \hline

    \multirow{4}{*}{\texttt{highMT-noVBFJet-VV*}} &
      $\mT\!>\!70$ GeV & \multirow{4}{*}{CR3} &
      \multirow{4}{*}{5.7} & \multirow{4}{*}{904.7} &
      \multirow{4}{*}{137.0} & \multirow{4}{*}{470.3} &
      \multirow{4}{*}{$3.4\times10^{-3}$} \\ \cline{2-2}
    & veto on \texttt{VBFJet} & & & & & & \\ \cline{2-2}
    & $\sum p_T\!<\!150$ GeV & & & & & & \\ \cline{2-2}
    & $p_T^H\!<\!20$ GeV & & & & & & \\ \hline

  \end{tabular}
  \caption{Summary of the event selections. Control regions are denoted by the “CR” label. Expected Poisson event yields, calculated as $\mathcal{L}\sigma$ with $\mathcal{L}$ representing the integrated luminosity, are provided for each process across different regions. The final column lists the corresponding signal-to-background ratios. Compared to Ref~\cite{Bhimji:2024bcd}, the total yields in the “inclusive” row are slightly reduced due to the stricter threshold requirements on $\pt^{\tau_\text{had}}$ imposed by the FAIR-HUC benchmark. $\hat p_{t\overline t}$ and $\hat p_{VV}$ refer to the predicted class probabilities for the $t\overline t$ and diboson processes, respectively, as assigned by the four-class classifier.}
  \label{table:interpolation}
\end{table}

 The expected Poisson event yields, computed as $\mathcal{L}\sigma$, span several orders of magnitude across different processes and regions. To accommodate the wide dynamic range and sparse occupancy in some bins, we adopt a region-specific adaptive binning strategy based on the distribution of the signal class output probability of the four-class classifier. The binning process begins with 100 uniformly spaced bins. The neighbouring bins are then iteratively merged in a greedy manner until each remaining bin contains at least ten effective weighted events from the $\PH \rightarrow \tau \tau$ signal process and all three major backgrounds: $\PZ\to\tau\tau$, $t\overline{t}$, and diboson. This procedure is applied independently for each region. As a result, the inclusive region ends up with 52 bins after merging, while CR1, CR2, and CR3 contain 14, 15, and 13 bins, respectively. This region-wise adaptive binning guarantees that the interpolated event yields are statistically meaningful and stable under systematic variations, while preserving the discriminating power in signal-sensitive regions. Additionally, in Appendix~\ref{sec:kin_CR}, we study the impact of defining the CR only based on kinematics and not using the classifier output $\hat p_{i}$.

\subsection{Parametrisation of Systematics and Event Rate Interpolation}
\label{parametrisation}

To ensure smoothness and avoid boundary effects in the likelihood surface, we reparameterize \(\boldsymbol{\alpha}\) in terms of an unconstrained vector of standard normal variables \(\boldsymbol{\nu} = (\nu_{\mathrm{tes}}, \nu_{\mathrm{jes}}, \nu_{\mathrm{met}}, \nu_{t\overline{t}}, \nu_{VV}, \nu_{\mathrm{bkg}})\), such that each component of \(\boldsymbol{\nu}\) follows a Gaussian prior centered at zero. The mapping from \(\boldsymbol{\nu}\) to \(\boldsymbol{\alpha}\) is defined as:
\[
\begin{aligned}
\alpha_{\mathrm{tes}} &= 1 + 0.01 \cdot \nu_{\mathrm{tes}}, \\
\alpha_{\mathrm{jes}} &= 1 + 0.01 \cdot \nu_{\mathrm{jes}}, \\
\alpha_{\mathrm{met}} &= e^{\nu_{\mathrm{met}}} - 1, \\
\alpha_{t\overline{t}} &= e^{\sigma_{t\overline{t}}\nu_{t\overline{t}}}, \\
\alpha_{VV} &= e^{\sigma_{\mathrm{VV}}\nu_{VV}}, \\
\alpha_{\mathrm{bkg}} &= e^{\sigma_{\mathrm{bkg}}\nu_{\mathrm{bkg}}}.
\end{aligned}
\]
where the prior widths are set to $\sigma_{t\overline{t}} = 0.02, \sigma_{\mathrm{VV}} = 0.25, \sigma_{\mathrm{bkg}} = 0.001$, reflecting the expected scale of variation in each process.

The statistical treatment of systematic uncertainties in this work is based on process-wise binned event yields, following the standard binned-likelihood approach used in LHC analyses, as implemented in statistical fitting frameworks such as CMS Combine~\cite{CMS:2024onh} and HistFactory~\cite{Cranmer:2012sba}. To propagate the effect of calibration uncertainties to the event yields, we construct dense three-dimensional interpolation grids in the physical space of $(\alpha_{\text{tes}}, \alpha_{\text{jes}}, \alpha_{\text{met}})$ using simulated predictions at fixed systematic shifts. Each region is associated with a table of shape \(17 \times 17 \times 41\), uniformly covering a hypercube extending to \(\pm4\sigma\) along each axis. At every grid point, we store the binned raw event yields for four individual processes: the signal (\(S_\text{raw}\)) and the three main backgrounds (\(Z_\text{raw}, t\overline{t}_\text{raw}, VV_\text{raw}\)). The total number of grid nodes per region is 11849.

It should be noted that while TES/JES/MET capture the dominant calibration-type variations in FAIR-HUC, the framework is not restricted to these three sources. Additional nuisance parameters can be incorporated depending on whether they modify kinematic shapes or act primarily as overall normalizations. Uncertainties whose effects are approximately uniform across the analysis regions and classifier-output bins, such as luminosity and genuinely normalization-only components of theoretical cross-section uncertainties, can enter the likelihood as multiplicative yield scalings without enlarging the interpolation grid. PDF and \(\alpha_s\) uncertainties, as well as renormalization- and factorization-scale variations, may also alter kinematic distributions and hence selection efficiencies, region populations, and classifier-score shapes, particularly in high-scale regions probing large Bjorken-\(x\). Given corresponding varied event weights or alternative simulated samples, the same event selection and trained classifier could be applied to each variation to construct process-specific binned score templates in every analysis region. The resulting rate and shape changes could then be interpolated and profiled as additional nuisance effects in the likelihood, preserving their correlations across bins and regions. Incorporating these variations into the uncertainty-aware training would additionally require introducing them into the training samples and revalidating the estimator. These theory variations are not included among the six FAIR-HUC nuisance parameters and are therefore outside the scope of the present numerical study. Given suitable variation templates, extending the grid-based surrogate to additional shape-type nuisances faces the usual curse of dimensionality, as a dense tensor grid scales as $\mathcal{O}(\prod_i n_i)$ and quickly becomes prohibitive beyond a few dimensions. In practice, this can be mitigated by exploiting the structure of shape-type effects. The response to each shape-type nuisance is always retained through a one-dimensional interpolation obtained by varying that parameter individually around the nominal point. Potential non-linear couplings are treated selectively: mixed terms are introduced only for nuisance pairs whose joint variation cannot be described by the sum of their individual interpolated responses, indicating a non-negligible interaction. In this way, all shape nuisances contribute at linear order, while higher-order effects are included only when necessary, avoiding the need for a fully dense high-dimensional grid. Within each retained low-dimensional subspace, dense tensor products can further be replaced by sparse-grid constructions, which achieve comparable interpolation accuracy for smooth response surfaces with substantially fewer evaluation points. Together, this combination provides a scalable and numerically controlled path to incorporating additional shape-type uncertainties. The dependence on grid density is further investigated in Appendix~\ref{sec:grid_density}, where we consider lighter configurations, including a medium grid ($9\times9\times21$) and a coarse grid ($5\times5\times11$).

Given a continuous nuisance configuration \(\boldsymbol{\nu}\), the corresponding physical scale factors \(\alpha\) are computed and used to interpolate event rates via trilinear interpolation:
\begin{equation}
(\alpha_{\text{tes}}, \alpha_{\text{jes}}, \alpha_{\text{met}}) \;\longrightarrow\;
\{ S_\text{raw},\; Z_\text{raw},\; t\overline{t}_\text{raw},\; VV_\text{raw} \},
\end{equation}
where \(S_\text{raw}, Z_\text{raw}, t\overline{t}_\text{raw}, VV_\text{raw}\) denote the interpolated templates for signal, $\PZ\to\tau\tau$, $t\overline{t}$, and diboson processes, respectively. This interpolation is performed independently for each analysis region, using region-specific bin edges defined by the adaptive binning procedure in Section~\ref{binning}.

The raw yields are then rescaled using the normalization parameters to construct the total expected counts in each bin. For the inclusive region, the expected count in bin \(b\) is given by:
\[
\begin{aligned}
\lambda_b^{\mathrm{full}}(\mu, \boldsymbol{\alpha}) =\ 
& \mu \cdot S_\text{raw}(\alpha_{\mathrm{tes}}, \alpha_{\mathrm{jes}}, \alpha_{\mathrm{met}})  +\ \alpha_{\mathrm{bkg}} \cdot Z_\text{raw}(\alpha_{\mathrm{tes}}, \alpha_{\mathrm{jes}}, \alpha_{\mathrm{met}})\\
& +\ \alpha_{t\overline{t}} \cdot \alpha_{\mathrm{bkg}} \cdot t\overline{t}_\text{raw}(\alpha_{\mathrm{tes}}, \alpha_{\mathrm{jes}}, \alpha_{\mathrm{met}}) 
+ \alpha_{VV} \cdot \alpha_{\mathrm{bkg}} \cdot VV_\text{raw}(\alpha_{\mathrm{tes}}, \alpha_{\mathrm{jes}}, \alpha_{\mathrm{met}}) ,
\end{aligned}
\]
In the three control regions, the signal yield is omitted, and only background components are retained:
\[
\lambda_b^{\mathrm{CR}}(\mu, \boldsymbol{\alpha}) = 
\alpha_{\mathrm{bkg}} \cdot Z_\text{raw}(\alpha_{\mathrm{tes}}, \alpha_{\mathrm{jes}}, \alpha_{\mathrm{met}})+\alpha_{t\overline{t}} \cdot \alpha_{bkg} \cdot t\overline{t}_\text{raw}(\alpha_{\mathrm{tes}}, \alpha_{\mathrm{jes}}, \alpha_{\mathrm{met}})
+ \alpha_{VV} \cdot \alpha_{bkg} \cdot VV_\text{raw}(\alpha_{\mathrm{tes}}, \alpha_{\mathrm{jes}}, \alpha_{\mathrm{met}}),
\]

Finally, the bin-wise predictions from all four regions are concatenated into a single event rate vector $\boldsymbol{\lambda}(\mu, \boldsymbol{\nu})$, which serves as the input to the likelihood function described in Section~\ref{profile}. This construction allows the likelihood to respond smoothly to variations in both the signal strength and the nuisance parameters, enabling reliable profiling and uncertainty estimation. Since using a fully parametrized classifier that explicitly depends on $\boldsymbol{\nu}$ would create a severe bottleneck during profiling—where the likelihood requires repeated evaluations across $(\mu,\boldsymbol{\nu})$—we instead adopt a single classifier that is valid across all nuisance configurations. To make this feasible, the classifier is trained to be robust against nuisance variations, such that any remaining $\boldsymbol{\nu}$-dependence can be handled efficiently through the interpolation tables.

\subsection{Construction of the Profile Likelihood}
\label{profile}
To extract the best-fit value of the signal strength parameter $\mu$ and quantify its associated uncertainty, we adopt the profile likelihood formalism. This approach enables the systematic treatment of nuisance parameters $\boldsymbol{\alpha}$, reparameterized in terms of $\boldsymbol{\nu}$, by profiling them out during inference. The profile likelihood is constructed by minimizing the negative log-likelihood (NLL) with respect to the nuisance parameters at fixed signal strength \(\mu\):
\begin{equation}
\mathrm{NLL}_\text{prof}(\mu) = \min_{\boldsymbol{\nu}}\, \mathrm{NLL}(\mu, \boldsymbol{\nu}).
\end{equation}

The full likelihood function is constructed as a product of independent Poisson terms over all analysis bins, together with Gaussian priors on each component of $\boldsymbol{\nu}$:
\begin{equation}
\mathcal{L}(\mu, \boldsymbol{\nu}) = \prod_{b=1}^{B} \mathrm{Poisson}(n_b(\boldsymbol{\nu}) \mid \lambda_b(\mu, \boldsymbol{\nu})) \cdot \prod_{i=1}^6 \mathcal{N}(\nu_i \mid 0, 1),
\end{equation}
where $n_b(\boldsymbol{\nu})$ is the observed count in bin $b$, and $\lambda_b(\mu, \boldsymbol{\nu})$ denotes the expected yield as defined in Section~\ref{parametrisation}. In log space, the NLL becomes:
\begin{equation}
\mathrm{NLL}(\mu, \boldsymbol{\nu}) =
\sum_{b=1}^{B} \left[ \lambda_b(\mu, \boldsymbol{\nu}) - n_b(\boldsymbol{\nu}) \log \lambda_b(\mu, \boldsymbol{\nu}) \right]
+ \frac{1}{2} \|\boldsymbol{\nu}\|^2 + \mathrm{const.}
\end{equation}

The second term penalizes large deviations of the nuisance parameters from their nominal values via a standard normal prior, enforcing a regularized fit.

\subsection{Signal Strength Estimation and Confidence Interval Extraction}
\label{CL}

To determine the most likely value of $\mu$, we first identify the global minimum of the profile NLL:
\begin{equation}
\hat\mu = \mathop{\mathrm{argmin}}_\mu \mathrm{NLL}_\text{prof}(\mu).
\end{equation}
Uncertainty estimation is then based on the profile likelihood ratio:
\begin{equation}
\Delta \mathrm{NLL}(\mu) = \mathrm{NLL}_\text{prof}(\mu) - \mathrm{NLL}_\text{prof}(\hat\mu),
\end{equation}
which serves as the test statistic for constructing confidence intervals. According to Wilks' theorem, $\Delta \mathrm{NLL}$ follows a $\chi^2$ distribution with one degree of freedom in the asymptotic regime. Therefore, the $68.3\%$ confidence interval on $\mu$ corresponds to $\Delta \mathrm{NLL}(\mu) \leq 0.5$.

The minimization over the nuisance parameters $\boldsymbol{\nu}$ at fixed signal strength $\mu$ is carried out using LBFGS optimizer in PyTorch with a strong Wolfe line search. For the first $\mu$ grid point, we obtain an initial seed via dual annealing and then refine it with LBFGS, while for all subsequent points we warm-start from the previous grid point’s profiled solution. To mitigate local minima, 5 additional LBFGS restarts from different initializations through random perturbations are performed and the best solution is retained. Box constraints are enforced by projecting the solution back to the allowed ranges after each LBFGS run. The bounds are $[-5,5]$ for all components except $\nu_{\text{met}}\in [\text{log}(1.001),\text{log}(6)]$, which ensures that $\alpha_{\text{met}}=e^{\nu_{\text{met}}}-1\in [0,5]$.

The core profile scan is performed on a uniform grid of 100 points in $\mu$ over $[0,3]$. At each grid point we profile $\boldsymbol{\nu}$ as described above, using warm starts to ensure a smooth and efficient profile curve. To provide stable gradients, the full likelihood evaluation is implemented in PyTorch with automatic differentiation. Once the profile curve is obtained, we refine the maximum-likelihood estimate by fitting a simple quadratic function to the seven points closest to the minimum. This gives a more accurate estimate of $\hat{\mu}$ between the grid points. We then \emph{re-profile} the nuisance parameters at this sub-grid $\hat{\mu}$ to obtain the true minimum $\mathrm{NLL}_\text{min}$ and the corresponding profiled parameters.

To determine the $68.3\%$ confidence interval, we solve for the roots of
\begin{equation}
\Delta \mathrm{NLL}(\mu) = \mathrm{NLL}_\text{prof}(\mu) - \mathrm{NLL}_\text{min} = 0.5,
\label{eq:deltaNLL}
\end{equation}
using Brent’s method applied to the scan grid with on-the-fly re-profiling of $\boldsymbol{\nu}$ at each function evaluation. In cases where a valid bracket is not found, cubic spline or quadratic fits are used as fallbacks to ensure stability. 

\section{Performance Validation with Pseudo-Experiments}
\label{sec:toy}

To quantitatively assess the statistical performance of our profile likelihood framework, we conduct three complementary studies: an Asimov study (Section~\ref{asimov}) and two sets of Monte Carlo studies (Sections~\ref{fix} and \ref{random}). The Asimov study uses the Asimov dataset to present the profiled likelihood curve for $\mu$ and the corresponding impact ranking of nuisance parameters, providing a baseline validation of the inference pipeline and parameter sensitivities. The two Monte Carlo studies are designed to evaluate the coverage, interval, and parameter estimation related to the extracted signal strength. The first set of studies (Section~\ref{fix}) is based on 9000 pseudo-experiments generated at nine fixed signal strengths \(\mu^\text{true} = 0.3, 0.6, \ldots, 2.7\), with 1000 pseudo-experiments per point. This setup allows for a direct evaluation of the coverage and the interval width associated with the constructed confidence intervals. The second set (Section~\ref{random}) involves 50000 pseudo-experiments with signal strengths sampled uniformly from a continuous range [0, 3]. For each pseudo-experiment, we extract the best-fit values of both \(\mu\) and the six nuisance parameters. Together, these studies provide a rigorous and comprehensive validation of the likelihood-based inference procedure under both controlled and fully randomized conditions, and can be performed efficiently in our framework, with the full inference for a single pseudo-experiment taking about 5 minutes on a CPU and 2 minutes on an NVIDIA Tesla V100 GPU. All validation studies in this section use the FAIR-HUC setup and its six prescribed nuisance parameters. The reported coverage is therefore conditional on this uncertainty model.

\subsection{Asimov Study}
\label{asimov}

\textbf{Asimov dataset construction.}
In our surrogate-likelihood setup, four Asimov datasets are built, one for each analysis region (inclusive, CR1, CR2, CR3), by querying the region-wise interpolation tables at the nominal calibration point
\[
\boldsymbol{\alpha}_\star \equiv (\alpha_{\text{tes}},\alpha_{\text{jes}},\alpha_{\text{met}})=(1,1,0),
\]
For each process \(k\in\{H\!\to\!\tau\tau,\,Z\!\to\!\tau\tau,\,t\overline t,\,VV\}\) and region \(R\),
we obtain the expected per-bin yields in the signal-class probabilities based on the classifier—using the adaptive binning of Section~\ref{binning}—directly from the interpolation tables (Section~\ref{parametrisation}) evaluated at \(\boldsymbol{\alpha}_\star\).
Let \(S_{b,R}(\boldsymbol{\alpha}_\star)\), \(Z_{b,R}(\boldsymbol{\alpha}_\star)\),
\(t\overline{t}_{b,R}(\boldsymbol{\alpha}_\star)\), and \(VV_{b,R}(\boldsymbol{\alpha}_\star)\) denote the resulting histograms for the signal and three backgrounds, respectively.
The Asimov bin contents are then defined by summing the process-wise histograms (i.e., without statistical fluctuations) as
\[
n^{\text{A}}_{b,\text{inclusive}}(\mu_A)
= \mu_A\, S_{b,\text{inclusive}}(\boldsymbol{\alpha}_\star)
  + Z_{b,\text{inclusive}}(\boldsymbol{\alpha}_\star)
  + t\overline{t}_{b,\text{inclusive}}(\boldsymbol{\alpha}_\star)
  + VV_{b,\text{inclusive}}(\boldsymbol{\alpha}_\star),
\]
and, for the three control regions (where the signal template is omitted by construction),
\[
n^{\text{A}}_{b,R} = Z_{b,R}(\boldsymbol{\alpha}_\star) + t\overline{t}_{b,R}(\boldsymbol{\alpha}_\star) + VV_{b,R}(\boldsymbol{\alpha}_\star),
\qquad R\in\{\text{CR1, CR2, CR3}\}.
\]
We take \(\mu_A=1\) and set all normalization-type nuisance parameters to their nominal values when forming the Asimov dataset.

\textbf{Profiled likelihood on Asimov.}
Using the four region-wise Asimov histograms, we profile the likelihood as described in Section~\ref{CL}.
For each grid point we minimize the NLL to obtain $\mathrm{NLL}_\mathrm{prof}(\mu)=\min_{\boldsymbol{\nu}}\mathrm{NLL}(\mu,\boldsymbol{\nu})$, and form the (asymptotic) profile-likelihood test statistic, which differs from Eq.~\eqref{eq:deltaNLL} only by a factor of 2:
\begin{equation}
\label{eq:qmu-asimov}
q_\mu(\mathcal{D}_\mathrm{A}) \;=\; 2\Big[\mathrm{NLL}_\mathrm{prof}(\mu)-\mathrm{NLL}_\mathrm{prof}(\hat\mu)\Big],
\end{equation}
where $\hat\mu$ is the best fit. Under Wilks’ theorem for one degree of freedom, $q_\mu$ is approximately $\chi^2_1$-distributed, so the intersections with $q_\mu=1,4,9,16,25$ correspond to the asymptotic $1\sigma$–$5\sigma$ levels.

\begin{figure}
    \centering
    \includegraphics[width=0.6\linewidth]{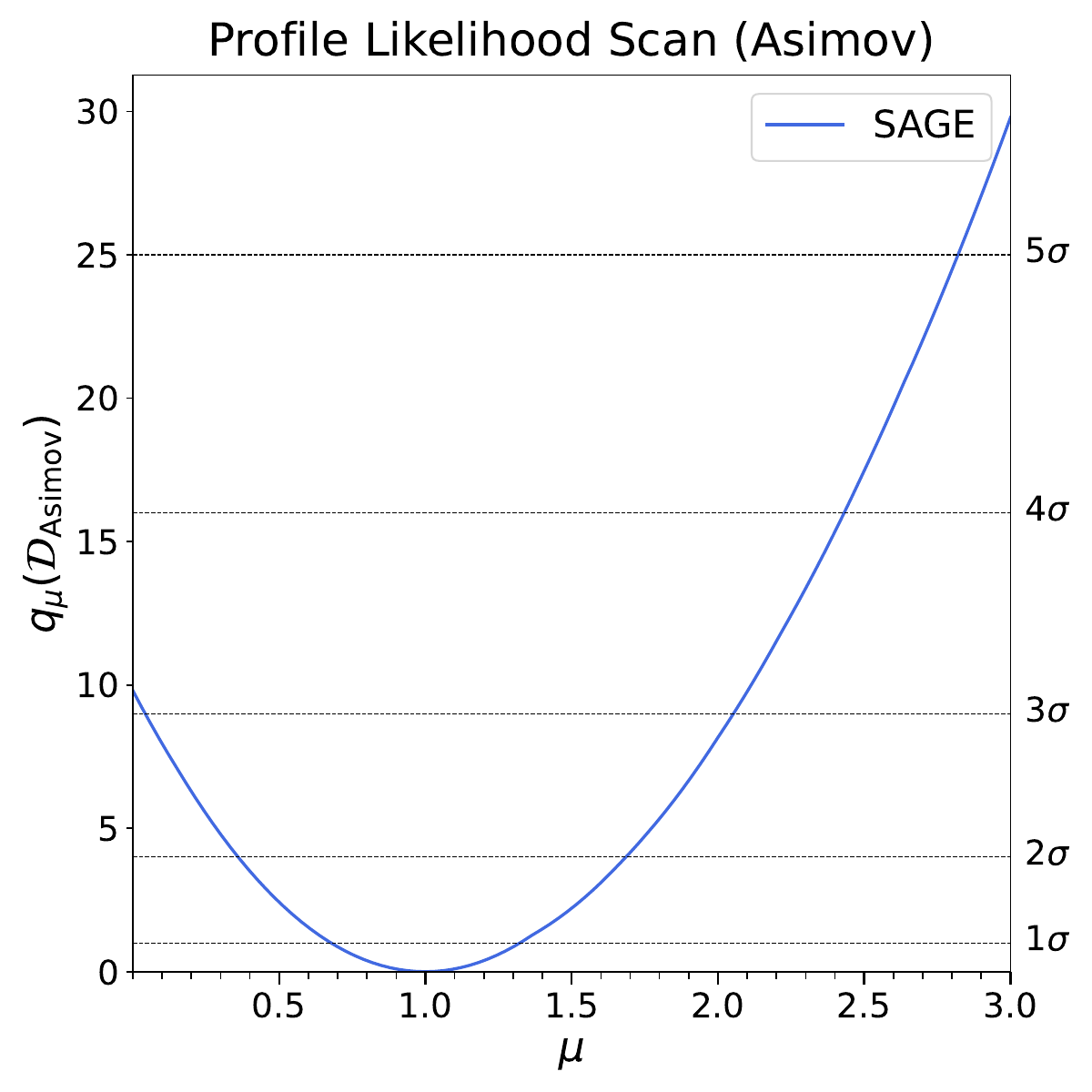}
    \caption{Asimov dataset fit scan of the signal strength parameter $\mu$, with the profile likelihood ratio $q_\mu$ shown as a function of $\mu$.}
    \label{fig:asimov-profile}
\end{figure}

Figure~\ref{fig:asimov-profile} shows the resulting $q_\mu(\mathcal{D}_\mathrm{A})$ curve.
As expected, the minimum occurs near $\mu\simeq 1$. We observe no noticeable bias in the location of the minimum: the profiled fit returns $(\hat{\mu},\hat{\boldsymbol{\nu}})=(1,\mathbf{0})$ with a numerical deviation below $10^{-3}$.
Repeating the construction for injected values $\mu_A\in[0.1,3]$ and re-fitting confirms this behavior.

\textbf{Postfit (Asimov).}
We start from the Asimov best-fit point $(\hat{\mu},\hat{\boldsymbol{\nu}})$ obtained at $\boldsymbol{\alpha}_\star=(1,1,0)$ with $\mu_A=1$.
For each nuisance $\nu_i$, the \emph{pre-fit} uncertainty is fixed to one standard deviation by the unit-width standard-normal prior.
The \emph{post-fit} $1\sigma$ interval is then determined on the Asimov dataset by scanning $\nu_i$ around its best-fit value while profiling all other parameters; the two endpoints are defined by the condition that the profiled likelihood increases by one unit in the usual $\Delta q=1$ convention. These posterior intervals are consistently narrower than the unit pre-fit widths, reflecting the information gained from the four regions.

\textbf{Nuisance-parameter impacts (Asimov).}
Impacts on the signal strength are evaluated using the \emph{posterior} $1\sigma$ (not the prior).
For each $\nu_i$ we freeze it at $\hat{\boldsymbol{\nu}}_i\pm\sigma_i^{\text{post}}$, re-profile all remaining nuisance parameters together with $\mu$, and record the corresponding shifts in the best-fit $\mu$.
The impact of $\nu_i$ is reported as the larger of the absolute shifts from the two variations.
For the one-sided \text{met} prior, only the $+1\sigma$ variation is considered both for defining the post-fit interval and for the impact.

\begin{figure}
    \centering
    \includegraphics[width=0.8\linewidth]{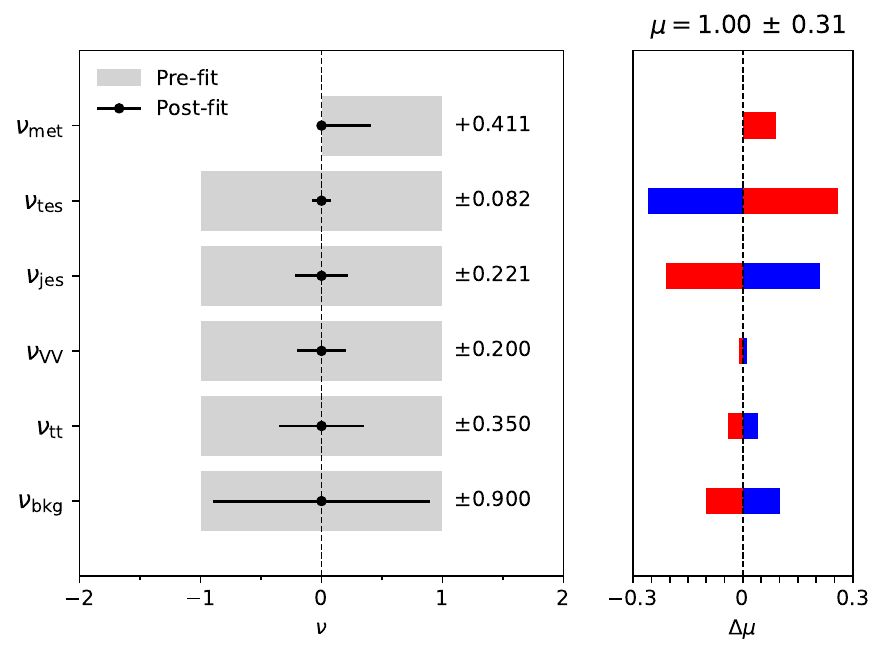}
    \caption{Post-fit and impact plot showing the effect of each nuisance parameter $\nu_i$ on the fitted value of $\mu$, including both pre-fit (shaded bands) and post-fit (error bars) uncertainties. The red (blue) bars on the right quantify the upward (downward) shift in $\mu$ when $\nu_i$ is varied by $+1\sigma$ ($-1\sigma$). For the soft MET nuisance parameter, the variation is shown as one-sided, reflecting the parameterization and bounded domain used in the likelihood profiling.}\label{fig:asimov-impacts}
\end{figure}

Figure~\ref{fig:asimov-impacts} shows the pre-fit (unit-width bands) and post-fit ($1\sigma$ error bars from the Asimov profile) for each nuisance parameter, together with the impacts computed as described above.
Calibration nuisance parameters ($\nu_\text{tes}$ and $\nu_\text{jes}$) are significantly constrained in the post-fit, and normalization nuisance parameters ($\nu_{t\overline{t}}$, $\nu_\text{VV}$ and $\nu_\text{bkg}$) also exhibit nontrivial reductions; the soft MET nuisance parameter $\nu_{\text{met}}$ appears one-sided, reflecting the parameterization and bounded domain used in the likelihood profiling. Overall, the ranking reflects the constraints provided jointly by the inclusive region and the three control regions in the surrogate-likelihood fit. Ordering the six nuisance parameters by absolute impact on $\mu$ reveals a clear hierarchy: $\nu_\text{tes}$ and $\nu_\text{jes}$ dominate, followed by $\nu_{t\overline{t}}$ and $\nu_\text{VV}$, while $\nu_\text{bkg}$ and the one-sided $\nu_\text{met}$ have the smallest impacts.

\subsection{Coverage Evaluation at Fixed Signal Strengths}
\label{fix}

To evaluate the coverage and interval width of our inference framework, we conduct a large-scale Monte Carlo study based on 9000 pseudo-experiments. Each pseudo-experiment is generated from one of nine fixed signal strengths,
\(\mu^\text{true} = 0.3, 0.6, \ldots, 2.7\), with 1000 pseudo datasets produced for each value. For every pseudo-experiment, we independently sample the six nuisance parameters from their respective priors defined in the FAIR-HUC benchmark.

Using the injected values of \(\mu^\text{true}\) and \(\boldsymbol{\alpha}^\text{true}\), a perturbed test dataset is constructed by applying the corresponding systematic shifts to the nominal features, including all derived kinematic and event-level observables.

For each pseudo-experiment, the trained classifier is applied to its events, and the resulting outputs are processed through the profile pipeline described in Section~\ref{sec:scan}. A profile likelihood curve is then constructed by scanning over a fixed \(\mu\) grid within the range \([0, 3]\). From this curve, we extract the maximum likelihood estimate \(\hat{\mu}\) and construct the 68.3\% confidence interval using the profile likelihood ratio criterion \(\Delta \mathrm{NLL}(\mu) \leq 0.5\).

The pseudo-experiments in this setup allow us to directly compute the empirical coverage of the constructed confidence intervals by checking how often \(\mu^\text{true}\) falls within the estimated interval \((\hat\mu_{16}, \hat\mu_{84})\). We also examine the typical widths of the intervals. These two metrics are quantified as follows:
\begin{align}
\label{eq:width}
    w &= \frac{1}{N_{\text{test}}} \sum_{i=1}^{N_{\text{test}}} \left| \mu_{84,i} - \mu_{16,i} \right|
\end{align}

\begin{align}
\label{eq:coverage}
    c &= \frac{1}{N_{\text{test}}} \sum_{i=1}^{N_{\text{test}}} \mathbb{I} \left[ \mu_{\text{true}, i} \in [\mu_{16,i}, \mu_{84,i}] \right] \;,
\end{align}
where \(w\) denotes the average interval width and \(c\) denotes the empirical coverage. Here, \(\mu_{16,i}\) and \(\mu_{84,i}\) represent the lower and upper bounds of the 68.3\% confidence interval for the \(i\)-th pseudo-experiment, and \(\mathbb{I}[\cdot]\) is the indicator function. 

\begin{figure*}
    \centering
    \includegraphics[width=0.32\linewidth]{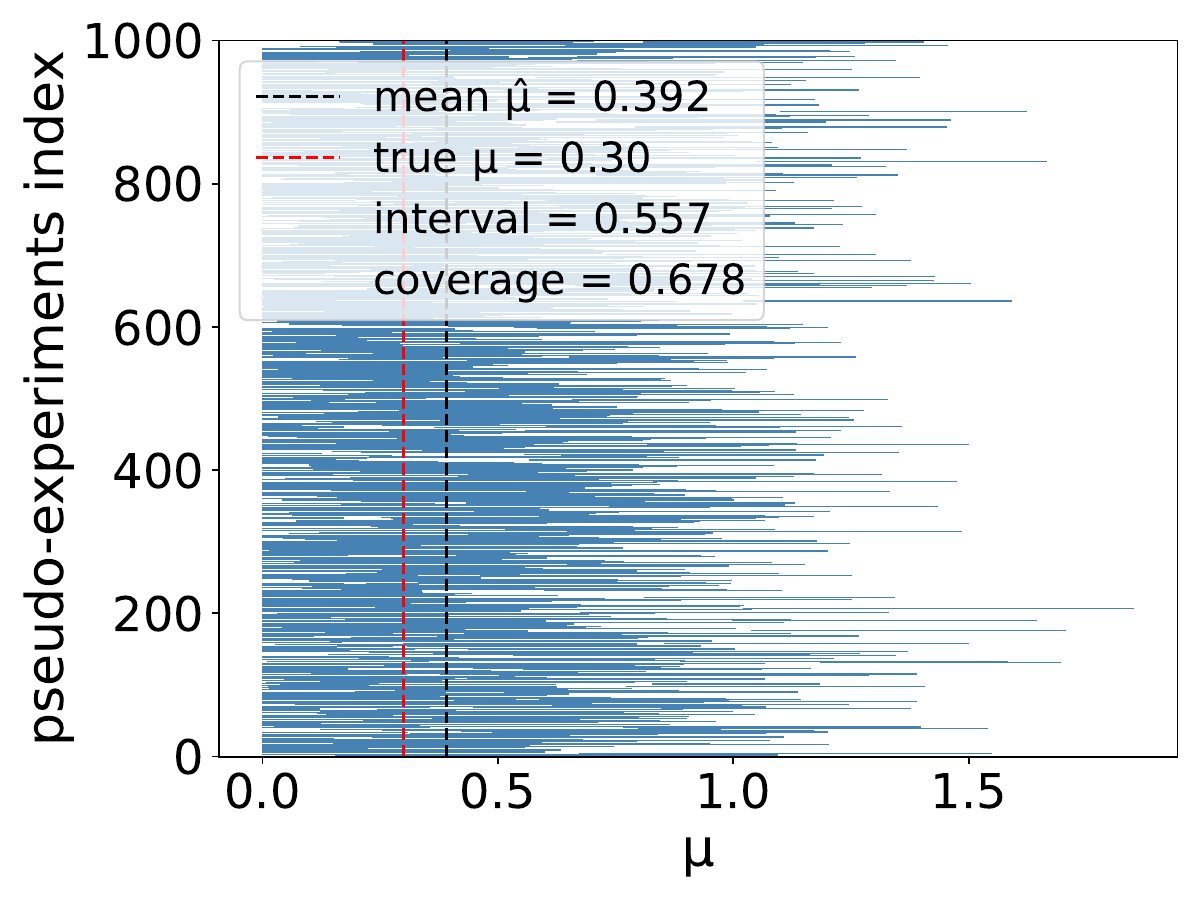}
    \includegraphics[width=0.32\linewidth]{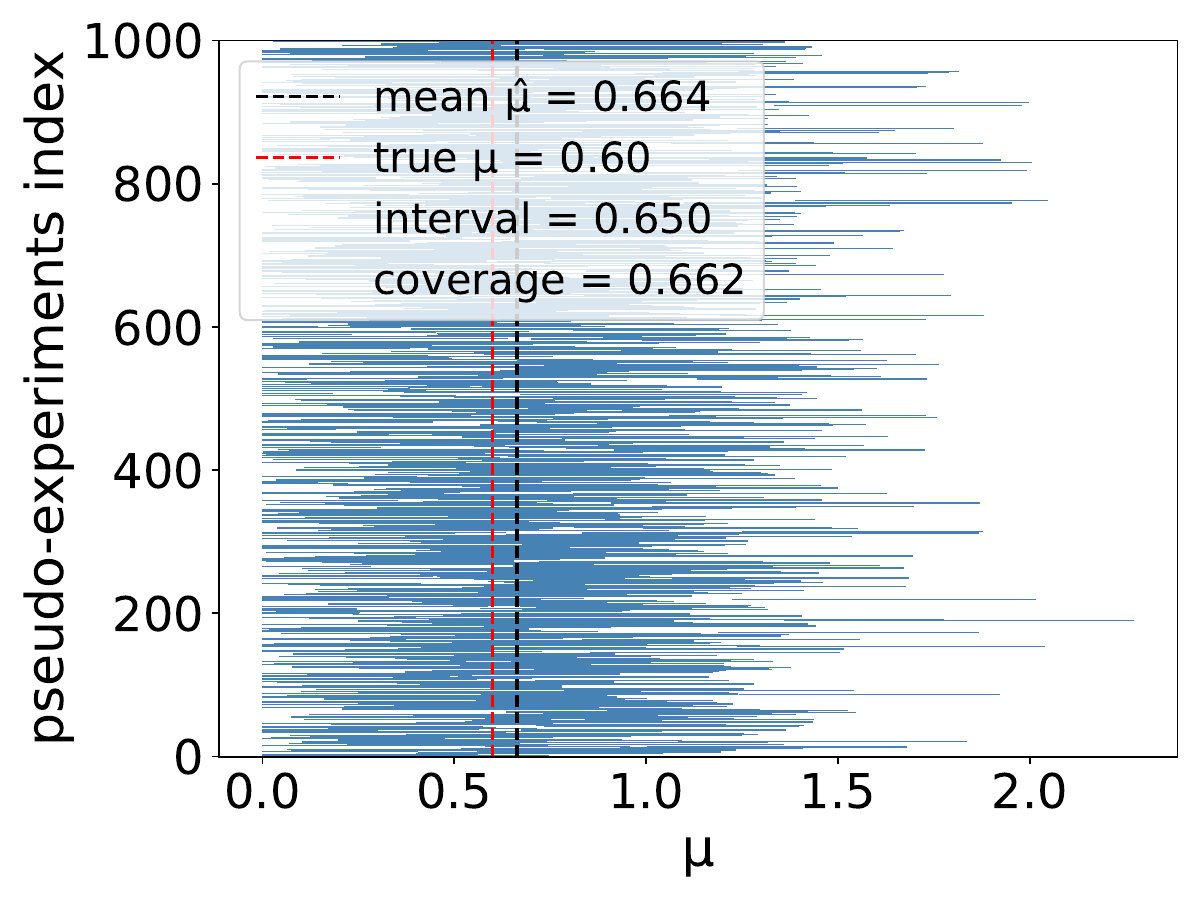}
    \includegraphics[width=0.32\linewidth]{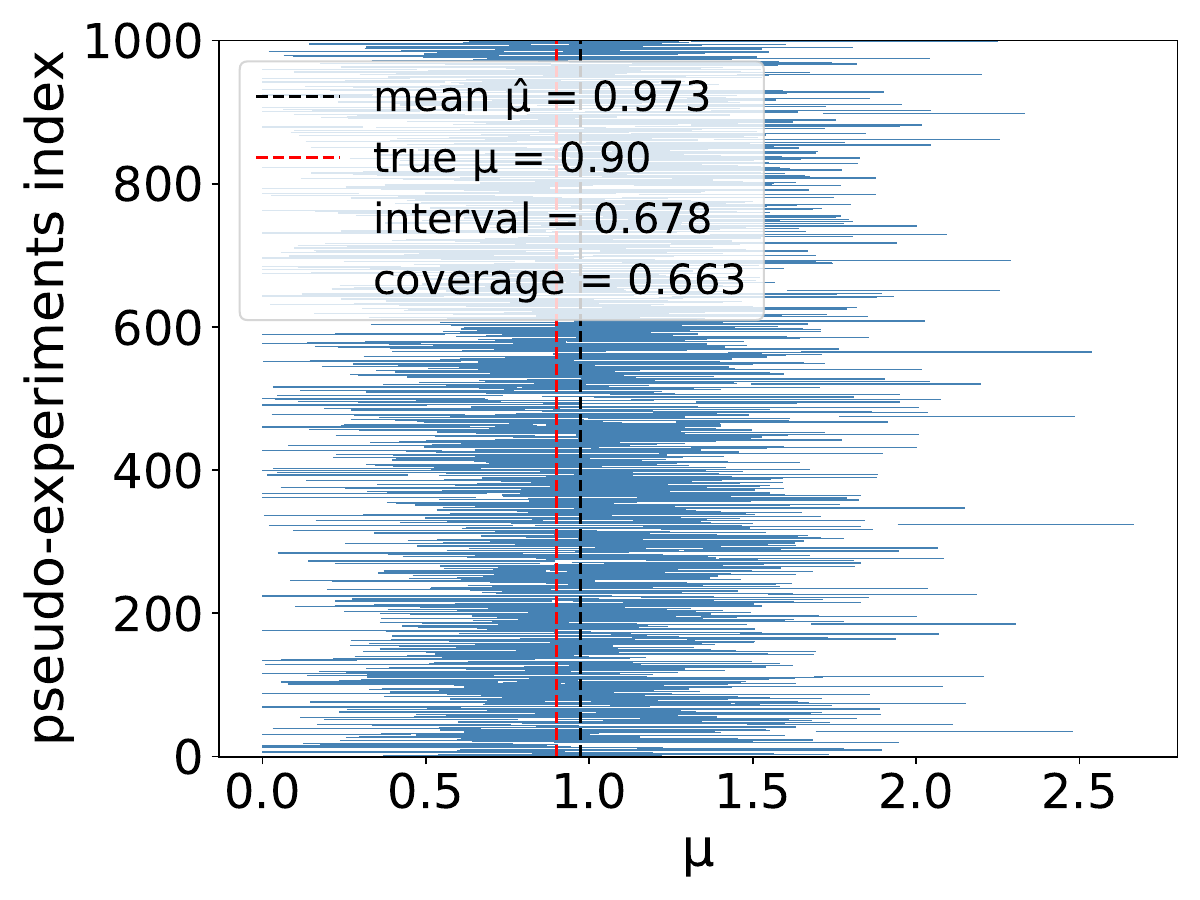}
    
    \includegraphics[width=0.32\linewidth]{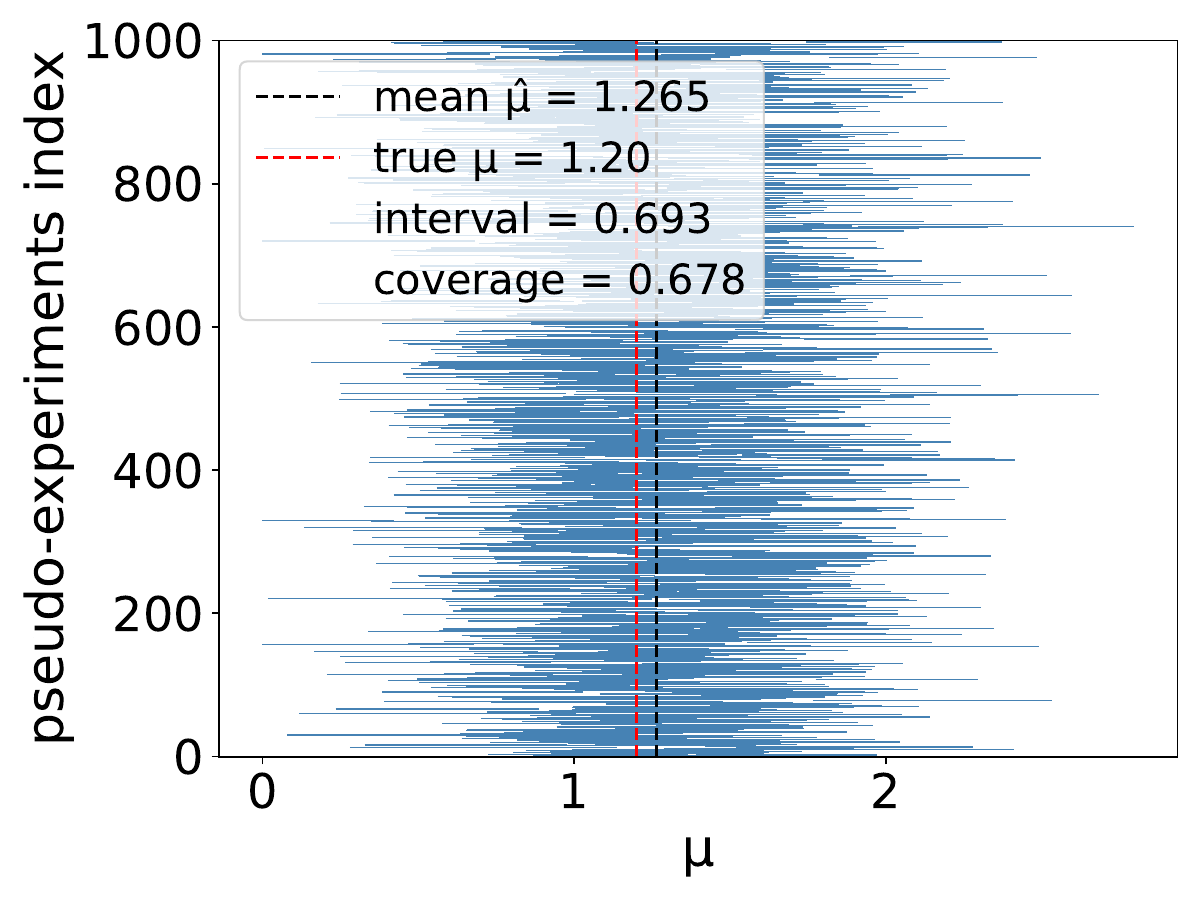}
    \includegraphics[width=0.32\linewidth]{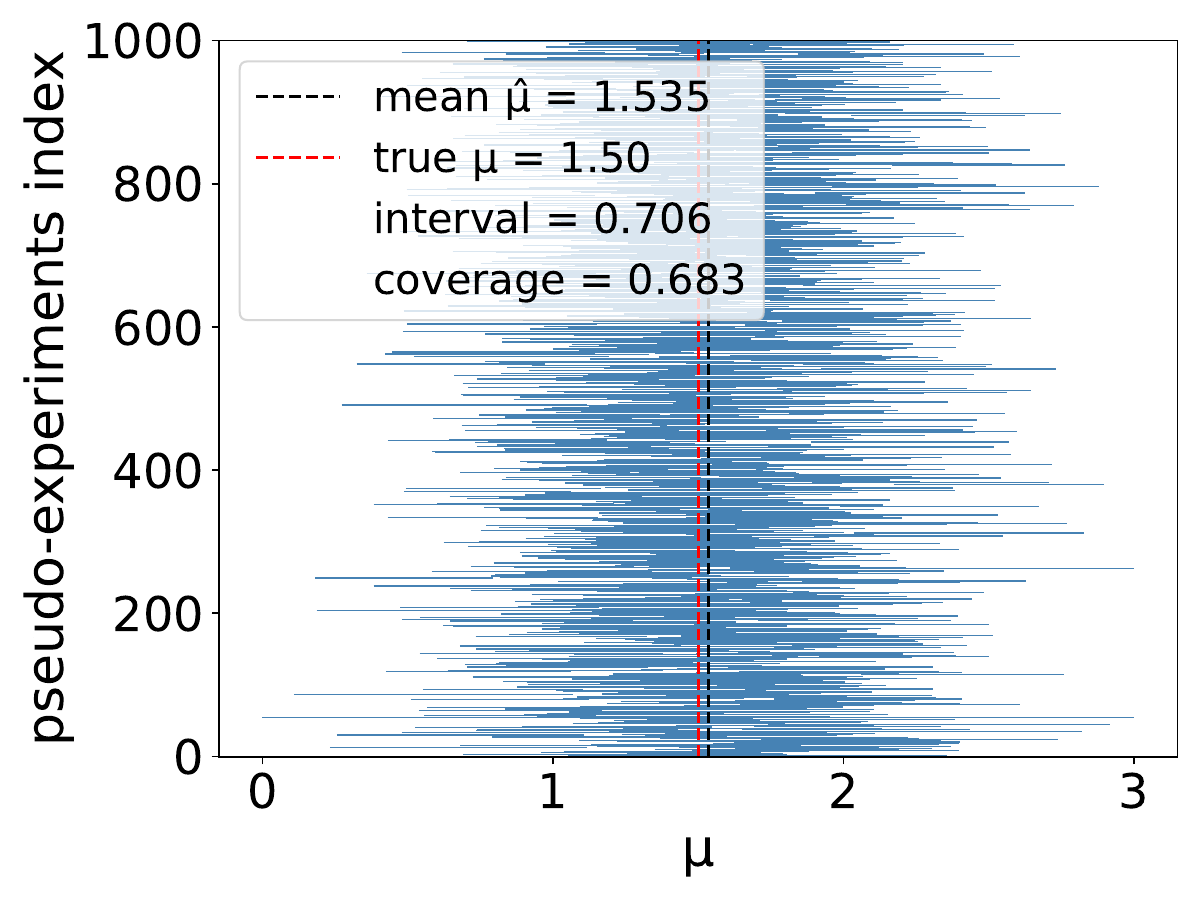}
    \includegraphics[width=0.32\linewidth]{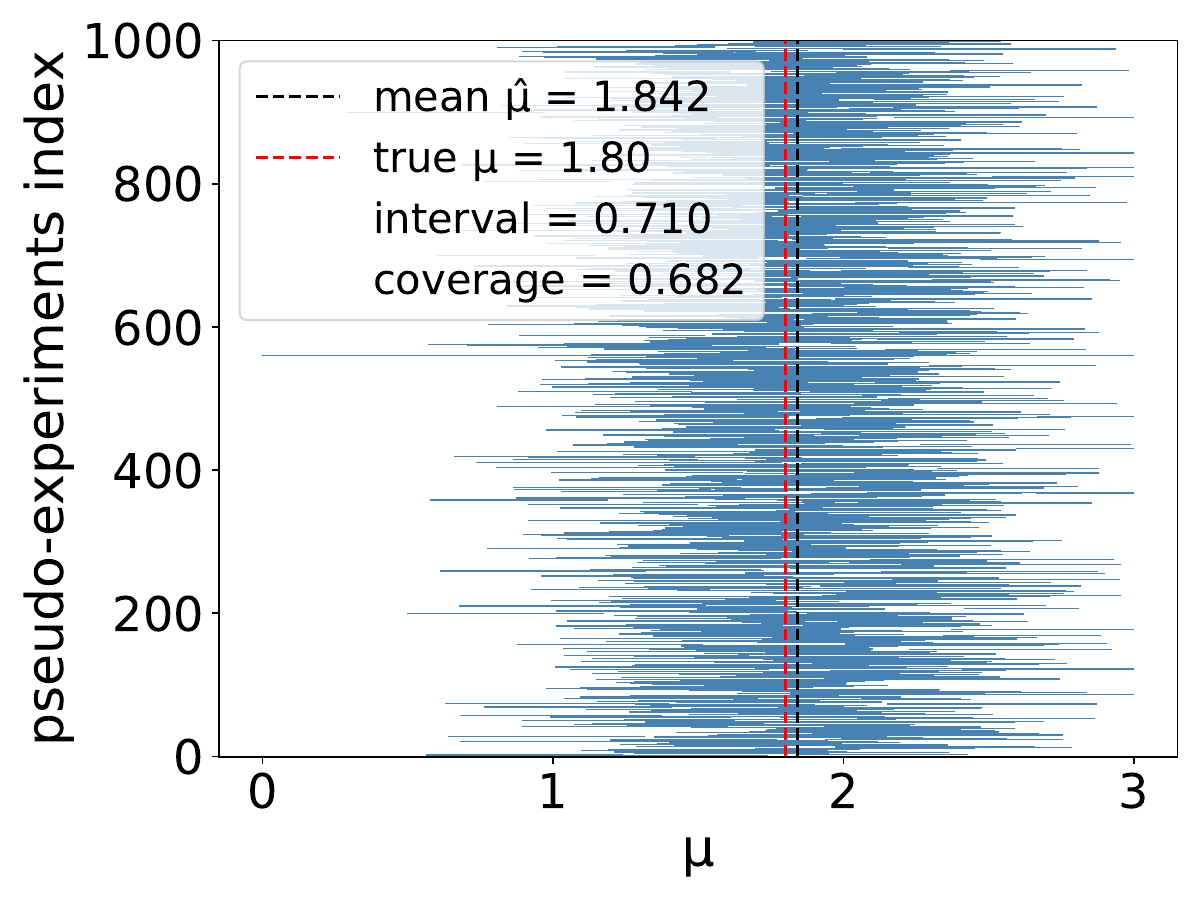}
    
    \includegraphics[width=0.32\linewidth]{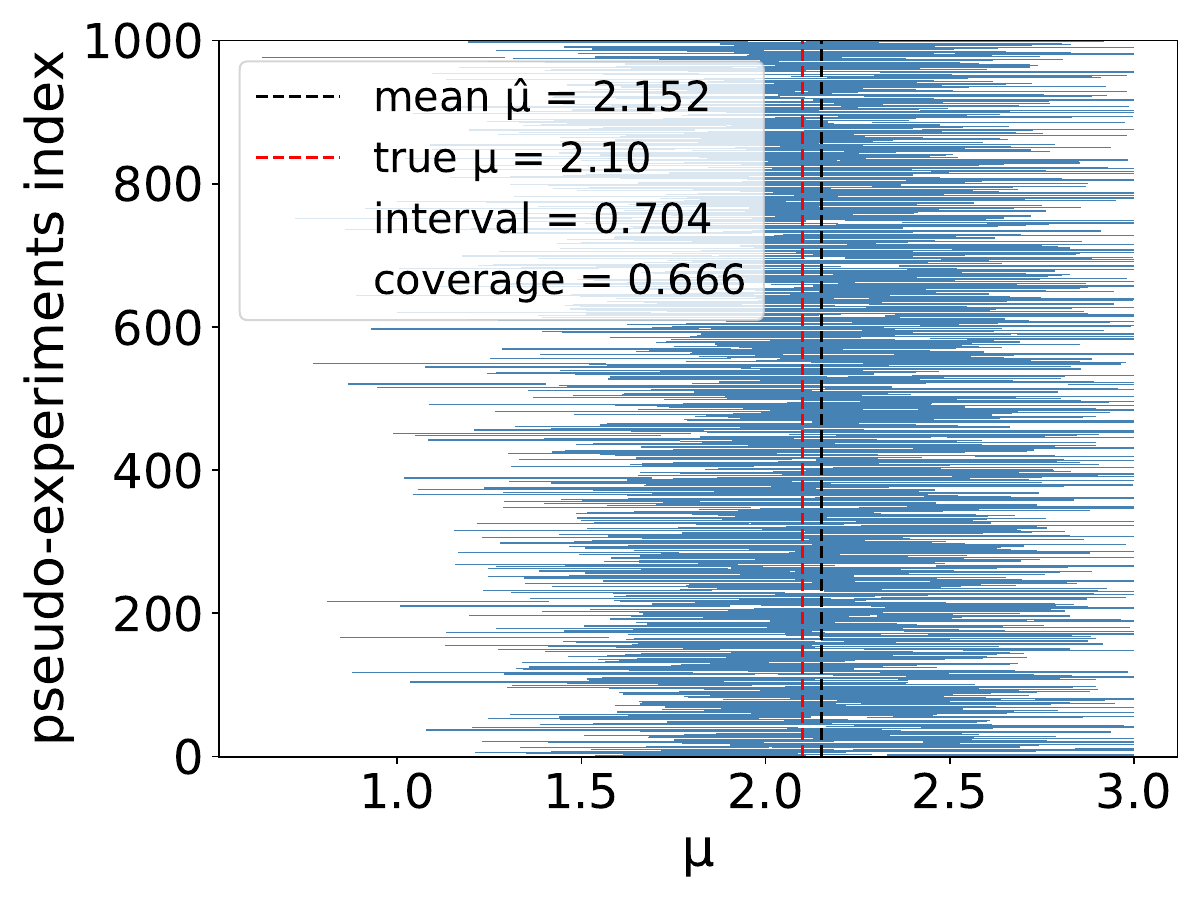}
    \includegraphics[width=0.32\linewidth]{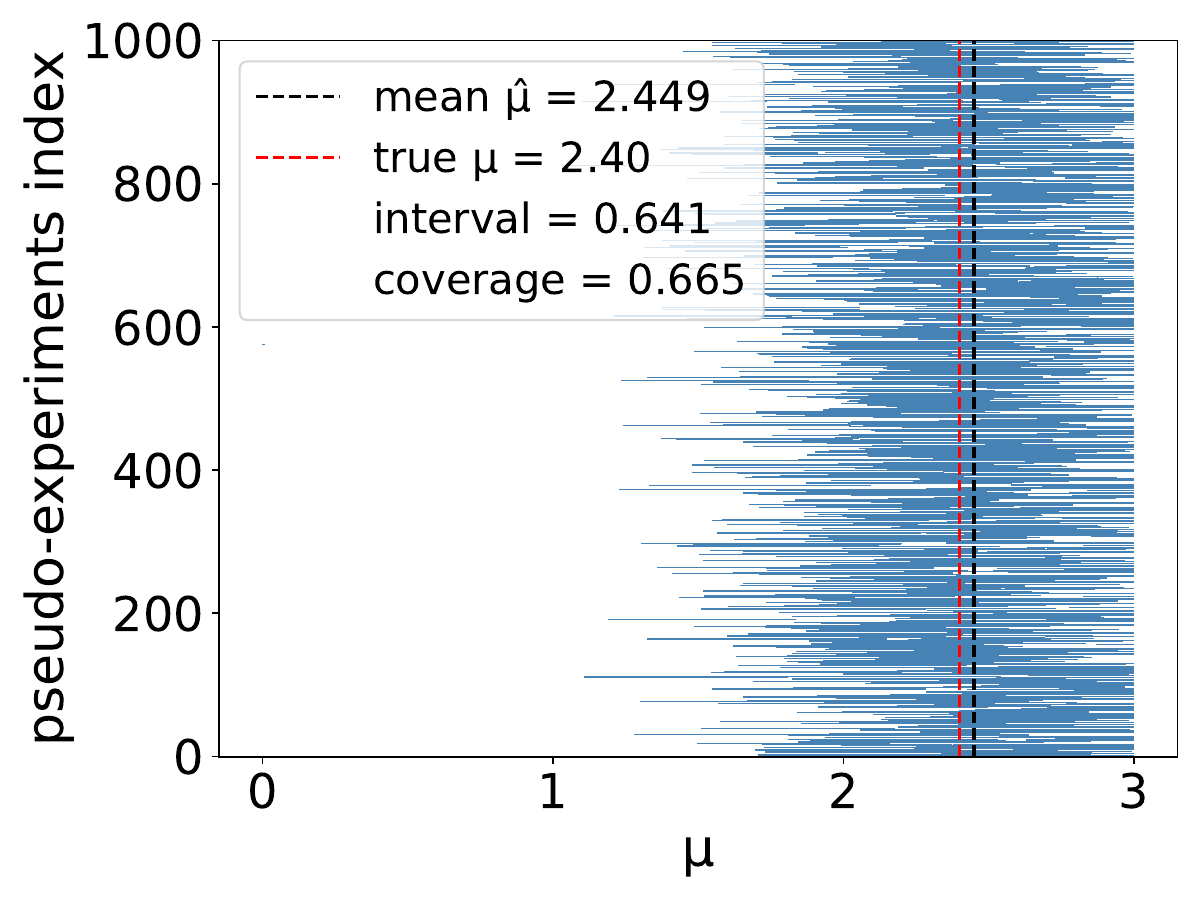}
    \includegraphics[width=0.32\linewidth]{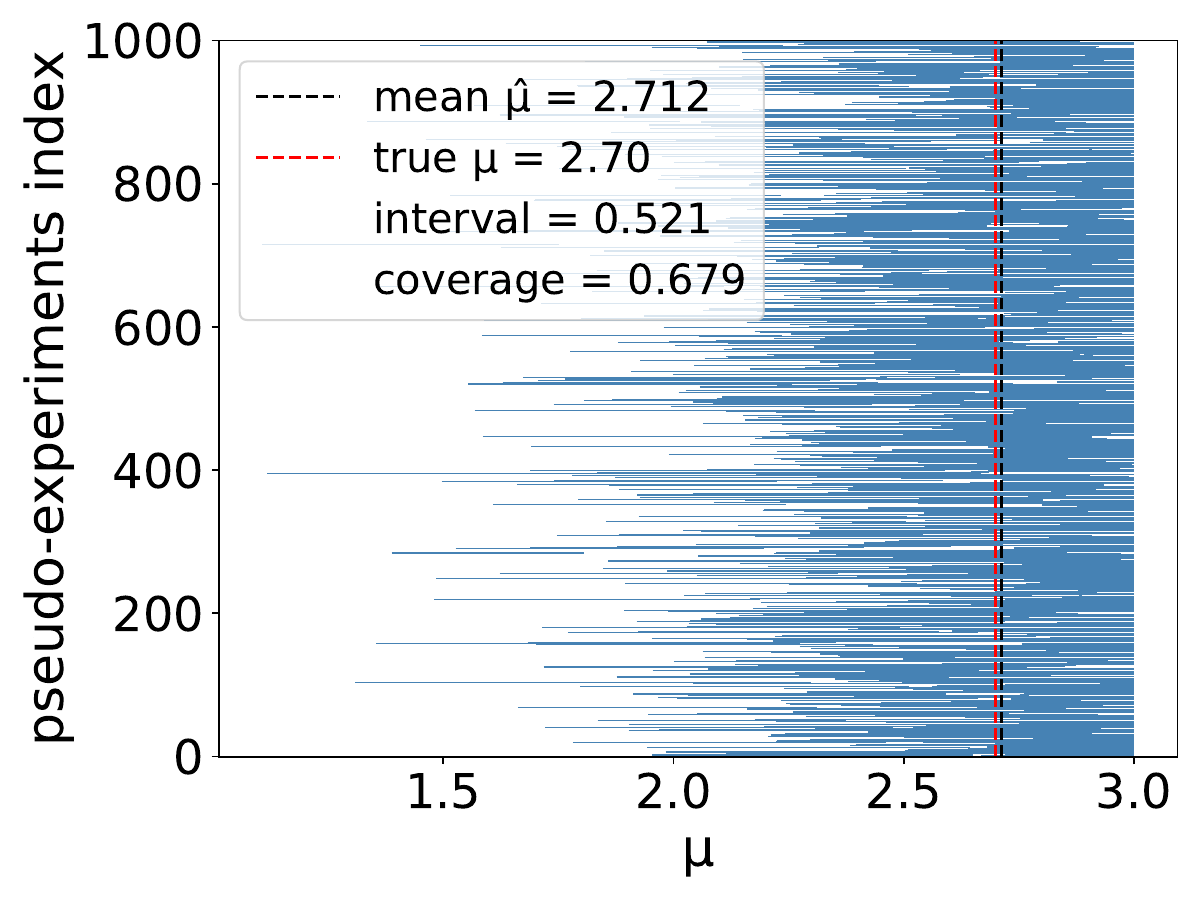}
    
    \caption{
    \label{fig:toy-scatter-nuisances-all}
    Visualization of 1000 pseudo-experiments for each of the signal strength values 
    $\mu_{\text{true}} = 0.3$, $0.6$, $0.9$, $1.2$, $1.5$, $1.8$, $2.1$, $2.4$, and $2.7$ 
    (from top-left to bottom-right). 
    For each pseudo-experiment, the predicted $68.27\%$ confidence interval for $\mu$ is shown as a horizontal blue line. 
    The vertical dotted line indicates the true value $\mu_{\text{true}}$, and the dashed black line marks the average of the maximum likelihood estimates $\hat{\mu}$. 
    The coverage is computed as the fraction of pseudo-experiments where $\mu_{\text{true}}$ lies within the predicted interval.
    }
\end{figure*}

The results shown in Figure~\ref{fig:toy-scatter-nuisances-all} demonstrate that the constructed confidence intervals maintain stable performance across the entire range of injected signal strengths. The average interval width increases from \(w = 0.557\) at \(\mu^\text{true} = 0.3\) to a maximum of \(w = 0.71\) at \(\mu^\text{true} = 1.8\), before decreasing to \(w = 0.521\) at \(\mu^\text{true} = 2.7\). This behavior can be attributed to two complementary effects. In the central region, where signal and background contributions are more comparable, statistical uncertainties are larger, leading to wider intervals. Towards the edges of the scan range, the likelihood function is truncated by the fixed bounds at \(\mu = 0\) and \(\mu = 3\), which artificially suppresses the width of the resulting intervals. Overall, the interval widths remain relatively narrow across all tested values of \(\mu^\text{true}\), indicating a high precision in signal strength estimation without sacrificing coverage. The empirical coverage remains close to the nominal $68.3\%$ target throughout, ranging from $c = 0.662$ to $c = 0.683$, with no significant under- or over-coverage observed. The slight fluctuations across different values of $\mu^\text{true}$ are consistent with statistical variability at the level of 1000 pseudo-experiments per point. These results confirm that the surrogate likelihood framework provides statistically reliable confidence intervals, with well-controlled coverage properties.

\subsection{Global Performance under Randomized Signal Strengths}
\label{random}

To further validate the robustness and reliability of our inference framework, we conduct a second Monte Carlo study using 50000 pseudo-experiments. In contrast to the fixed-\(\mu\) setup in Section~\ref{fix}, here the true signal strength \(\mu^\text{true}\) is drawn uniformly from the continuous range \([0, 3]\), allowing us to probe the global behavior of the profile likelihood across a wide spectrum of scenarios.

\begin{figure}
    \centering
    \includegraphics[width=0.8\linewidth]{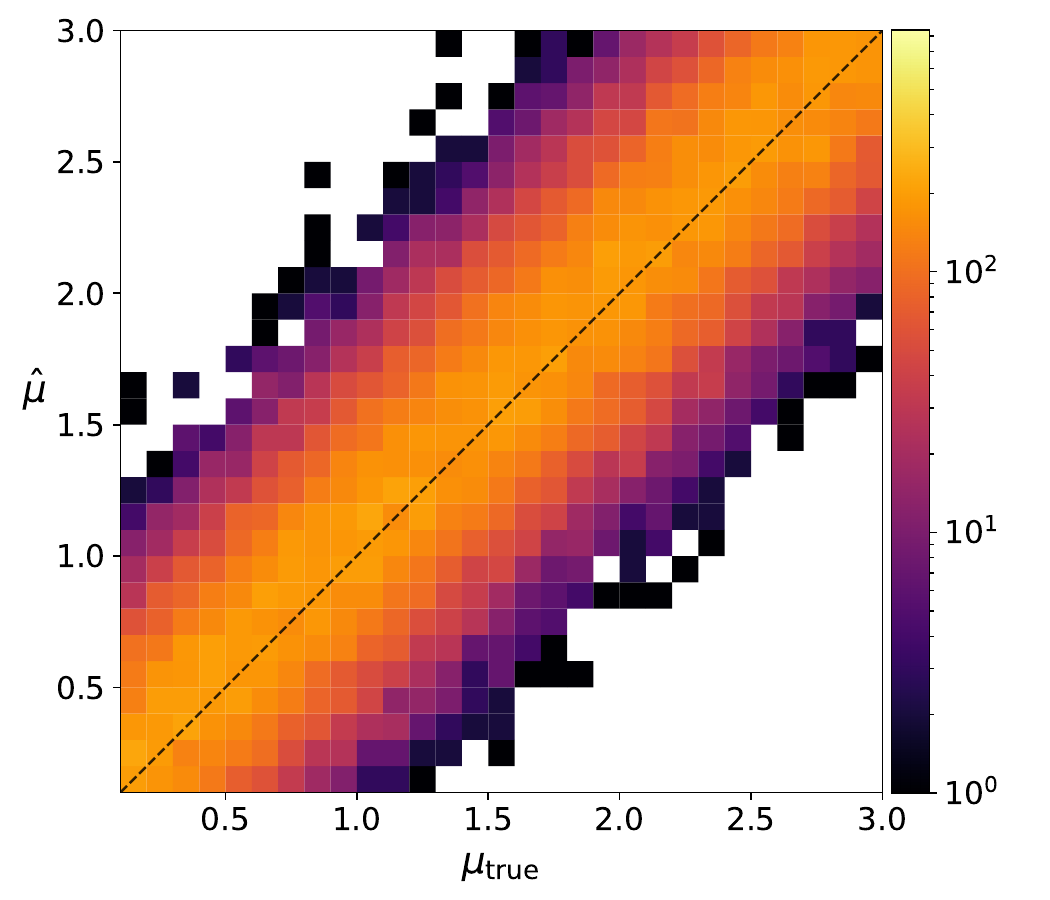}
    \caption{\label{fig:toy-scatter-mu} Scatter plot of $\mu_\text{true}$ versus the maximum likelihood estimate $\hat{\mu}$ from $5 \cdot 10^4$ pseudo-experiments. }\label{fig:mu-corr}
\end{figure}

Figure~\ref{fig:mu-corr} shows a scatter plot of \(\hat{\mu}\) versus \(\mu_\text{true}\), with points closely aligned along the diagonal and no significant bias or outliers. This indicates that the surrogate likelihood yields stable and accurate point estimates across the full range of signal strengths, with fluctuations consistent with statistical uncertainty.

\begin{figure*}
    \centering
    \includegraphics[width=0.32\linewidth]{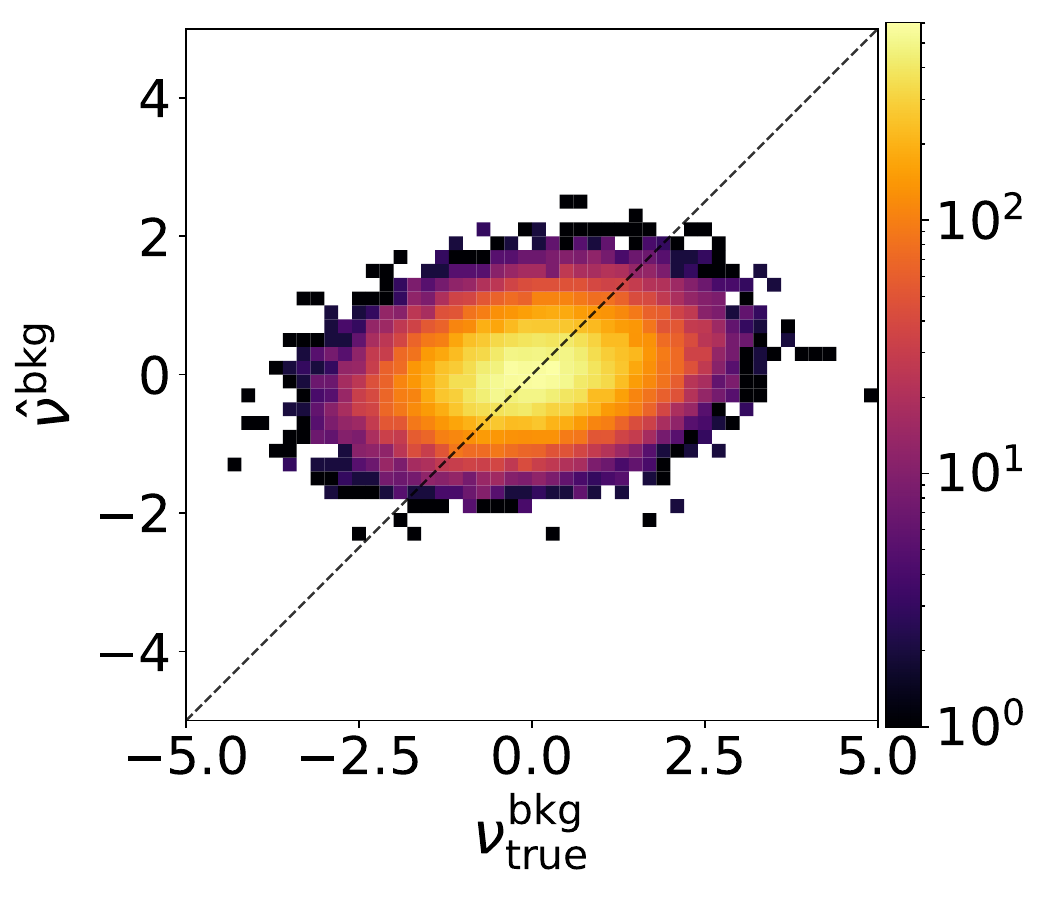}
    \includegraphics[width=0.32\linewidth]{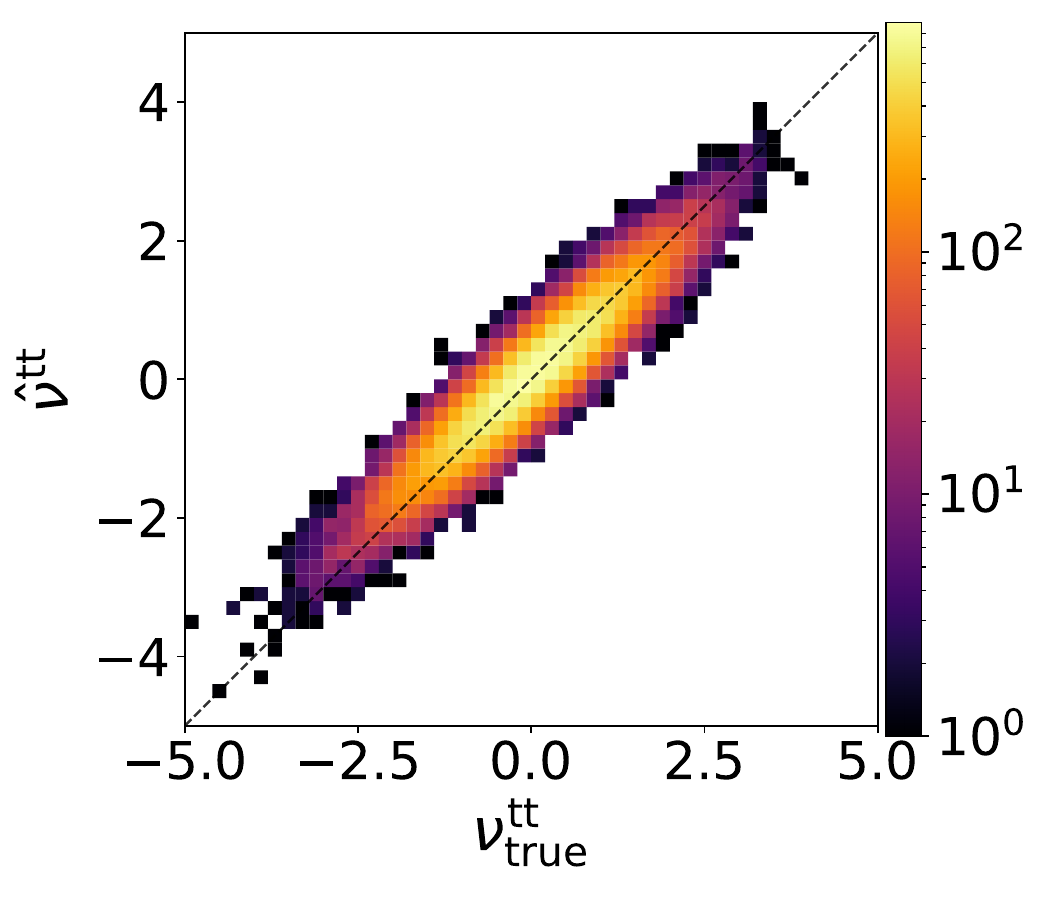}
    \includegraphics[width=0.32\linewidth]{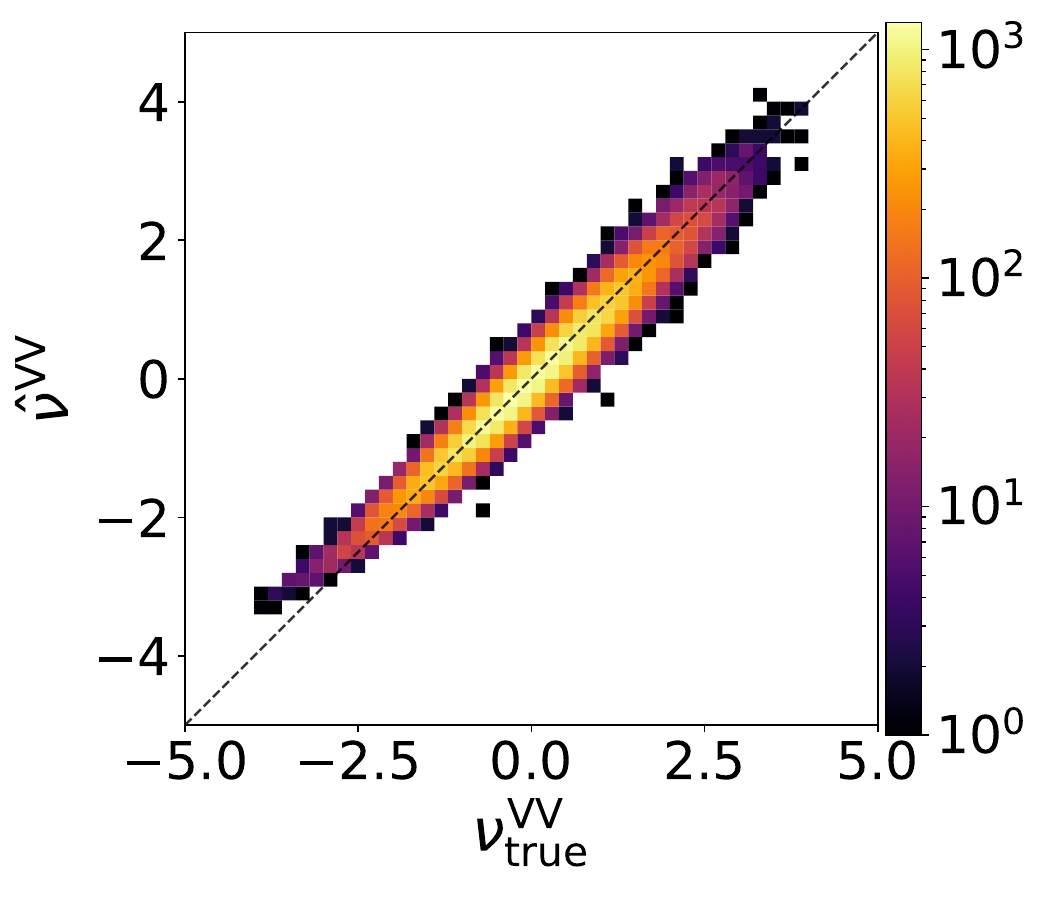} \\
    \includegraphics[width=0.32\linewidth]{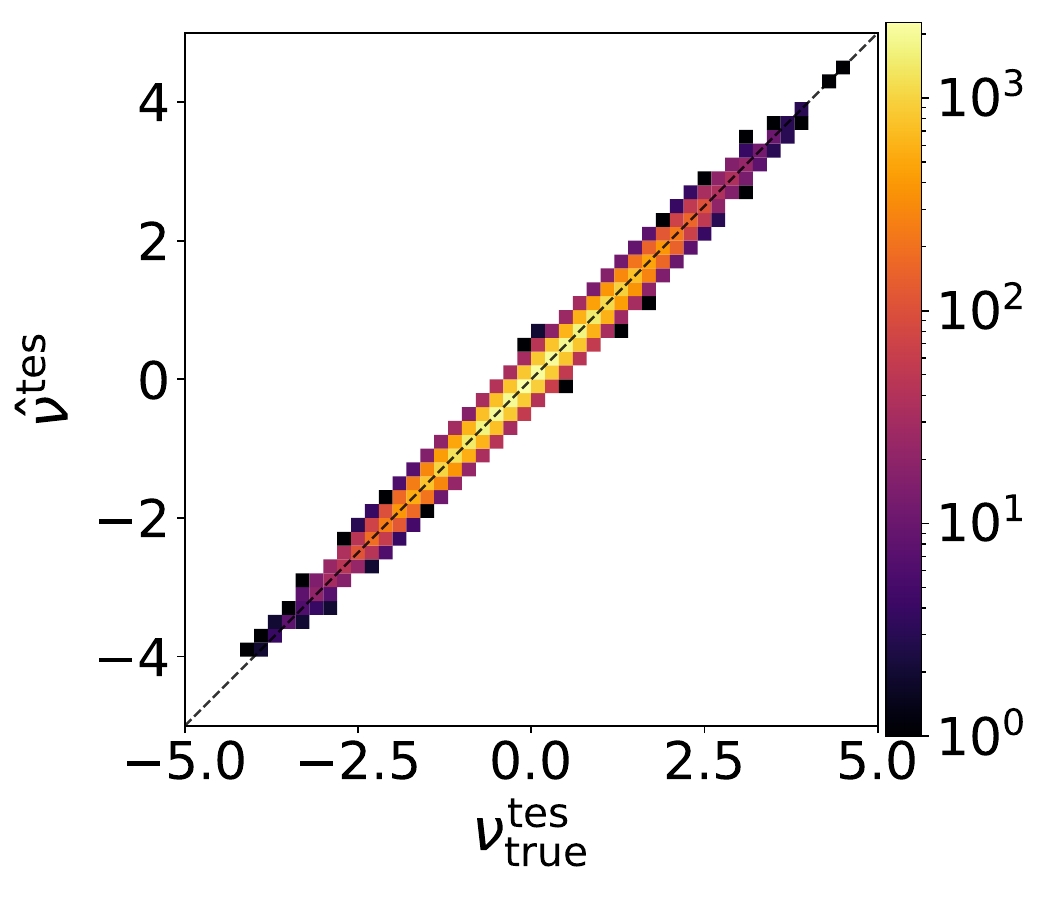}
    \includegraphics[width=0.32\linewidth]{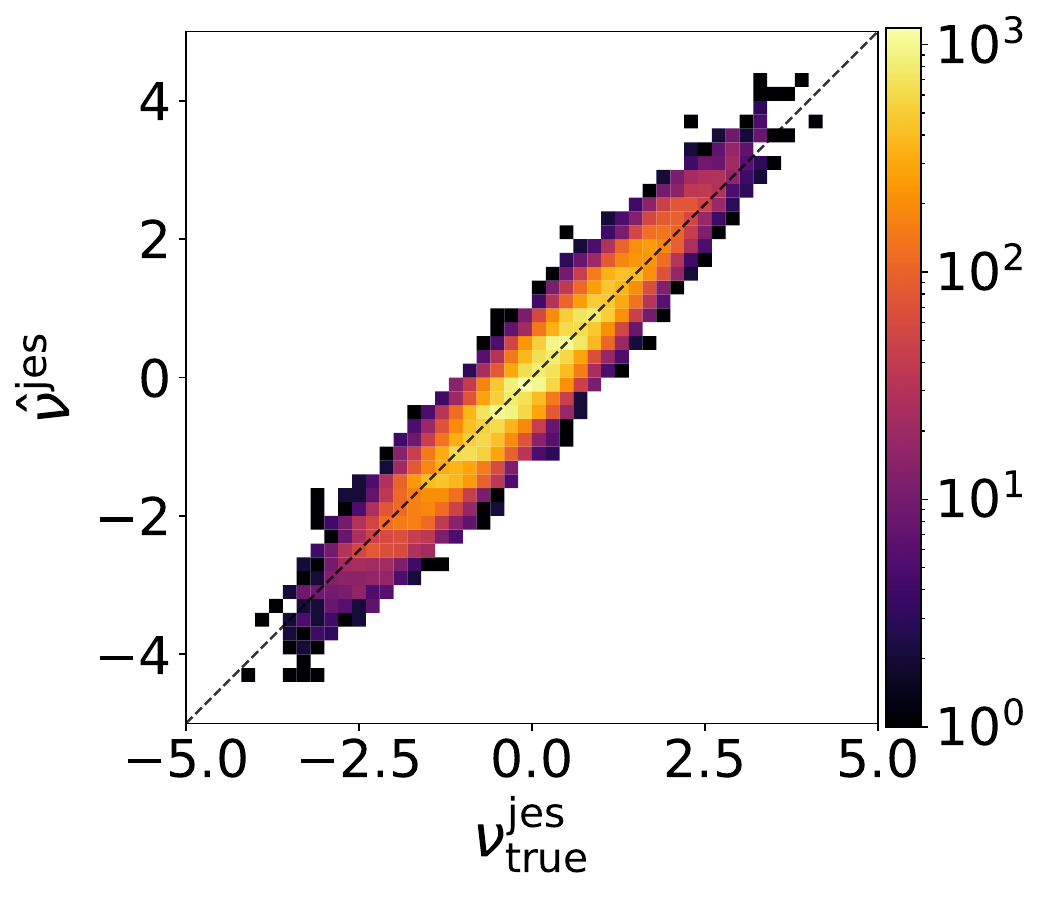}
    \includegraphics[width=0.32\linewidth]{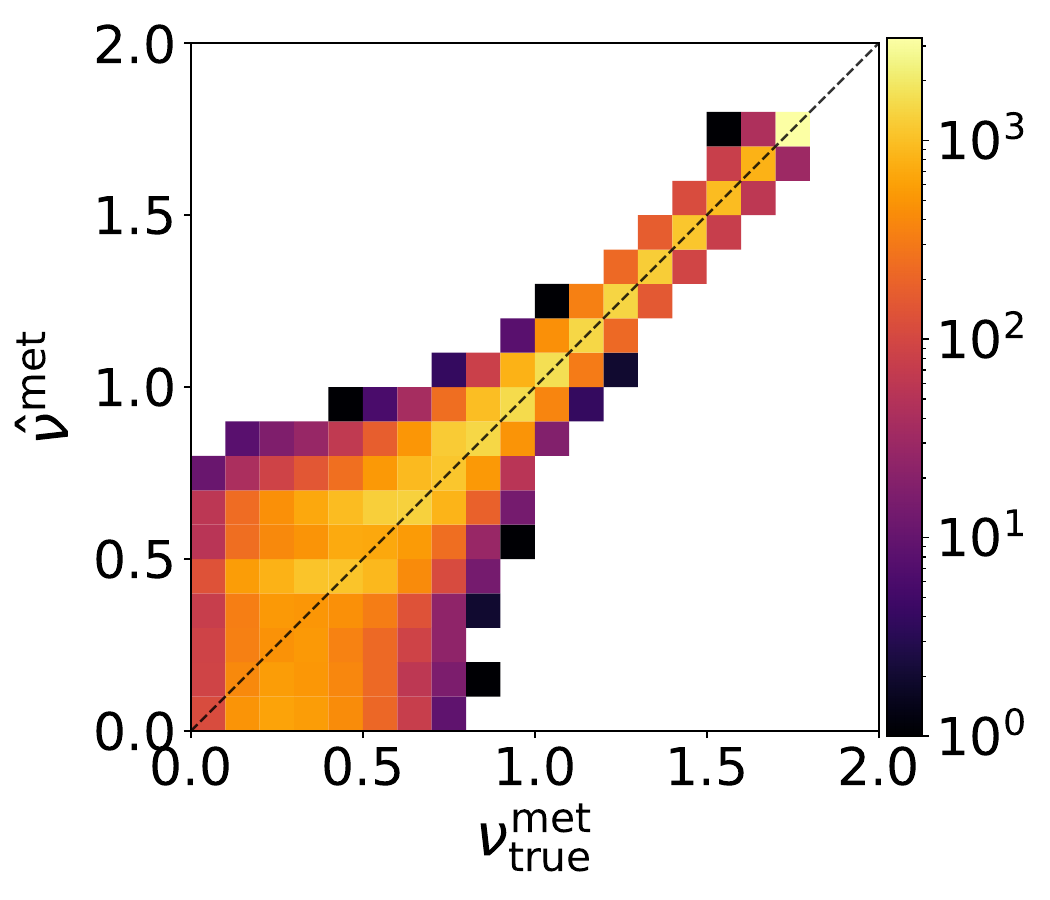}
    
    \caption{
        \label{fig:scatter-nuis-all}
        Same as Fig.~\ref{fig:mu-corr}, but for all six nuisance parameters: 
        $\nu_\text{bkg}$, $\nu_{\ttbar}$, and $\nu_{\VV}$ (top row, left to right), 
        and $\nu_\text{tes}$, $\nu_\text{jes}$, and $\nu_\text{met}$ (bottom row, left to right).
    }
\end{figure*}

Figures~\ref{fig:scatter-nuis-all} displays the scatter plots of the maximum likelihood estimates \(\hat{\boldsymbol{\nu}}\) versus the injected true values \(\boldsymbol{\nu}_\text{true}\) for all six nuisance parameters. Overall, the reconstructed values align well with the truth, clustering around the diagonal with very few outliers, which confirms the stability and reliability of the inference procedure across all systematic directions. Among the calibration-type parameters, \(\nu_{\text{tes}}\) exhibits the best performance, with tight correlation to the injected values. This superior behavior arises from the presence of a \(\tau\)-jet transverse momentum threshold at 26~GeV in the original dataset, which makes the total event yield highly sensitive to variations in \(\nu_{\text{tes}}\). Consequently, the surrogate likelihood naturally favors accurate adjustment of this parameter during profiling. In comparison, \(\nu_{\text{jes}}\) is also well constrained but primarily affects the shape rather than the overall event yield, resulting in slightly less sensitivity and a broader spread than \(\nu_{\text{tes}}\). For the normalization-type nuisances, \(\nu_{\text{bkg}}\) is poorly constrained and shows little correlation with the true value. This is expected: in the inclusive region, the dominant background process is \(\PZ\to\tau\tau\), and even large variations in \(\nu_{\text{bkg}}\) (up to 5 standard deviations) translate to only about a 0.5\% change in total yield. In contrast, \(\nu_{\ttbar}\) and \(\nu_{\VV}\) dominate in their respective control regions and can induce 10\% and even 100\% variations in event yields, respectively, making them far more identifiable in the profiling process. As a result, \(\nu_{\VV}\) is especially well reconstructed, followed by \(\nu_{\ttbar}\). Finally, \(\nu_{\text{met}}\) exhibits slightly degraded performance at the low end of its range, where variations in the parameter induce only negligible shifts in the event yields. In this regime, the profile likelihood becomes nearly flat along the \(\nu_{\text{met}}\) direction, making it difficult to determine a precise MLE and leading to the observed bias and increased spread. By contrast, at larger values of \(\nu_{\text{met}}\), the same variations result in more pronounced yield changes, which steepen the likelihood curve and allow for a more precise estimation.

\begin{figure*}
    \centering
    \includegraphics[width=0.48\linewidth]{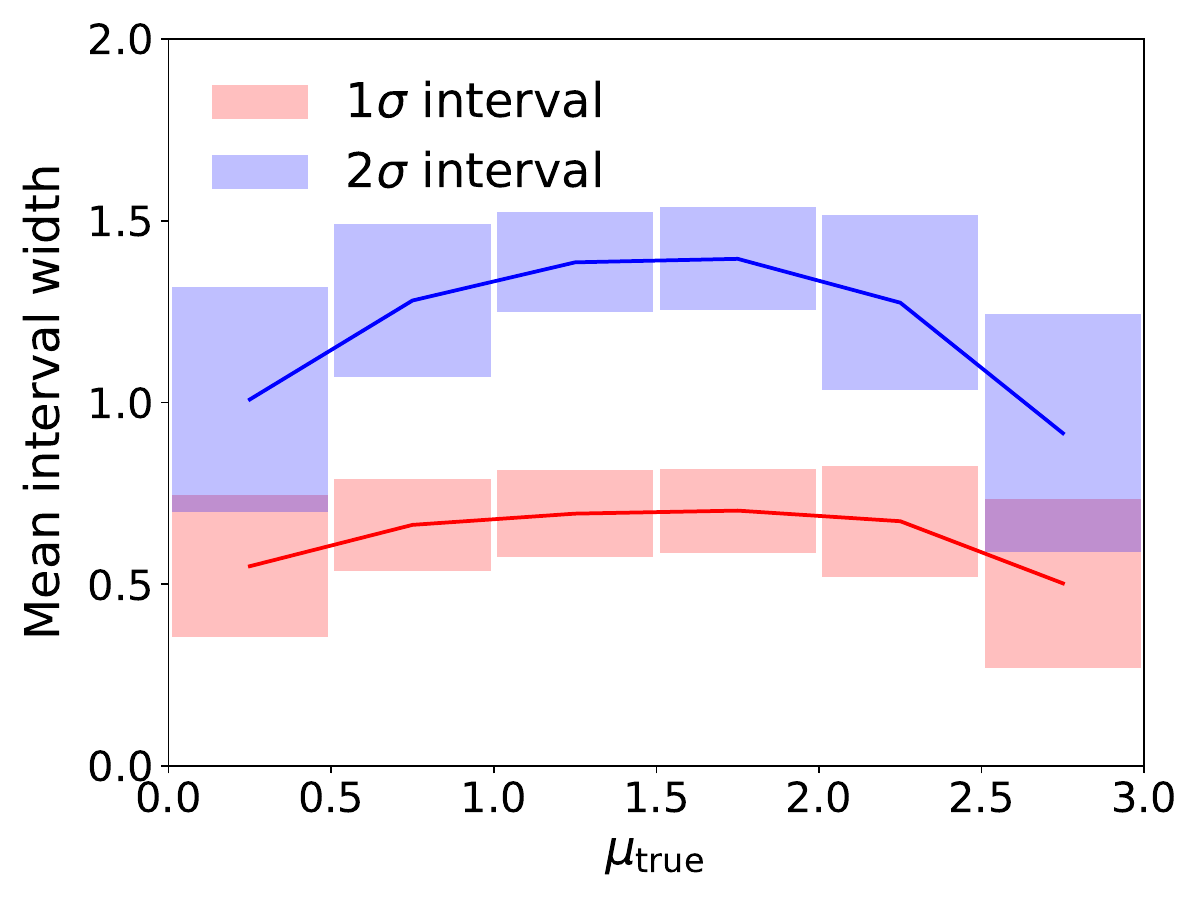}\hfill
    \includegraphics[width=0.48\linewidth]{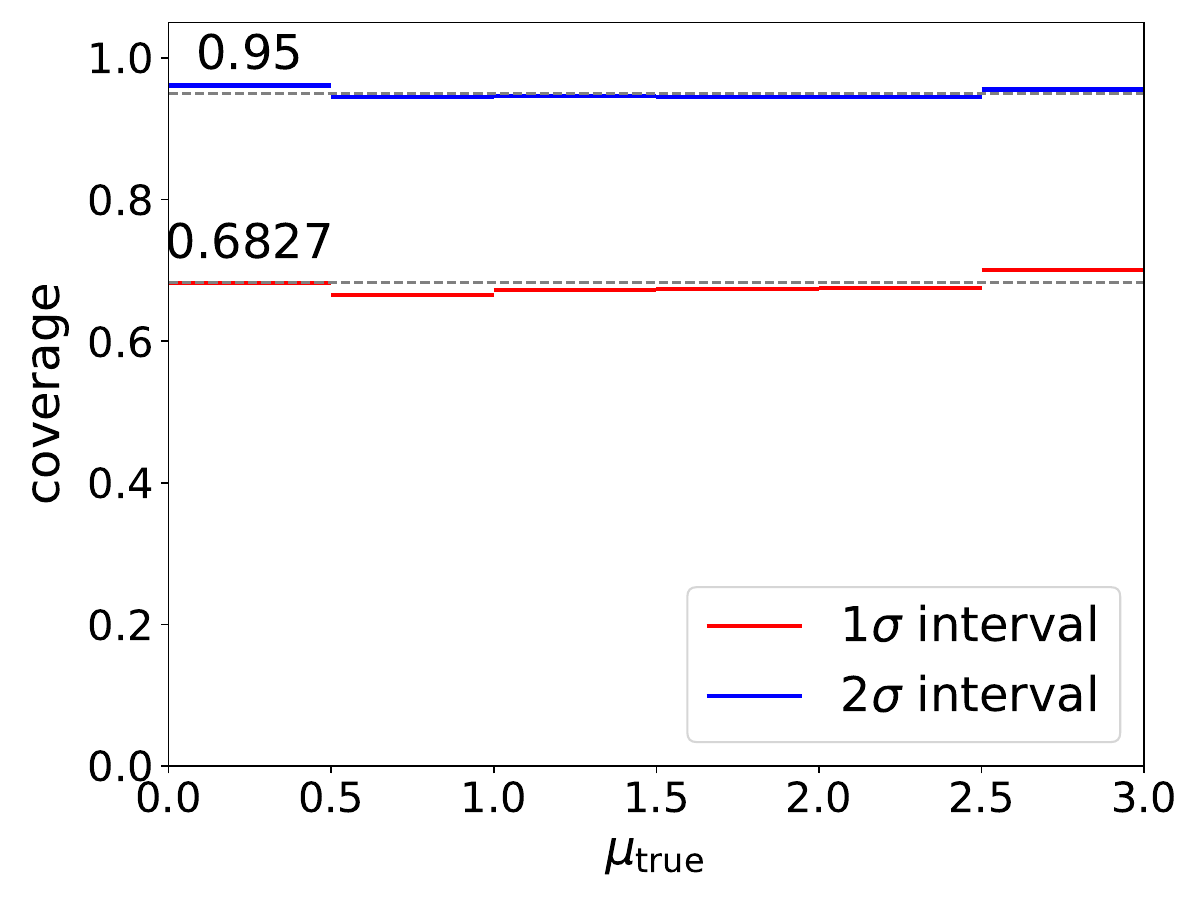}\\
    \caption{\label{fig:toy-coverage} The average width of the confidence interval for the signal strength $\mu$ (left) and the corresponding coverage (right) are shown as functions of the true value $\mu^\text{true}$, based on results from 50000 pseudo-experiments. The average interval length is 0.6342 and the coverage is 0.6697.}
\end{figure*}

Figure~\ref{fig:toy-coverage} presents the average widths of the 68.3 \% (red) and 95 \% (blue) confidence intervals given by $\mu_{84}-\mu_{16}$ and $\mu_{97.5}-\mu_{2.5}$, obtained from 50000 pseudo-experiments with \(\mu_\text{true}\) drawn uniformly in \([0,3]\). The shaded bands indicate the corresponding \(1\sigma\) spread.  The interval width is smallest near the scan boundaries and reaches its maximum around \(\mu^\text{true}\approx1.5\).  The edge–wise narrowing is a consequence of the fixed scan limits at \(\mu=0\) and \(\mu=3\). The right panel shows the empirical coverage, defined in Eq.~\eqref{eq:coverage}. Both the 68.3\% and 95\% intervals demonstrate excellent agreement with their nominal targets across the full range, with only small fluctuations consistent with the expected sampling noise from 50000 pseudo-experiments.

On the FAIR Universe Codabench final evaluation (100$\times$100 values of $\mu$, each bootstrapped 1000 times for precision), the two top-performing submissions report average coverages and interval lengths of $(\text{coverage},\,\text{interval})=(0.6683,\,0.4599)$ for Reference~\cite{Benato:2025rgo} and $(0.6698,\,0.4974)$ for Reference~\cite{Elsharkawy:2025six}.
Using our protocol (50000 pseudo-experiments with $\mu\sim \mathcal{U}[0,3]$), we obtain an average coverage of $0.6697$ and an average interval length of $0.6342$.
While these protocols are not identical and a like-for-like re-evaluation would be required for a rigorous comparison, the coverage is comparable to the leaderboard methods, with moderately wider intervals. An additional methodological difference is that the leading FAIR-HUC solution employs an unbinned likelihood fit, whereas our approach constructs a binned likelihood from the classifier output. The unbinned approach avoids the potential information loss associated with binning and may therefore offer additional gains in statistical precision. The moderately wider intervals are expected consequence of our training strategy. Our model is trained using replica-averaged injections of nuisance parameters, where for each training batch multiple nuisance configurations are sampled and the corresponding losses are averaged prior to the parameter update. Optimising the model parameters against this averaged objective effectively introduces an implicit marginalisation over nuisance parameters at the training stage, leading to a more conservative response to systematic variations and a mildly reduced curvature of the profiled likelihood in the signal-strength direction. A practical advantage of our approach is its ease of extension to more complex final states: graph-based representations naturally accommodate high object multiplicities and intricate topologies, allowing the model to exploit multi-object correlations that become increasingly salient in such processes.

\section{Conclusions}
\label{sec:conclusions}

We have introduced a dedicated graph neural network framework for Higgs signal strength estimation under systematic uncertainties. The architecture combines a deterministic GNN and an uncertainty-aware GNN, enabling separate treatment of stable and perturbed input features. Systematic variations are explicitly injected during training via dynamic sampling of multiple nuisance parameter configurations in each epoch, encouraging robustness in the classifier outputs. These outputs are propagated through a template morphing pipeline to construct a surrogate likelihood function that supports efficient profile likelihood inference.

Validated on the FAIR-HUC benchmark, our method delivers accurate maximum likelihood estimates of the signal strength $\mu$ and its associated confidence intervals. In large-scale pseudo-experiments, the resulting intervals exhibit excellent coverage and controlled widths across both fixed and randomized signal strengths. The nuisance parameters are consistently and stably reconstructed, with fitted values closely tracking their injected ground truths. In this sense, our framework explicitly operates in the regime of ``known unknowns'', where the relevant sources of uncertainty are identified but their exact magnitudes must be inferred.

Our publicly available implementation, \textsc{SAGE}~\cite{SAGE}, provides a flexible and scalable solution for likelihood-based inference, combining deep learning with statistical rigor. The framework could be applied to precision measurements at the LHC once analysis-specific experimental and theoretical uncertainties, including differential shape effects, are incorporated and validated. It could also be extended to other domains where systematic uncertainties are important, particularly to processes with complex topologies or high object multiplicities for which graph-based representations can capture intricate correlations. Looking ahead, we note that the method scales naturally with the number of nuisance parameters: the network performance itself is largely insensitive to the number of nuisance parameters included, although the construction of the interpolation table becomes increasingly challenging. Addressing this will require more efficient strategies, such as block-wise surrogate modeling combined with sparse-grid interpolation, to control the computational scaling in higher-dimensional nuisance spaces.

\appendix

\section{Re-training with Class-Frequency Weighting}

All results in the main text use uniform per-event weights. However, the FAIR-HUC training set is strongly class-imbalanced—the four classes ($H\!\to\!\tau\tau$, $Z\!\to\!\tau\tau$, $t\overline t$, $VV$) differ in event counts by orders of magnitude. As summarized in Table~\ref{tab:event-stats}, the raw event counts in the FAIR-HUC dataset also differ substantially from the physics-expected event yields based on cross section and luminosity, since the signal class is oversampled. Consequently, the equal-weight scheme used in the main analysis neither corresponds to luminosity-scaled training nor equalizes class contributions.

In this appendix we evaluate the effect of introducing class weights: we re-train the classifier with the \textsc{scikit-learn} ``balanced'' scheme. For class \(k\) with \(n_k\) training events, total \(N=\sum_k n_k\), and \(K=4\) classes, the class weight is
\(w_k = \tfrac{N}{K\,n_k}\). We then compare the re-trained model to the uniform-weight baseline. The training/validation splits and model hyperparameters are kept unchanged, and the full surrogate-likelihood pipeline (interpolation, profiling, and interval extraction) is repeated. This appendix summarizes the resulting changes and cross-checks: (i) unweighted and weighted signal-class probability distributions under nuisance variations; (ii) confusion matrices comparing the original (equal-weight) and re-trained (class-weighted) classifiers; and (iii) the final profile-likelihood intervals and coverage based on 50000 pseudo-experiments. These results are intended as a comparative study of class weighting and do not supersede the baseline numbers reported in the main text.

\begin{figure}[ht]
    \includegraphics[width=8cm,height=6cm]{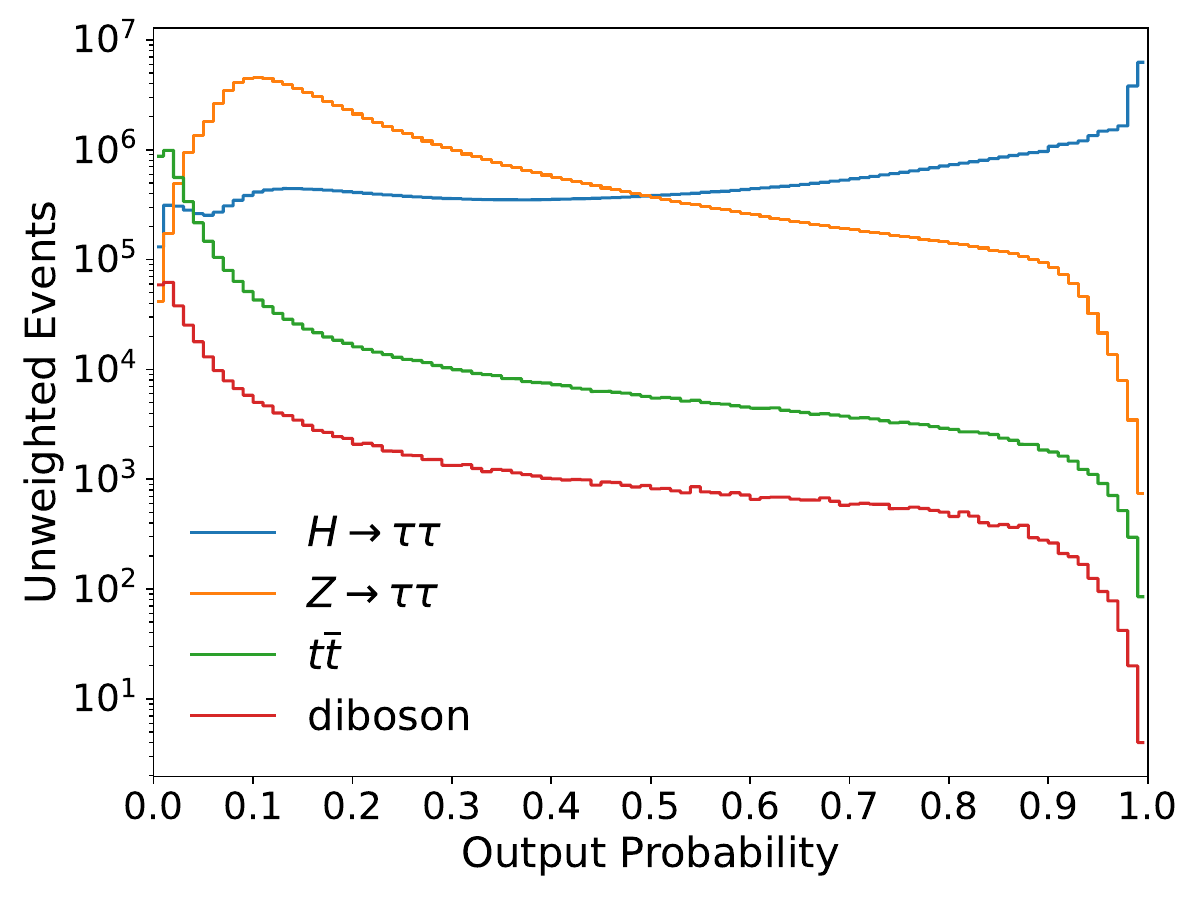}
    \includegraphics[width=8cm,height=6cm]{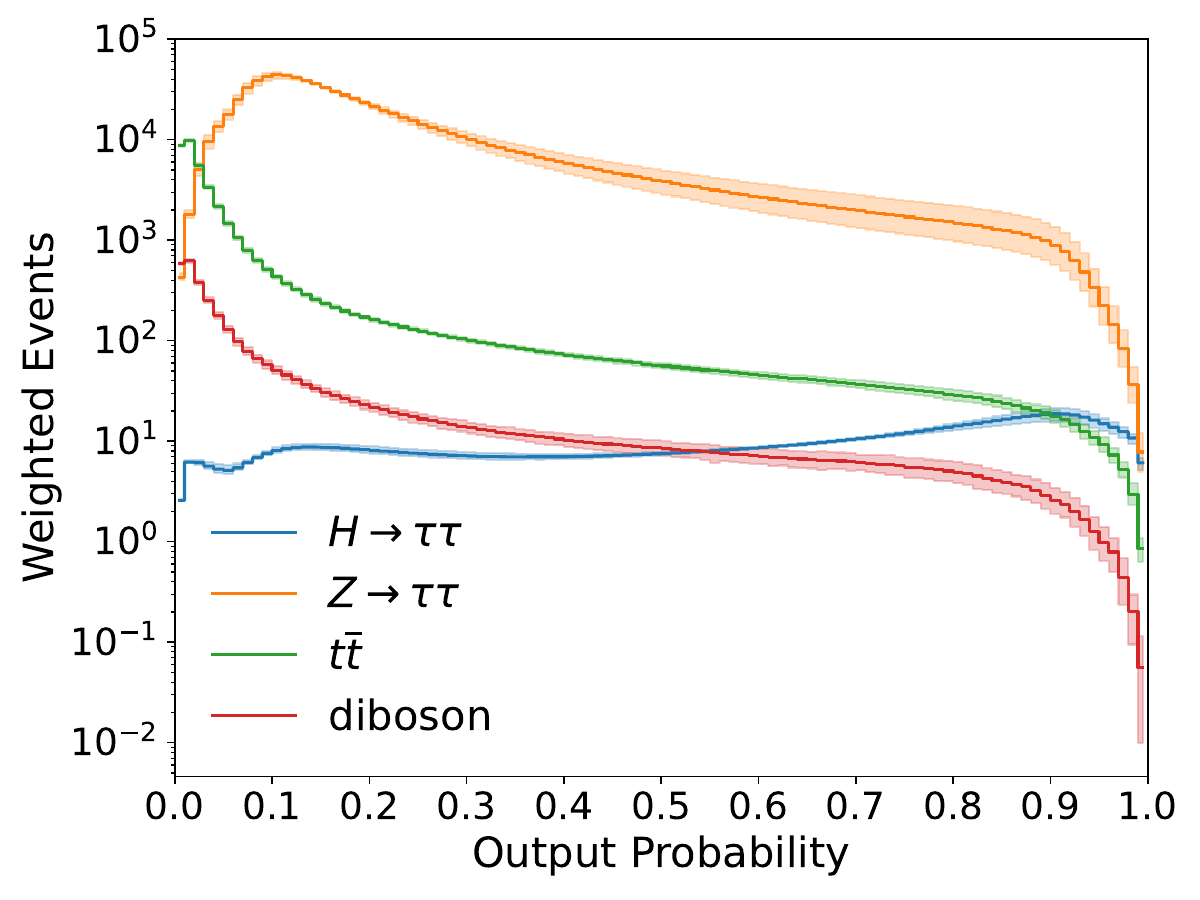}
    \caption{Distributions of the predicted signal-class scores from the re-trained (class-weighted) four-class classifier. Left: unweighted per-event histograms. Right: expected yields after applying per-class weights; curves are averaged over the same grid of 11849 fixed nuisance parameter settings as in the main text, and bands show the min–max envelope. Relative to Fig.~\ref{fig:signal_dist}, the signal shape is nearly unchanged, while $Z\!\to\!\tau\tau$ flattens and $t\overline t$/$VV$ become steeper due to class-frequency weighting.}
    \label{fig:signal_dist2}
\end{figure}

In Fig.~\ref{fig:signal_dist2}, we present the distributions of the predicted signal-class probabilities for the four processes based on the output of the retrained four-class classifier. Compared to the baseline in Fig.~\ref{fig:signal_dist}, the re-trained classifier (with class-balanced loss) shows only marginal changes for the signal class, whereas the background shapes respond more visibly once process weights are applied (right panel of Fig.~\ref{fig:signal_dist2}). In the weighted plot, the $Z\!\to\!\tau\tau$ distribution becomes noticeably flatter, while the $t\overline t$ and diboson distributions become steeper and concentrate more strongly toward the background-like score region. This pattern follows directly from the class-frequency weighting used during training and in the yield construction: as the dominant background, $Z\!\to\!\tau\tau$ receives the smallest weight, whereas the rarer $t\overline t$ and diboson processes are up-weighted, boosting their effective contributions and sharpening their weighted profiles. The left panel shows the corresponding unweighted score distributions from the re-trained classifier for reference. As in the main text, the weighted panel accounts for systematics by averaging expected bin counts over 11849 fixed nuisance configurations; the shaded bands indicate the full envelope across this grid and remain narrow, confirming that the score distributions are stable under nuisance variations even after class balancing.

\begin{figure*}
    \centering
    \includegraphics[width=0.5\linewidth]{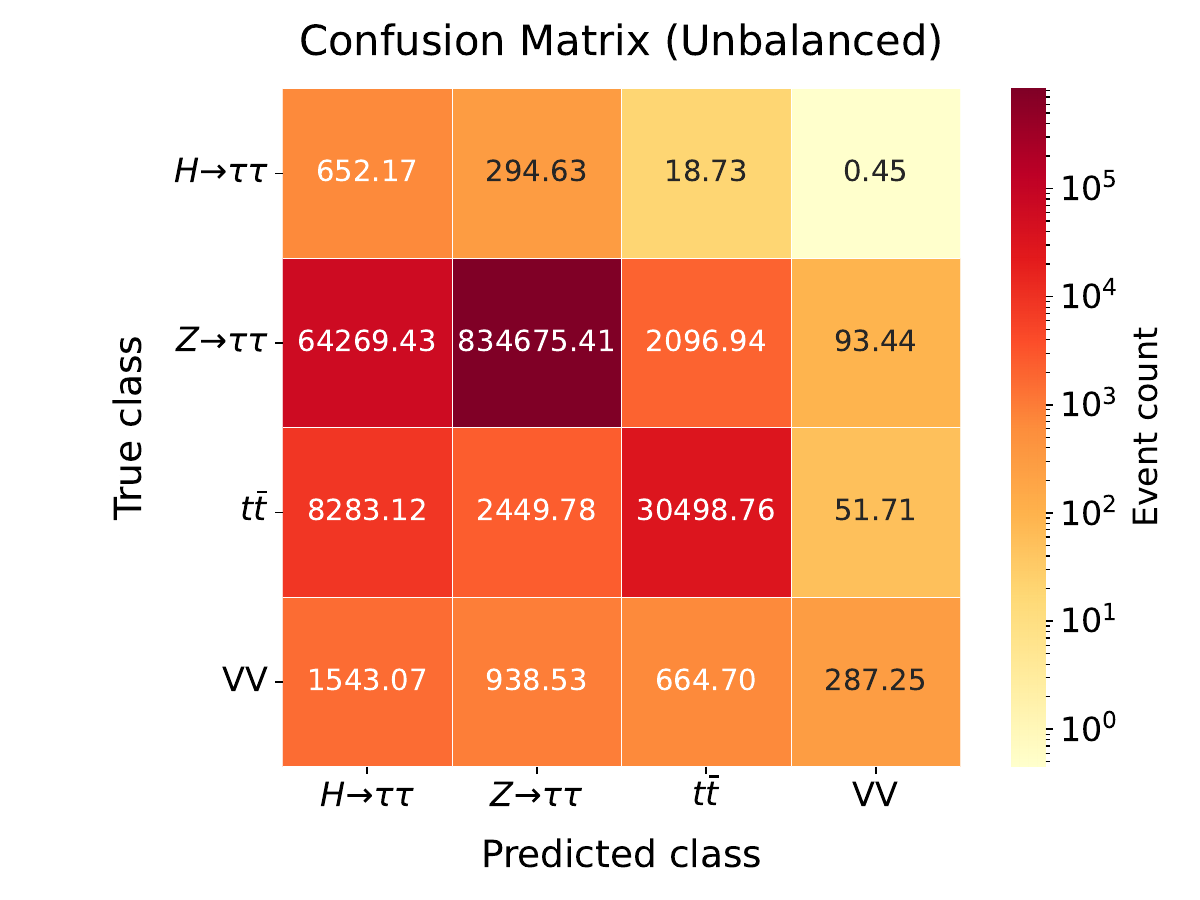}\hfill
    \includegraphics[width=0.5\linewidth]{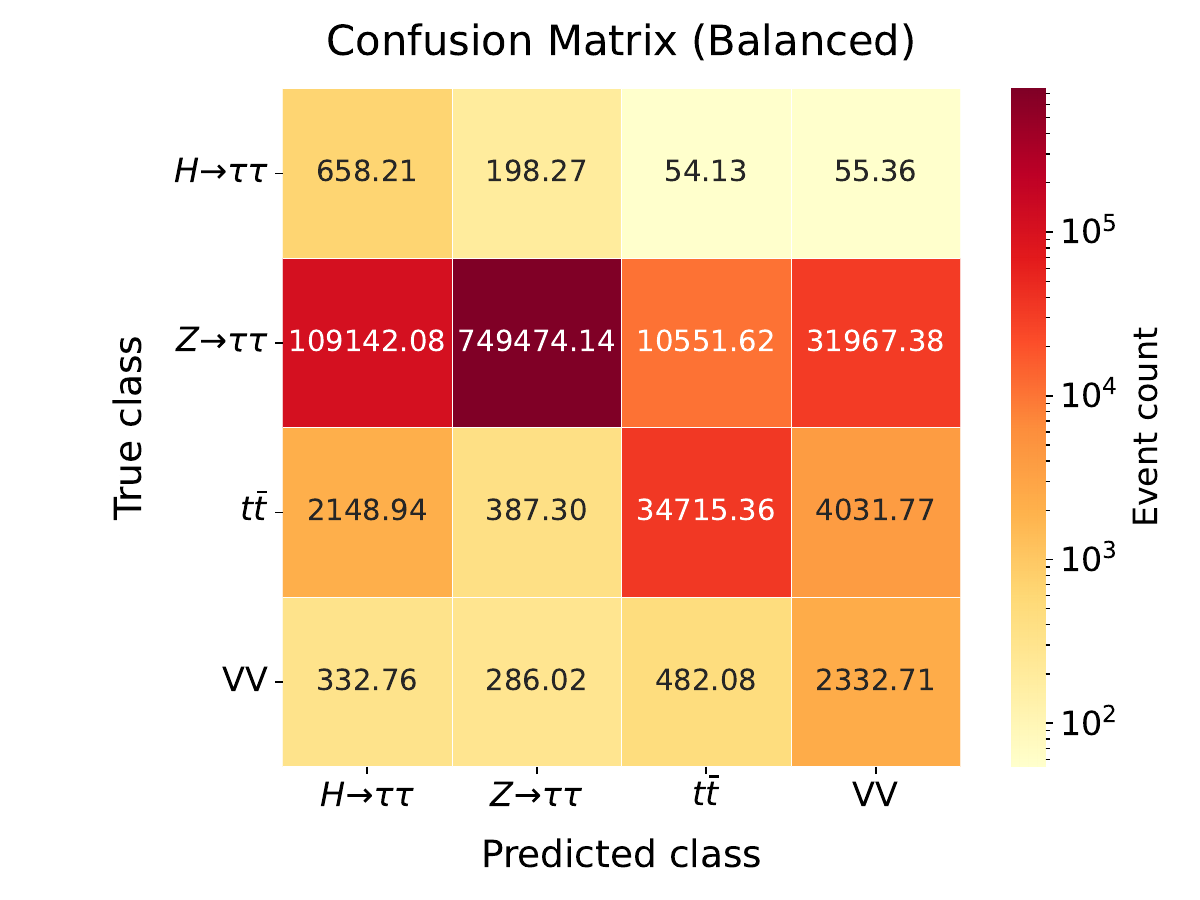}\\
    \caption{Confusion matrices for two four-class classifiers, shown on a logarithmic color scale. Rows indicate the true class and columns the predicted class; entries are weighted event yields. \textbf{Left:} classifier trained with uniform per-event weighting across all classes. \textbf{Right:} classifier trained with class-frequency weighting.}
\label{fig:cm}
\end{figure*}

In Fig.~\ref{fig:cm}, we present the confusion matrices for the two four-class classifiers. The vertical axis is the \emph{true class} and the horizontal axis is the \emph{predicted class}; the colour scale is logarithmic.
Each cell shows the \emph{weighted} event count (physical event yield) accumulated in that category.
The left panel corresponds to the classifier trained with uniform per-event weights, and the right panel to the classifier trained with class-frequency weighting. Comparing the two, the patterns shift as expected from the class weights
($w_{\text{VV}} > w_{\text{tt}} > w_{H\to\tau\tau} > w_{Z\to\tau\tau}$):

\begin{itemize}
  \item \textbf{$H\!\to\!\tau\tau$ row.}
  The correct classification rate remains similar, while misidentified signal events migrate more often to $t\overline{t}$ and VV, and less often to $Z\!\to\!\tau\tau$.
  \item \textbf{$Z\!\to\!\tau\tau$ row.}
  Because $Z\!\to\!\tau\tau$ receives the smallest weight, the weighted counts of events misidentified as $H\!\to\!\tau\tau$/$t\overline{t}$/VV increase markedly in the balanced model; the diagonal entry correspondingly decreases in weighted yield.
  \item \textbf{$t\overline{t}$ row.}
  The weighted number of correctly classified $t\overline{t}$ events increases noticeably; misidentifications into VV also increase, reflecting the up-weighting of both processes.
  \item \textbf{\texttt{VV} row.}
  With the largest class weight, the correctly classified \texttt{VV}$\!\to\!\texttt{VV}$ yield grows substantially in the balanced model, while misidentifications into other classes are relatively suppressed in weighted yield.
\end{itemize}

Overall, class-frequency weighting amplifies the effective contributions of the rarer $t\overline{t}$ and VV processes and suppresses the dominant $Z\!\to\!\tau\tau$, which explains the row-wise changes observed between the two confusion matrices.

Table~\ref{table:interpolation2} summarizes the selections and expected yields obtained with the re-trained, class-balanced classifier.
The requirements for the \texttt{inclusive}, \texttt{highMT-VBFJet} (CR1), and \texttt{highMT-noVBFJet-tt} (CR2) regions are \emph{identical} to those in Table~\ref{table:interpolation}.
However, because the classifier decision boundaries shift under class-frequency weighting, the flow of events into the \texttt{highMT-noVBFJet-VV} control region (CR3) changes appreciably.
In particular, the large class weight assigned to \texttt{VV} increases the weighted misidentification rates into the \texttt{VV} class from the other three processes. To preserve a \texttt{VV}-dominated sample and obtain a tight constraint on the normalization parameter $\nu_{\mathrm{VV}}$, we therefore tighten the CR3 classifier requirement to $\hat p_{VV}>0.9$ (last row), while keeping all kinematic cuts unchanged.
The table reflects the resulting \emph{weighted} Poisson yields per region and process and the corresponding $S/B$; the inclusive and CR1 yields remain consistent with Table~\ref{table:interpolation}, whereas CR2 and CR3 show the expected shifts due to the updated classifier decision boundaries.

\begin{table}[htbp]
  \footnotesize
  \centering
  \renewcommand{\arraystretch}{1.25}
  \setlength{\tabcolsep}{5pt}
  \begin{tabular}{|l|l|c|c|c|c|c|c|}
    \hline
    \multirow{2}{*}{\textbf{Region}} &
    \multirow{2}{*}{\textbf{Requirements}} &
    \multirow{2}{*}{\textbf{Type}} &
    \multicolumn{4}{c|}{\textbf{Poisson yield} $\mathcal{L}\sigma$} &
    \multirow{2}{*}{$S/B$} \\
    \cline{4-7}
    & & & $H\!\to\!\tau\tau$ & $Z\!\to\!\tau\tau$ & $t\overline t$ & VV & \\ \hline
    \texttt{inclusive} & -- & -- &
    966.0 & 901\,137.5 & 41\,283.4 & 3\,433.5 & $1.02\times10^{-3}$ \\ \hline
    \multirow{3}{*}{\texttt{highMT-VBFJet}} &
      $\pt^{j_1}\!>\!50$ GeV & \multirow{3}{*}{CR1} &
      \multirow{3}{*}{14.7} & \multirow{3}{*}{721.7} & \multirow{3}{*}{16\,768.6} & \multirow{3}{*}{193.2} & \multirow{3}{*}{$8.30\times10^{-4}$} \\ \cline{2-2}
    & $\pt^{j_2}\!>\!30$ GeV & & & & & & \\ \cline{2-2}
    & $\mT\!>\!70$ GeV      & & & & & & \\ \hline
    \multirow{3}{*}{\texttt{highMT-noVBFJet-tt}} &
      $\mT\!>\!70$ GeV & \multirow{3}{*}{CR2} &
      \multirow{3}{*}{3.0} & \multirow{3}{*}{304.0} &
      \multirow{3}{*}{3381.5} & \multirow{3}{*}{141.9} &
      \multirow{3}{*}{$7.96\times10^{-4}$} \\ \cline{2-2}
    & veto on \texttt{VBFJet} & & & & & & \\ \cline{2-2}
    & $\hat p_{t\overline t}\!>\!0.4$ & & & & & & \\ \hline
    \multirow{3}{*}{\texttt{highMT-noVBFJet-VV}} &
      $\mT\!>\!70$ GeV & \multirow{3}{*}{CR3} &
      \multirow{3}{*}{3.2} & \multirow{3}{*}{464.1} &
      \multirow{3}{*}{198.5} & \multirow{3}{*}{725.6} &
      \multirow{3}{*}{$2.3\times10^{-3}$} \\ \cline{2-2}
    & veto on \texttt{VBFJet} & & & & & & \\ \cline{2-2}
    & $\hat p_{VV}\!>\!0.9$ & & & & & & \\ \hline
  \end{tabular}
  \caption{Event selections and expected yields with the re-trained (class-frequency–weighted) four-class classifier. The \texttt{inclusive}, \texttt{highMT-VBFJet} (CR1), and \texttt{highMT-noVBFJet-tt} (CR2) requirements are unchanged from Table~\ref{table:interpolation}; CR3 keeps the same kinematic cuts but uses a tighter classifier threshold, $\hat p_{VV}>0.9$, to preserve \texttt{VV} dominance under class weighting. Entries are expected Poisson yields ($\mathcal{L}\sigma$) per process; the last column reports $S/B$. Differences relative to Table~\ref{table:interpolation} reflect shifts in decision boundaries.}
  \label{table:interpolation2}
\end{table}

\begin{figure*}
    \centering
    \includegraphics[width=0.48\linewidth]{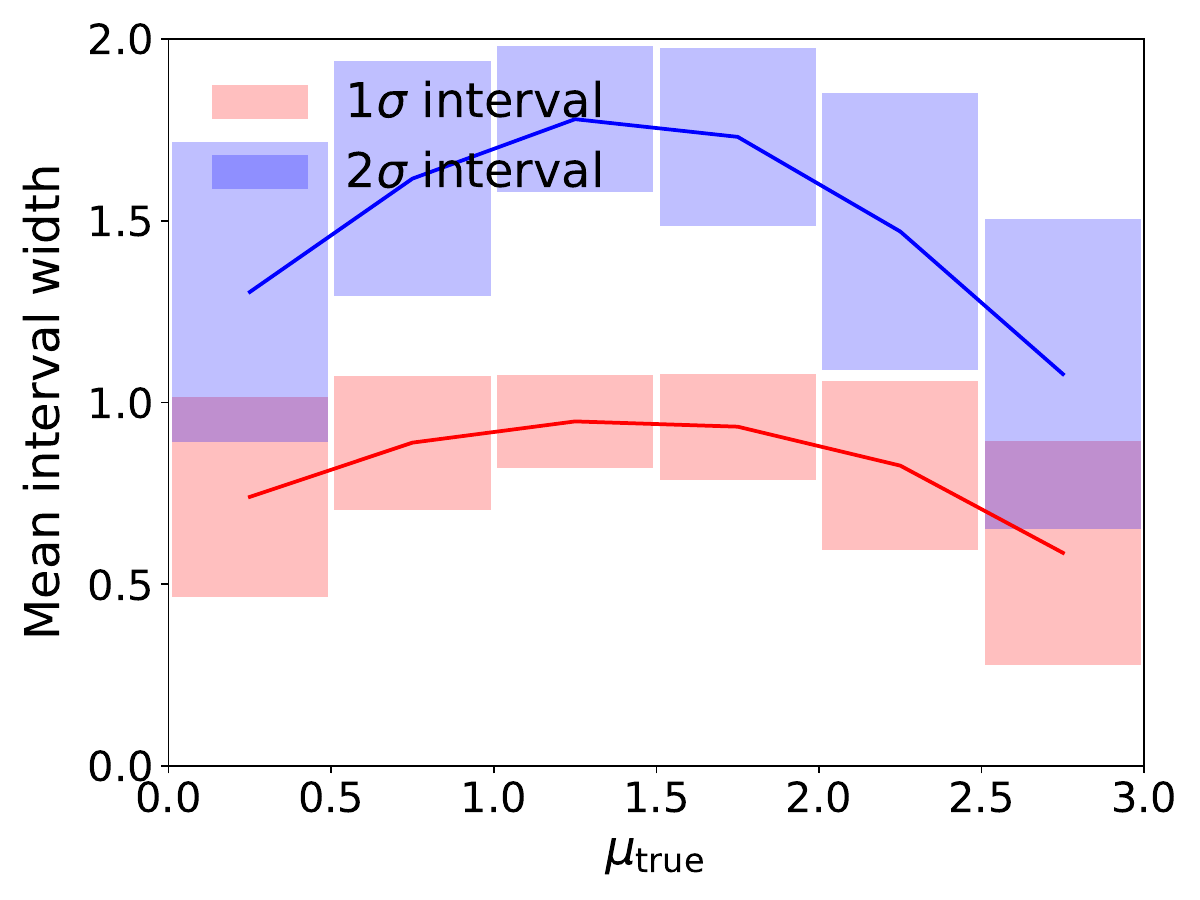}\hfill
    \includegraphics[width=0.48\linewidth]{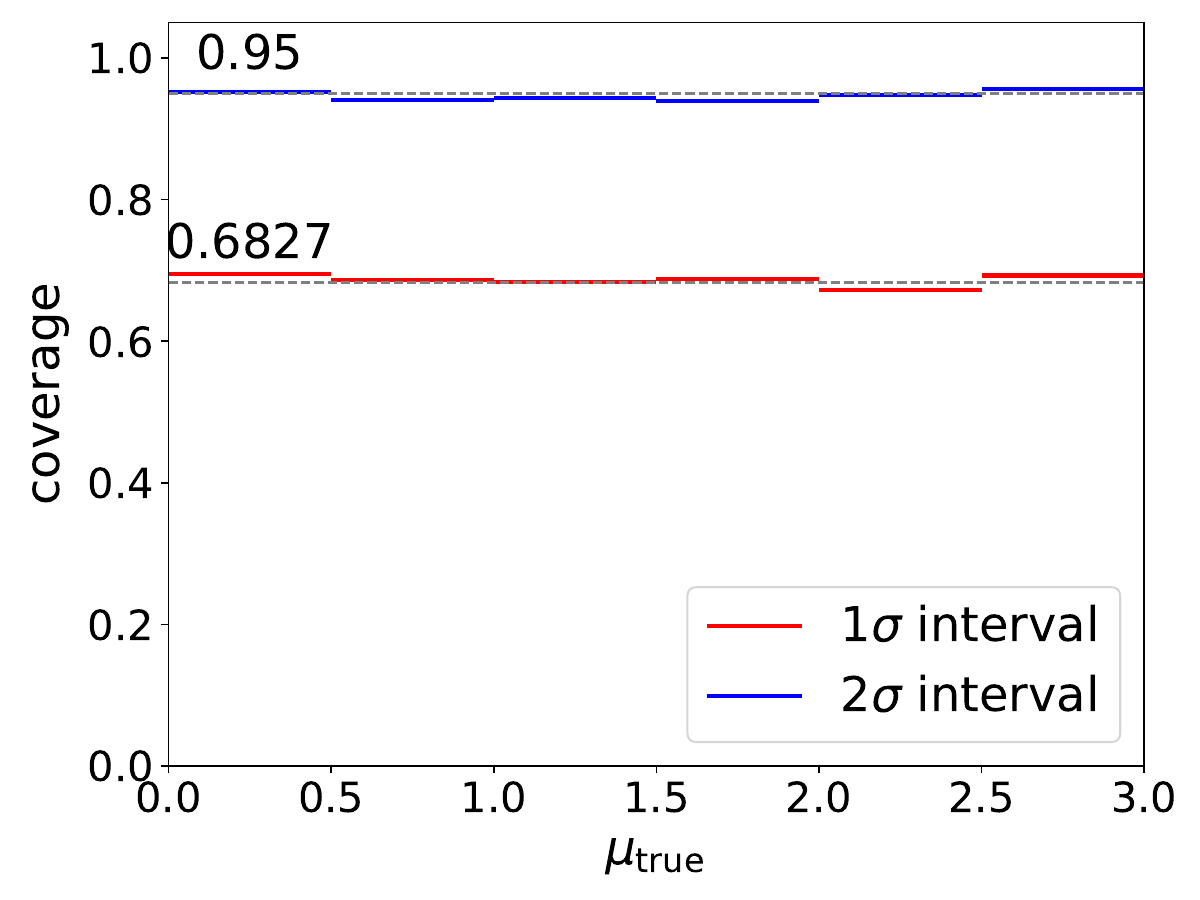}\\
    \caption{\label{fig:toy-coverage2} Results obtained with the re-trained (class-frequency–weighted) classifier; the full surrogate-likelihood pipeline matches the main-text setup. 
Left: average widths of the 68.3\% (red) and 95\% (blue) intervals as functions of $\mu^\text{true}$. 
Right: empirical coverage of the corresponding intervals. 
All curves are derived from 50000 pseudo-experiments.}
\end{figure*}

In Fig~\ref{fig:toy-coverage2}, we show the average width of the confidence interval for the signal strength $\mu$ and the corresponding coverage based on retrained classifier from 50000 pseudo-experiments. Relative to the baseline results in Fig.~\ref{fig:toy-coverage}, the \emph{coverage} obtained with the class-balanced classifier remains essentially unchanged across the full range of $\mu^\text{true}$, with fluctuations compatible with the Monte Carlo statistics. In contrast, the \emph{intervals} are systematically wider. The broadening is driven by how class-frequency weighting reshapes the effective composition of the signal region: after re-training, misidentifications of $t\overline t$ and VV into the signal class are reduced, while misidentifications of the dominant $Z\!\to\!\tau\tau$ background into the signal class increase. The net effect---visible both in the weighted score distributions of Fig.~\ref{fig:signal_dist2} and in the region yields of Table~\ref{table:interpolation2}---is a higher background contamination in the signal-enriched region, which reduces the sensitivity to variations in $\mu$ and therefore leads to wider confidence intervals.

\section{Robustness to Kinematics-Only Control Regions}
\label{sec:kin_CR}

To assess the robustness of the inference with respect to the definition of control regions, we repeat the full pseudo-experiment study using control regions defined exclusively by kinematic selections, without relying on classifier outputs. The resulting control-region definitions and their expected yields are summarized in 
Table~\ref{table:interpolation3}. Compared to the baseline setup, the alternative selections lead to control regions in which the $t\overline t$ and diboson processes remain dominant but with slightly reduced purity, reflecting the absence of explicit classifier-based information. Note that for CR3, the kinematic selections are chosen to enhance the diboson fraction as much as possible while retaining adequate statistics.

\begin{table}[htbp]
  \footnotesize
  \centering
  \renewcommand{\arraystretch}{1.25}
  \setlength{\tabcolsep}{5pt}
  \begin{tabular}{|l|l|c|c|c|c|c|c|}
    \hline
    \multirow{2}{*}{\textbf{Region}} &
    \multirow{2}{*}{\textbf{Requirements}} &
    \multirow{2}{*}{\textbf{Type}} &
    \multicolumn{4}{c|}{\textbf{Poisson yield} $\mathcal{L}\sigma$} &
    \multirow{2}{*}{$S/B$} \\
    \cline{4-7}
    & & & $H\!\to\!\tau\tau$ & $Z\!\to\!\tau\tau$ & $t\overline t$ & VV & \\ \hline
    \multirow{4}{*}{\texttt{highMT-noVBFJet-tt}} &
      $\mT\!>\!70$ GeV & \multirow{4}{*}{CR2} &
      \multirow{4}{*}{4.3} & \multirow{4}{*}{384.7} &
      \multirow{4}{*}{2306.5} & \multirow{4}{*}{172.9} &
      \multirow{4}{*}{$1.50\times10^{-3}$} \\ \cline{2-2}
    & veto on \texttt{VBFJet} & & & & & & \\ \cline{2-2}
    & $\sum p_T\!>\!200$ GeV & & & & & & \\ \cline{2-2}
    & $p_T^H\!>\!29$ GeV & & & & & & \\ \hline

    \multirow{4}{*}{\texttt{highMT-noVBFJet-VV}} &
      $\mT\!>\!70$ GeV & \multirow{4}{*}{CR3} &
      \multirow{4}{*}{5.7} & \multirow{4}{*}{904.7} &
      \multirow{4}{*}{137.0} & \multirow{4}{*}{470.3} &
      \multirow{4}{*}{$3.4\times10^{-3}$} \\ \cline{2-2}
    & veto on \texttt{VBFJet} & & & & & & \\ \cline{2-2}
    & $\sum p_T\!<\!150$ GeV & & & & & & \\ \cline{2-2}
    & $p_T^H\!<\!20$ GeV & & & & & & \\ \hline

  \end{tabular}
  \caption{Summary of the event selections. Control regions are denoted by the ``CR'' label. Expected Poisson event yields, calculated as $\mathcal{L}\sigma$ with $\mathcal{L}$ representing the integrated luminosity, are provided for each process across different regions. The final column lists the corresponding signal-to-background ratios.}
  \label{table:interpolation3}
\end{table}

\begin{figure*}
    \centering
    \includegraphics[width=0.48\linewidth]{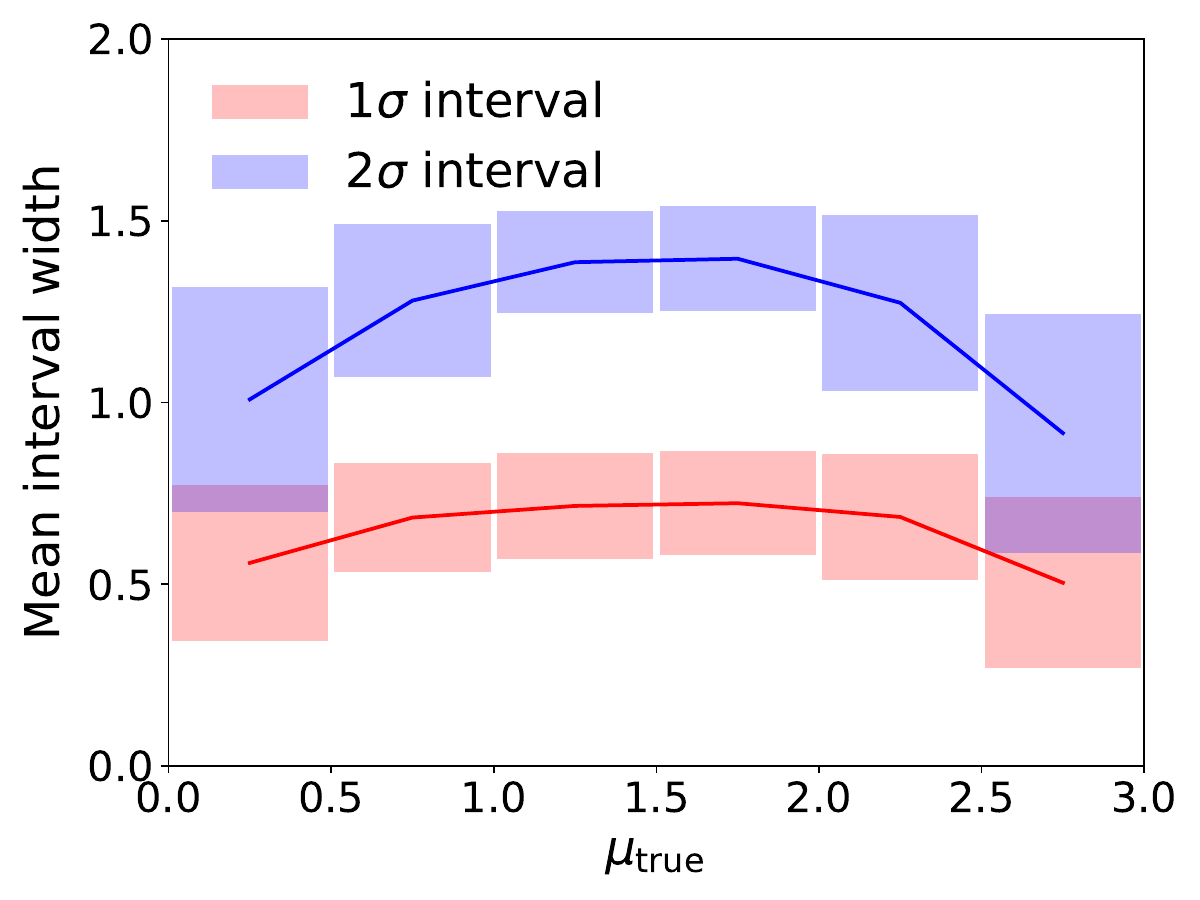}\hfill
    \includegraphics[width=0.48\linewidth]{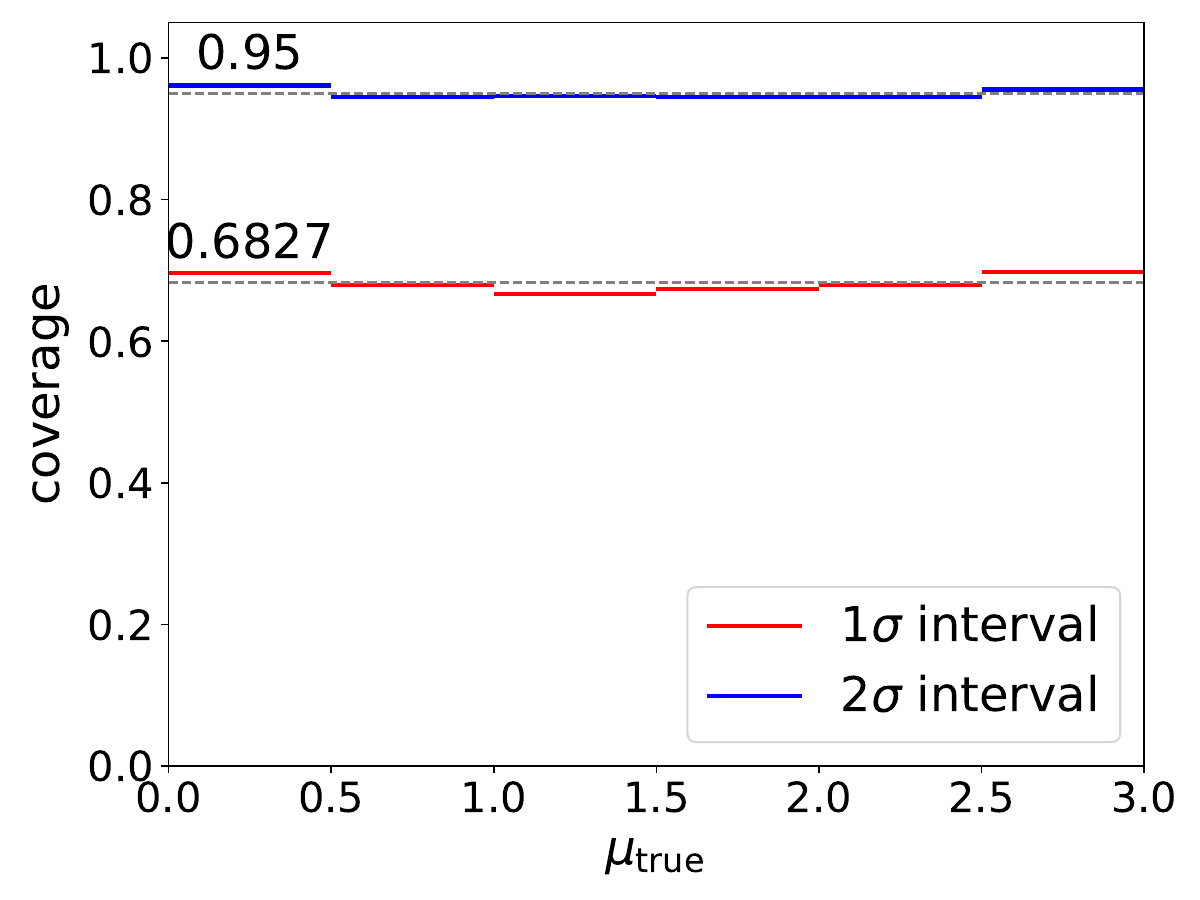}\\
    \caption{\label{fig:toy-coverageCR} The average width of the confidence interval for the signal strength $\mu$ (left) and the corresponding coverage (right) are shown as functions of the true value $\mu^\text{true}$, based on results from 50000 pseudo-experiments. These results correspond to the dedicated control-region study using control regions defined purely by kinematic selections. The average interval length is 0.6822 and the coverage is 0.6523.}
\end{figure*}

\begin{figure*}
    \centering
    \includegraphics[width=0.48\linewidth]{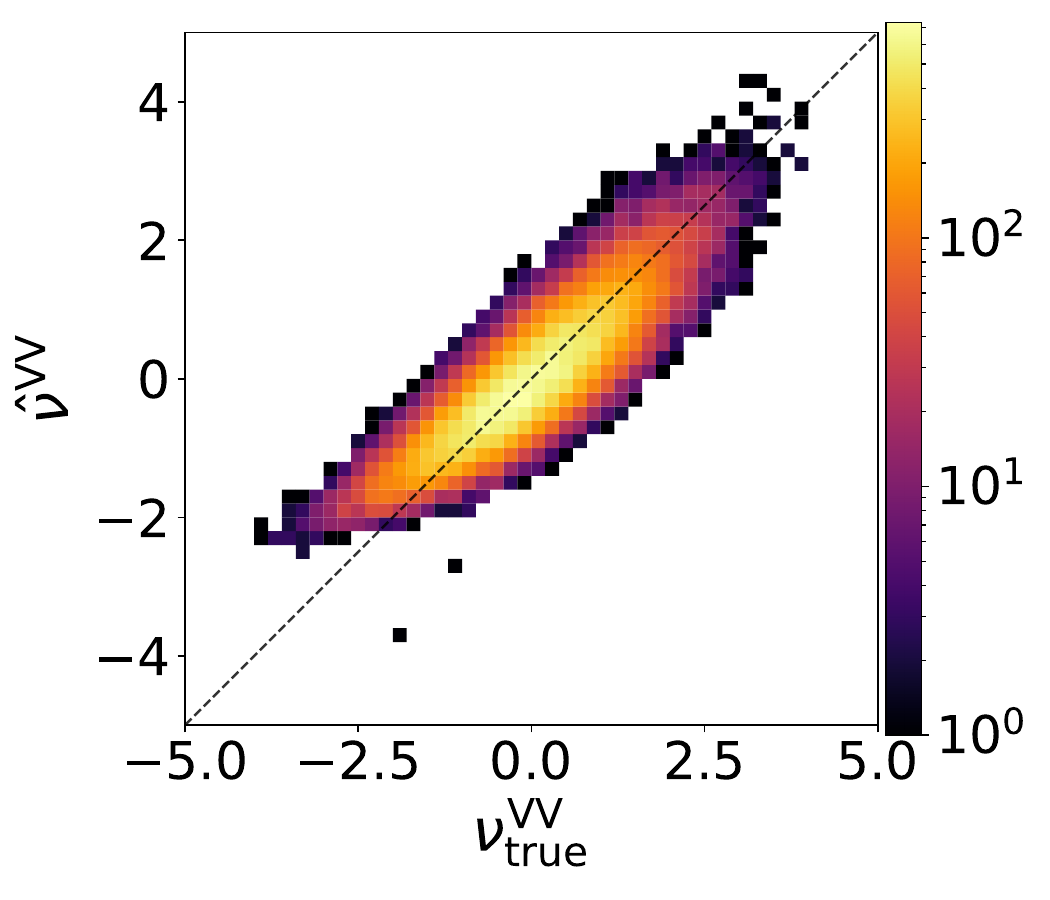}\hfill
    \includegraphics[width=0.48\linewidth]{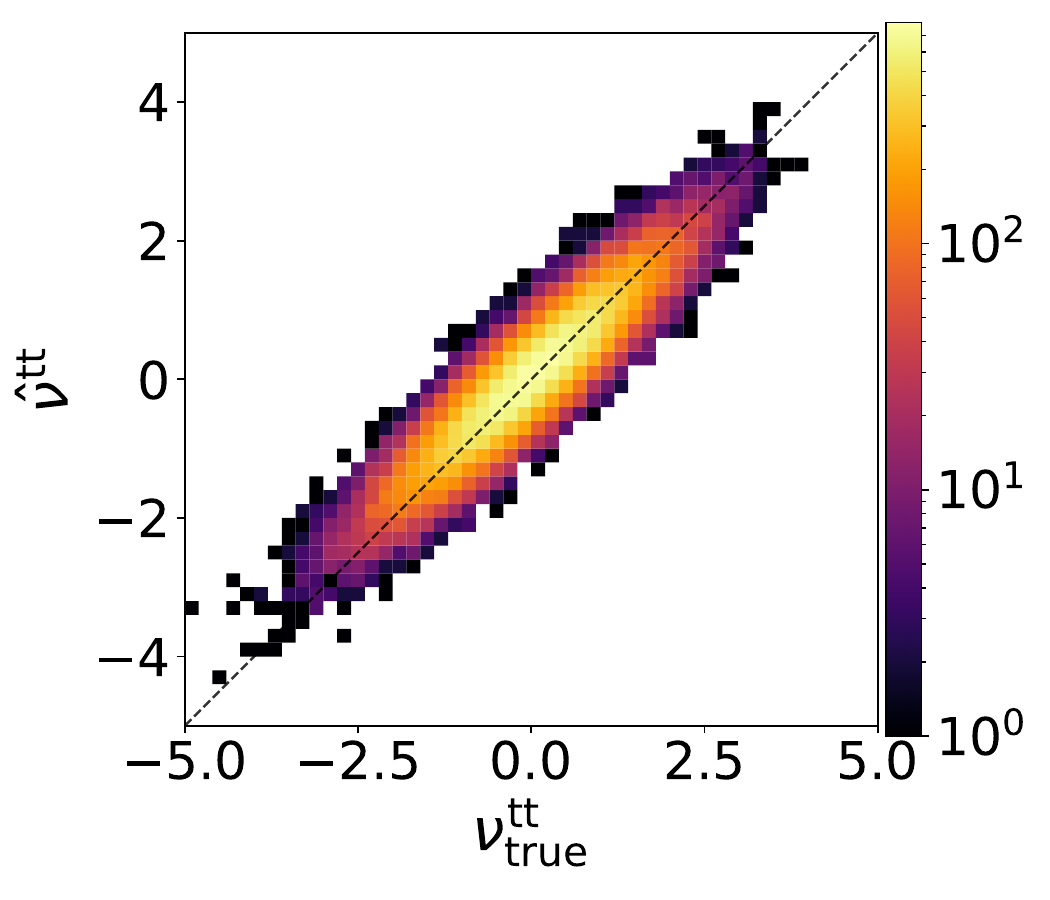}\\
    \caption{\label{fig:toy-scatter2} Scatter plot of $\hat{\nu}^{\ttbar}$, and $\hat{\nu}^{\VV}$ versus $\nu^{\ttbar}_{true}$, and $\nu^{\VV}_{true}$ from 50000 pseudo-experiments. }
\end{figure*}

The impact of this modification on the signal-strength inference is shown in Fig.~\ref{fig:toy-coverageCR}. We observe that both the confidence-interval width and the coverage remain largely stable 
across the full range of $\mu^\text{true}$. Relative to the baseline analysis, the average interval width exhibits a mild increase, while the coverage shows a small downward shift. Both effects are consistent with the expectation that the kinematics-only control regions 
provide slightly weaker constraints on the background normalization than their classifier-defined counterparts, yet without inducing any qualitative change in the statistical performance of the $\mu$ extraction.

As anticipated, the primary impact of the modified control regions is observed in the normalization nuisance parameters that they are designed to constrain. Figure~\ref{fig:toy-scatter2} shows the fitted values of $\nu^{t\overline t}$ and $\nu^{VV}$ as functions of their true values. Compared to the baseline setup, the correlation bands become moderately broader, indicating a relaxation of the constraints on these nuisance parameters. Importantly, the fitted values remain centered around the diagonal, and no systematic bias or instability is observed. This behavior is consistent with the observed small changes in interval width and coverage, and confirms that the signal-strength inference remains robust under alternative, purely kinematic control-region definitions.

\section{Robustness with Respect to the Interpolation Grid Density}
\label{sec:grid_density}

In the construction of the surrogate likelihood, the expected event yields are obtained through multi-dimensional interpolation over a discrete grid of nuisance parameters. The choice of the interpolation grid density therefore represents a potential source of numerical uncertainty. To assess the robustness of the inference with respect to this choice, we repeat the full pseudo-experiment study using three different interpolation grids of decreasing resolution: a baseline grid ($17\times17\times41$), a medium grid ($9\times9\times21$), and a coarse grid ($5\times5\times11$).

\begin{figure*}
    \centering
    \includegraphics[width=0.48\linewidth]{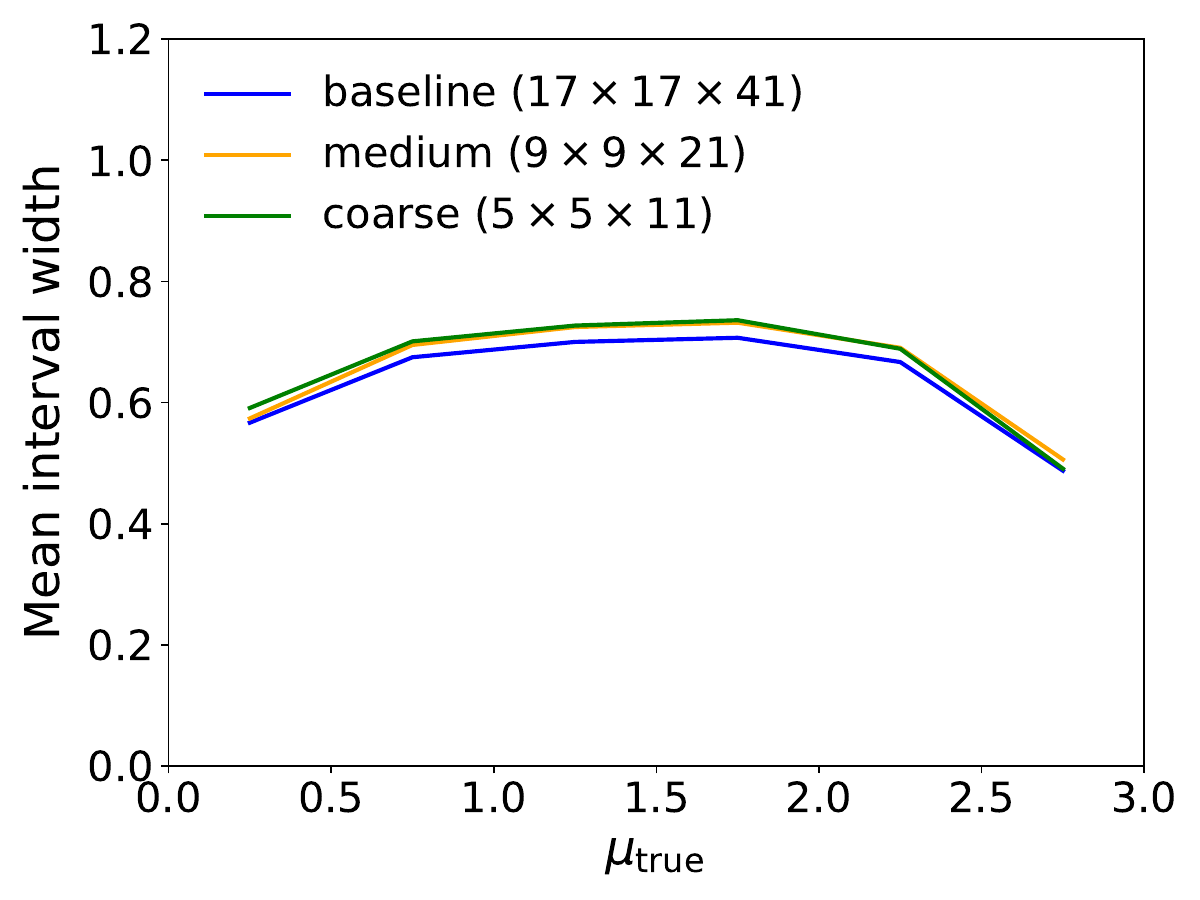}\hfill
    \includegraphics[width=0.48\linewidth]{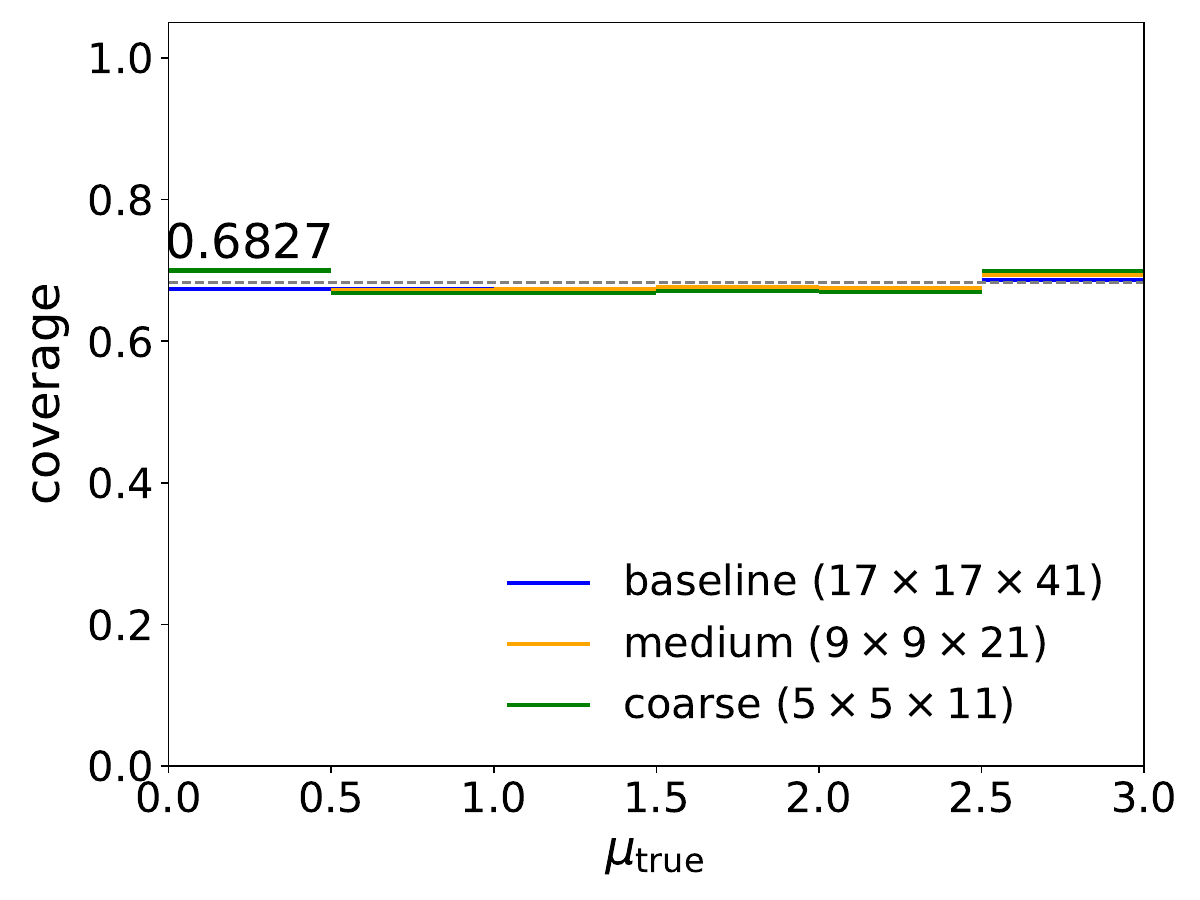}\\
    \caption{\label{fig:toy-coverage-grid}
    Dependence of the signal-strength inference on the interpolation grid density. The left panel shows the mean width of the $1\sigma$ confidence interval for the signal strength $\mu$, while the right panel shows the corresponding empirical coverage, both as functions of the true value $\mu^{\mathrm{true}}$. Results are obtained from 50000 pseudo-experiments for three different interpolation grids: baseline ($17\times17\times41$), medium ($9\times9\times21$), and coarse ($5\times5\times11$). The average interval widths are 0.634 (baseline), 0.690 (medium), and 0.691 (coarse), with corresponding coverages of 0.670, 0.658, and 0.657, respectively. The close agreement among the three configurations demonstrates the stability of the inference with respect to the interpolation grid density.
    }
\end{figure*}

Figure~\ref{fig:toy-coverage-grid} summarizes the impact of the grid density on the signal-strength inference. The mean width of the $1\sigma$ confidence interval and the corresponding empirical coverage
are shown as functions of the true signal strength $\mu^{\mathrm{true}}$, based on 50000 pseudo-experiments for each configuration. We observe that both quantities remain largely stable across the full $\mu^{\mathrm{true}}$ range, with only modest variations as the grid resolution is reduced. The slightly larger interval widths and mildly reduced coverages observed for the coarser
grids are consistent with the expected loss of numerical precision in the interpolation. Importantly, these effects remain small and do not lead to any qualitative change in the statistical performance of the signal-strength extraction. We therefore conclude that the results presented in this work are robust with respect to the choice of interpolation grid density.

\section{Ablation Studies}
\subsection{Robustness to the Number of Nuisance Replicas}

We study the robustness of the dual-branch GNN architecture with respect to the number of nuisance replicas used during training. The replica count controls the Monte Carlo approximation employed to enforce robustness against systematic variations, and thus represents a relevant training hyperparameter. To quantify the impact of this choice in a model-agnostic manner, we evaluate the stability of the classifier output under nuisance variations using the 1D Wasserstein distance. For each trained model, we compute the Wasserstein distance between the signal-class output distribution at each nuisance grid point and the corresponding nominal distribution, separately for each process. The distances are then averaged over all 11848 nuisance points which are uniformly sampled over the range of $\alpha_{\text{tes}} \in$ [0.96, 1.04], $\alpha_{\text{jes}} \in$ [0.96, 1.04], $\alpha_{\text{met}} \in$ [0, 5], providing a single robustness metric per process.

\begin{table}[t]
\centering
\small
\renewcommand{\arraystretch}{1.35}
\setlength{\tabcolsep}{10pt}
\begin{tabular}{lcccc}
\hline
\textbf{Replica} 
& $H\!\to\!\tau\tau$ 
& $Z\!\to\!\tau\tau$ 
& $t\overline{t}$ 
& Diboson \\
\hline
20  & 0.01294  & 0.00937  & 0.00102  & 0.00254 \\
50  & 0.01292  & 0.00931  & 0.00101  & 0.00237 \\
80  & 0.01284  & 0.00929  & 0.00098  & 0.00233 \\
100 & 0.01282  & 0.00925  & 0.00098  & 0.00231 \\
\hline
\end{tabular}
\caption{
Average 1D Wasserstein distance between the signal-class output distribution at each nuisance point and the nominal distribution, averaged over 11848 nuisance points. Results are shown separately for each process.
}
\label{tab:wasserstein_replicas}
\end{table}

The results are summarized in Table~\ref{tab:wasserstein_replicas}. As the number of replicas increases from 20 to 100, the average Wasserstein distance decreases monotonically and gradually approaches a stable plateau for all processes. This behavior indicates that the classifier response to nuisance variations becomes increasingly smooth and that the robustness of the model saturates for sufficiently large replica counts. In particular, the changes between 50, 80, and 100 replicas are small compared to the overall scale of the distances. Based on this observation, we adopt a replica count of 100 in the nominal training configuration. This choice lies in the plateau regime of the robustness metric while remaining compatible with the available GPU memory on NVIDIA V100 hardware, thereby providing a stable and computationally feasible setup.

\subsection{Robustness to Feature Partitioning and Architectural Variants}
\label{Robustness-Architecture}

In this subsection, we investigate the robustness of the proposed dual-branch GNN with respect to both the partitioning of input features into deterministic and uncertainty-aware branches and the specific architectural components used to handle nuisance variations. This study addresses the question of where the observed robustness gains originate and assesses the sensitivity of the method to the assumed feature categorization. We consider four architectural variants. The ``Baseline'' corresponds to the full dual-branch GNN. In ``Baseline + UA($\Delta R(\tau_\text{had},\ell)$)'', the angular separation $\Delta R(\tau_\text{had},\ell)$ is moved from the deterministic branch to the uncertainty-aware branch, testing the robustness of the method to a modified feature partition. The ``Baseline w/o UA-attention'' variant removes the attention and gating mechanism from the uncertainty-aware branch while keeping the rest of the architecture unchanged. Finally, the ``Single UA'' model corresponds to a single-branch uncertainty-aware architecture in which all systematically affected input features are processed without a separate deterministic branch.

\begin{table}[t]
\centering
\small
\renewcommand{\arraystretch}{1.35}
\setlength{\tabcolsep}{9pt}
\begin{tabular}{lcccc}
\hline
\textbf{Model} 
& $H\!\to\!\tau\tau$ 
& $Z\!\to\!\tau\tau$ 
& $t\overline{t}$ 
& Diboson \\
\hline
Baseline 
& 0.01282 & 0.00925 & 0.00098 & 0.00231 \\

Baseline + UA($\Delta R(\tau_\text{had},\ell)$) 
& 0.01286 & 0.00988 & 0.00098 & 0.00237 \\

Baseline w/o UA-attention 
& 0.01285 & 0.00987 & 0.00103 & 0.00234 \\

Single UA 
& 0.01189 & 0.00796 & 0.00105 & 0.00232 \\

Baseline + nominal training
& 0.01296 & 0.00965 & 0.00106 & 0.00232 \\

Single GNN + nominal training
& 0.01291 & 0.00963 & 0.00102 & 0.00236 \\
\hline
\end{tabular}
\caption{
Average 1D Wasserstein distance between the signal-class output distribution at each nuisance point and the nominal distribution, averaged over 11848 nuisance configurations.
The Baseline corresponds to the full dual-branch GNN trained with nuisance variations.
In ``Baseline + UA($\Delta R$)'', the feature $\Delta R(\tau_\text{had},\ell)$ is moved from the deterministic branch to the uncertainty-aware branch.
``Baseline w/o UA-attention'' denotes the full model with the attention mechanism removed from the uncertainty-aware branch.
``Single UA'' refers to a single-branch uncertainty-aware GNN in which all systematically affected input features are processed without a separate deterministic branch.
``Baseline + nominal training'' uses the same dual-branch architecture as the Baseline but is trained exclusively on nominal samples.
``Single GNN + nominal training'' processes all 35 input features within a single GNN branch and is likewise trained exclusively on nominal samples.
}
\label{tab:wasserstein_architecture_ablation}
\end{table}

The robustness of each model is quantified using the average 1D Wasserstein distance between the signal-class output distribution at each nuisance configuration and the nominal distribution, evaluated over 11848 nuisance points and shown in Table~\ref{tab:wasserstein_architecture_ablation}. We observe that the Baseline model exhibits a stable and consistently small Wasserstein distance across all processes. Moving the key deterministic feature $\Delta R(\tau_\text{had},\ell)$ to the uncertainty-aware branch or removing the uncertainty-branch attention mechanism leads to only minor changes in the robustness metric, indicating that the method is not strongly sensitive to these architectural variations. The Single UA model yields the smallest Wasserstein distance for the signal and $Z\to\tau\tau$ processes, although, as discussed below, this increased stability is accompanied by a substantial loss in nominal discrimination power.

To further disentangle the effect of nuisance-aware training from that of the network architecture, we additionally evaluate two models trained exclusively on nominal samples. The first retains the original dual-branch architecture (``Baseline + nominal training''), while the second processes all 35 input features jointly within a single GNN branch (``Single GNN + nominal training''). For the $H\to\tau\tau$ process, both nominally trained models exhibit slightly larger Wasserstein distances than the nuisance-aware Baseline, with values of 0.01296 and 0.01291, respectively, compared with 0.01282 for the Baseline. For the background processes, the differences remain modest. Moreover, the two nominal-training configurations yield very similar Wasserstein distances across all four processes, suggesting that replacing the dual-branch architecture with a single GNN does not by itself lead to a substantial change in the sensitivity of the classifier output to nuisance variations.

\begin{table}[t]
\centering
\small
\renewcommand{\arraystretch}{1.3}
\setlength{\tabcolsep}{12pt}
\begin{tabular}{lc}
\hline
\textbf{Model} & \textbf{AUC (Signal vs Rest)} \\
\hline
Baseline 
& 0.8901 \\

Baseline + UA($\Delta R(\tau_\text{had},\ell)$)
& 0.8892 \\

Baseline w/o UA-attention
& 0.8887 \\

Single UA
& 0.8436 \\

Baseline + nominal training
& 0.8883 \\

Single GNN + nominal training
& 0.8911 \\
\hline
\end{tabular}
\caption{
Area under the ROC curve (AUC) for $H\rightarrow\tau\tau$ signal versus all background processes, evaluated on the nominal test sample, for the architectural and training variants considered in this study.
}
\label{tab:auc_architecture_ablation}
\end{table}

To assess the impact of these architectural and training choices on nominal classification performance, we report the corresponding AUC values for $H\rightarrow\tau\tau$ signal versus all backgrounds in Table~\ref{tab:auc_architecture_ablation}. While the Single UA model exhibits enhanced robustness, it shows a substantially reduced discrimination power. In contrast, the Baseline and its mild architectural variants retain high AUC values while maintaining comparable robustness.

The two nominally trained models also retain essentially the same nominal discrimination performance as the nuisance-aware Baseline. The dual-branch model trained exclusively on nominal samples achieves an AUC of 0.8883, while the single-GNN nominal-training configuration yields 0.8911, compared with 0.8901 for the nuisance-aware Baseline. Thus, processing all 35 input features jointly within a single GNN does not lead to a substantial improvement in nominal classification performance. Taken together with the Wasserstein-distance results in Table~\ref{tab:wasserstein_architecture_ablation}, these results indicate that the proposed dual-branch nuisance-aware model preserves nominal signal--background discrimination while maintaining a similarly stable response to nuisance variations. In particular, the single-GNN nominal baseline exhibits only a marginal gain in nominal AUC, and its robustness remains comparable to that of the nominally trained dual-branch model.

\section{Comparison of Systematic-Aware Training to a Classical Template-Morphing Treatment}

To explicitly quantify the benefit of incorporating systematic uncertainties during training, it is essential to compare the proposed systematic-aware approach to a classical treatment that reflects standard analysis practice. The comparison presented in this appendix is designed to isolate the impact of systematic-aware training by adopting such a classical template-based treatment as a baseline, while keeping the network architecture and inference procedure otherwise identical. For the classical treatment of systematic uncertainties, we construct binned templates from the network outputs and incorporate systematic effects through a standard template morphing procedure at the likelihood level. The same dual-branch network architecture as used in the systematic-aware approach is employed, but trained exclusively on nominal samples. No systematic variations are injected during training.

For each control region, the network output is histogrammed and rebinned following exactly the same binning strategy and statistical requirements as described in the main text, ensuring a consistent likelihood construction between the classical baseline and the systematic-aware approach. Systematic uncertainties are modeled using a quadratic template morphing scheme with cross terms. Three nuisance parameters are considered in this study, namely TES, JES and MET. For each nuisance parameter, templates are produced at one nominal point and two symmetric variations. In addition, mixed “up–up” templates are included to capture correlations between different sources of uncertainty. In total, ten systematic states are used to construct the morphing model.

For each process and each rebinned bin, the expected yield $T(\boldsymbol{\alpha})$ is parameterized as a second-order polynomial in the nuisance parameters,
\[
T(\boldsymbol{\alpha}) = T_0
+ \sum_i a_i \alpha_i
+ \sum_i b_{ii} \alpha_i^2
+ \sum_{i<j} b_{ij} \alpha_i \alpha_j .
\]
where $T_0$ denotes the nominal template, the linear and quadratic coefficients are determined from the single-parameter variations, and the cross terms are fixed using the mixed variation templates. This construction ensures an exact interpolation at all provided systematic points and reproduces the standard quadratic morphing prescription commonly used in experimental analyses. The resulting interpolated templates are then used in a profile likelihood fit, where the nuisance parameters are treated as constrained parameters and profiled simultaneously with the signal strength $\mu$. This approach serves as the classical baseline against which the systematic-aware training strategy is compared.

\begin{table}[t]
\centering
\small
\renewcommand{\arraystretch}{1.3}
\setlength{\tabcolsep}{10pt}
\begin{tabular}{lcccc}
\hline
\textbf{Model} 
& Baseline 
& Nominal-training \\
\hline
AUC (Signal vs Rest) 
& 0.8901 
& 0.8883  \\
\hline
\end{tabular}
\caption{
Area under the ROC curve (AUC) for $H\rightarrow{\tau\tau}$ signal versus all background processes, evaluated on the nominal test sample, comparing nominal training and the baseline configuration.
}
\label{tab:auc_architecture_ablationx}
\end{table}

\begin{figure*}
    \centering
    \includegraphics[width=0.48\linewidth]{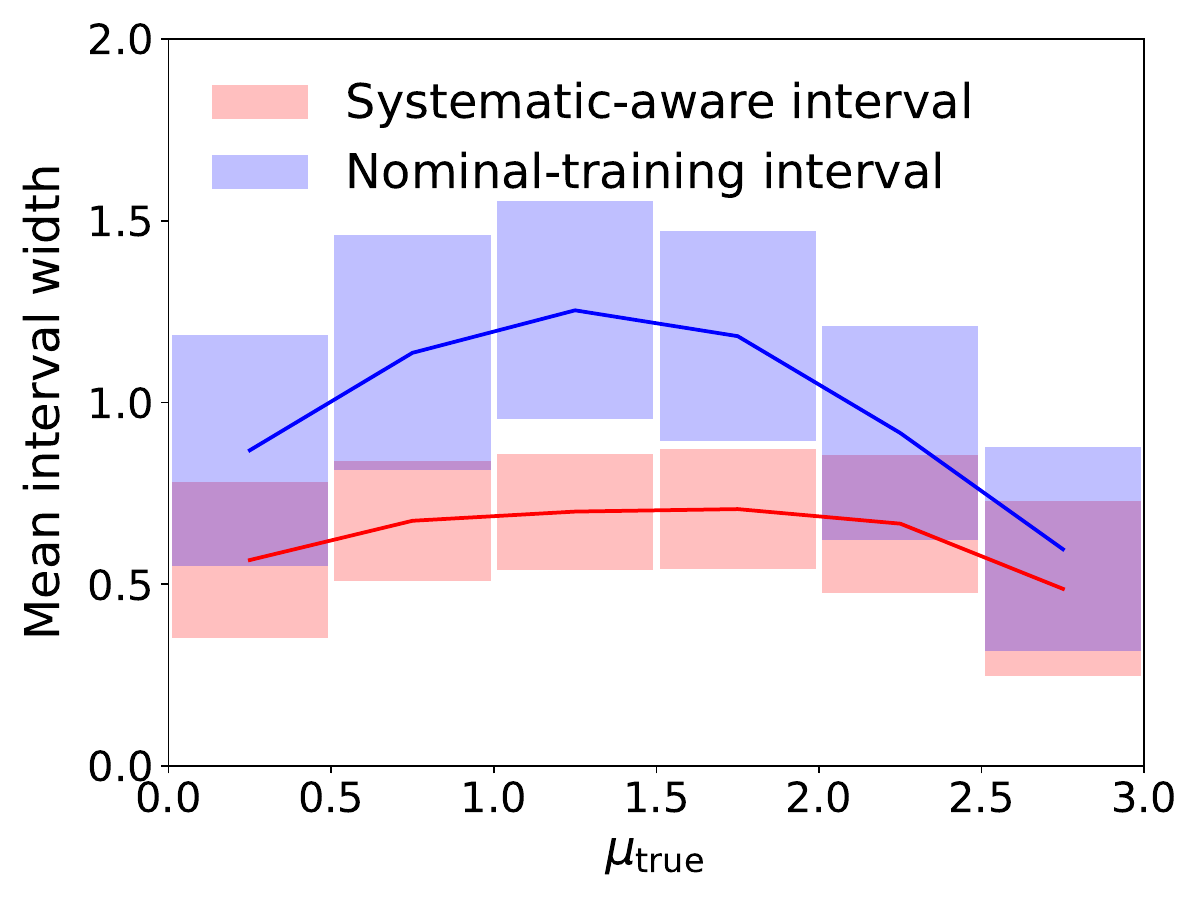}\hfill
    \includegraphics[width=0.48\linewidth]{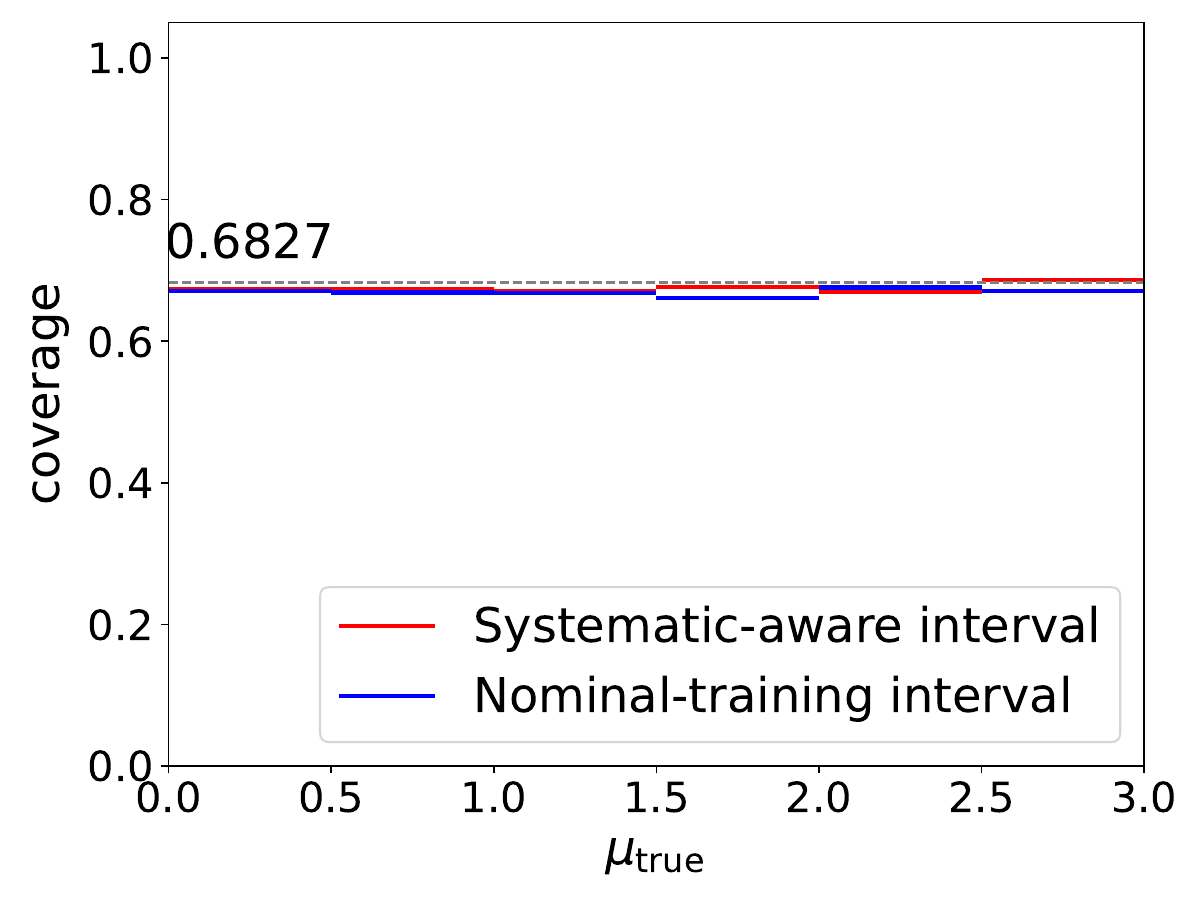}\\
    \caption{\label{fig:toy-coverage3} The average width of the confidence interval for the signal strength $\mu$ (left) and the corresponding coverage (right) are shown as functions of the true value $\mu^\text{true}$, based on 50000 pseudo-experiments. Results are compared between the proposed systematic-aware training and a classical treatment, where the same dual-branch network architecture is trained using only nominal samples, and systematic uncertainties are incorporated through template morphing in the final likelihood fit (“Nominal training”). The systematic-aware approach yields a significantly shorter average interval length of 0.6342, compared to 0.8968 for the nominal training, while achieving a comparable coverage of 0.6697 versus 0.6523.}
\end{figure*}

Table~\ref{tab:auc_architecture_ablationx} shows that the nominal-training model achieves an area under the ROC curve (AUC) of 0.8883 on the nominal test sample, which is very close to the baseline value of 0.8901. This indicates that training without systematic variations does not lead to a significant degradation in nominal signal–background discrimination performance. Figure~\ref{fig:toy-coverage3} shows the average width of the confidence interval for the signal strength $\mu$ and the corresponding coverage as functions of the true value $\mu^\text{true}$, obtained from 50000 pseudo-experiments. While the classical template-morphing baseline yields substantially wider confidence intervals across the full range of $\mu^\text{true}$, indicating a reduced statistical constraining power when systematic effects are handled exclusively at the likelihood level, the systematic-aware training consistently results in shorter confidence intervals with a comparable coverage. Since the network architecture is identical in both cases, this behavior reflects the different treatment of systematic uncertainties, encompassing both the systematic-aware training strategy and the corresponding interpolation scheme used to model nuisance parameter dependencies in the likelihood fit.

\section*{Acknowledgments}
We would like to thank the organizers of the FAIR-HUC challenge for the organization and technical support. We also would like to thank Robert Schöfbeck for fruitful discussions. The computational results were obtained using the CLIP computing cluster of the Austrian Academy of Sciences  at~\url{https://www.clip.science/}.

\bibliography{references} 

@article{DAgnolo:2018cun,
    author = "D'Agnolo, Raffaele Tito and Wulzer, Andrea",
    title = "{Learning New Physics from a Machine}",
    eprint = "1806.02350",
    archivePrefix = "arXiv",
    primaryClass = "hep-ph",
    doi = "10.1103/PhysRevD.99.015014",
    journal = "Phys. Rev. D",
    volume = "99",
    number = "1",
    pages = "015014",
    year = "2019"
}

@article{loshchilov2019decoupled,
  title   = {Decoupled Weight Decay Regularization},
  author  = {Loshchilov, Ilya and Hutter, Frank},
  journal = {arXiv preprint arXiv:1711.05101},
  year    = {2019}
}

@article{Ghosh:2021roe,
    author = "Ghosh, Aishik and Nachman, Benjamin and Whiteson, Daniel",
    title = "{Uncertainty-aware machine learning for high energy physics}",
    eprint = "2105.08742",
    archivePrefix = "arXiv",
    primaryClass = "physics.data-an",
    doi = "10.1103/PhysRevD.104.056026",
    journal = "Phys. Rev. D",
    volume = "104",
    number = "5",
    pages = "056026",
    year = "2021"
}

@inproceedings{Chen:2022pzc,
    author = "Chen, Thomas Y. and Dey, Biprateep and Ghosh, Aishik and Kagan, Michael and Nord, Brian and Ramachandra, Nesar",
    title = "{Interpretable Uncertainty Quantification in AI for HEP}",
    booktitle = "{Snowmass 2021}",
    eprint = "2208.03284",
    archivePrefix = "arXiv",
    primaryClass = "hep-ex",
    reportNumber = "FERMILAB-FN-1179-SCD, FERMILAB-FN-1179-SCD; arXiv:2208.03284 oai:inspirehep.net:2132723",
    doi = "10.2172/1886020",
    month = "8",
    year = "2022"
}

@article{Mastandrea:2024irf,
    author = "Mastandrea, Radha and Nachman, Benjamin and Plehn, Tilman",
    title = "{Constraining the Higgs Potential with Neural Simulation-based Inference for Di-Higgs Production}",
    eprint = "2405.15847",
    journal="Phys.Rev.D",
    volume="110",
    pages="056004",
    doi="10.1103/PhysRevD.110.056004",
    archivePrefix = "arXiv",
    primaryClass = "hep-ph",
    month = "5",
    year = "2024"
}

@article{Bahl:2021dnc,
    author = "Bahl, Henning and Brass, Simon",
    title = "{Constraining $ \mathcal{CP} $-violation in the Higgs-top-quark interaction using machine-learning-based inference}",
    eprint = "2110.10177",
    archivePrefix = "arXiv",
    primaryClass = "hep-ph",
    reportNumber = "DESY 21-160",
    doi = "10.1007/JHEP03(2022)017",
    journal = "JHEP",
    volume = "03",
    pages = "017",
    year = "2022"
}

@article{Barman:2021yfh,
    author = "Barman, Rahool Kumar and Gon\c{c}alves, Dorival and Kling, Felix",
    title = "{Machine Learning the Higgs-Top CP Phase}",
    eprint = "2110.07635",
      journal="Phys.Rev.D",
      volume="105",
      pages="035023",
      doi="10.1103/PhysRevD.105.035023",
    archivePrefix = "arXiv",
    primaryClass = "hep-ph",
    reportNumber = "DESY 21-161",
    month = "10",
    year = "2021"
}

@article{CMS:2024ddc,
    author = "{CMS Collaboration}",
    title = "{Measurement of boosted Higgs bosons produced via vector boson fusion or gluon fusion in the H $\to$$\mathrm{b\bar{b}}$ decay mode using LHC proton-proton collision data at $\sqrt{s}$ = 13 TeV}",
    eprint = "2407.08012",
    journal="JHEP",
    volume="12",
    pages="035",
    doi="10.1007/JHEP12(2024)035",
    archivePrefix = "arXiv",
    primaryClass = "hep-ex",
    reportNumber = "CMS-HIG-21-020, CERN-EP-2024-162",
    month = "7",
    year = "2024"
}

@article{ATLAS:2024itc,
    author = "{ATLAS Collaboration}",
    title = "{ATLAS searches for additional scalars and exotic Higgs boson decays with the LHC Run 2 dataset}",
    eprint = "2405.04914",
    archivePrefix = "arXiv",
    primaryClass = "hep-ex",
    reportNumber = "CERN-EP-2024-094",
    month = "5",
    year = "2024"
}

@article{ATLAS:2024auw,
    author = "{ATLAS Collaboration}",
    title = "{Search for a resonance decaying into a scalar particle and a Higgs boson in the final state with two bottom quarks and two photons in proton-proton collisions at a center of mass energy of 13 TeV with the ATLAS detector}",
    eprint = "2404.12915",
    journal="JHEP",
    volume="11",
    pages="047",
    doi="10.1007/JHEP11(2024)047",
    archivePrefix = "arXiv",
    primaryClass = "hep-ex",
    reportNumber = "CERN-EP-2024-072",
    month = "4",
    year = "2024"
}

@article{CMS:2024fkb,
    author = "{CMS Collaboration}",
    title = "{Search for Higgs Boson Pair Production with One Associated Vector Boson in Proton-Proton Collisions at $\sqrt{s}$ = 13 TeV}",
    eprint = "2404.08462",
    journal="JHEP",
    volume="10",
    pages="061",
    doi="10.1007/JHEP10(2024)061",
    archivePrefix = "arXiv",
    primaryClass = "hep-ex",
    reportNumber = "CMS-HIG-22-006, CERN-EP-2024-064",
    month = "4",
    year = "2024"
}

@article{ATLAS:2023dnm,
    author = "{ATLAS Collaboration}",
    title = "{Evidence of off-shell Higgs boson production from $ZZ$ leptonic decay channels and constraints on its total width with the ATLAS detector}",
    eprint = "2304.01532",
      journal="Phys.Lett.B",
      volume="846",
      pages="138223",
      doi="10.1016/j.physletb.2023.138223",
    archivePrefix = "arXiv",
    primaryClass = "hep-ex",
    reportNumber = "CERN-EP-2023-023",
    month = "4",
    year = "2023"
}

@article{CMS:2022fxs,
    author = "{CMS Collaboration}",
    title = "{Search for Higgs Boson and Observation of Z Boson through their Decay into a Charm Quark-Antiquark Pair in Boosted Topologies in Proton-Proton Collisions at s=13\,\,TeV}",
    eprint = "2211.14181",
    archivePrefix = "arXiv",
    primaryClass = "hep-ex",
    reportNumber = "CMS-HIG-21-012, CERN-EP-2022-233",
    doi = "10.1103/PhysRevLett.131.041801",
    journal = "Phys. Rev. Lett.",
    volume = "131",
    number = "4",
    pages = "041801",
    year = "2023"
}

@article{Sirunyan:2020hwz,
    author = "{CMS Collaboration}",
    title = "{Inclusive search for highly boosted Higgs bosons decaying to bottom quark-antiquark pairs in proton-proton collisions at $\sqrt{s} =$ 13 TeV}",
    eprint = "2006.13251",
    archivePrefix = "arXiv",
    primaryClass = "hep-ex",
    reportNumber = "CMS-HIG-19-003, CERN-EP-2020-107",
    doi = "10.1007/JHEP12(2020)085",
    journal = "JHEP",
    volume = "12",
    pages = "085",
    year = "2020"
}

@article{Aad:2020hzm,
    author = "{ATLAS Collaboration}",
    archivePrefix = "arXiv",
    eprint = "2004.01678",
      doi="10.1103/PhysRevLett.125.221802",
    month = "4",
    primaryClass = "hep-ex",
    reportNumber = "CERN-EP-2020-033",
    title = "{Search for Higgs boson decays into a $Z$ boson and a light hadronically decaying resonance using 13 TeV $pp$ collision data from the ATLAS detector}",
    year = "2020"
}

@article{Aad:2019yxi,
    author = "{ATLAS Collaboration}",
    archivePrefix = "arXiv",
    doi = "10.1016/j.physletb.2019.135145",
    eprint = "1908.06765",
    journal = "Phys. Lett. B",
    pages = "135145",
    primaryClass = "hep-ex",
    reportNumber = "CERN-EP-2019-143",
    title = "{Search for non-resonant Higgs boson pair production in the $bb\ell\nu\ell\nu$ final state with the ATLAS detector in $pp$ collisions at $\sqrt{s} = 13$ TeV}",
    volume = "801",
    year = "2020"
}

@article{Andreassen:2019nnm,
    author = "Andreassen, Anders and Nachman, Benjamin",
    title = "{Neural Networks for Full Phase-space Reweighting and Parameter Tuning}",
    eprint = "1907.08209",
    archivePrefix = "arXiv",
    primaryClass = "hep-ph",
    doi = "10.1103/PhysRevD.101.091901",
    journal = "Phys. Rev. D",
    volume = "101",
    number = "9",
    pages = "091901",
    year = "2020"
}

@article{Stoye:2018ovl,
      author         = "Stoye, Markus and Brehmer, Johann and Louppe, Gilles and
                        Pavez, Juan and Cranmer, Kyle",
      title          = "{Likelihood-free inference with an improved cross-entropy estimator}",
      year           = "2018",
      eprint         = "1808.00973",
      archivePrefix  = "arXiv",
      primaryClass   = "stat.ML",
      SLACcitation   = "%%CITATION = ARXIV:1808.00973;%%"
}

@article{Hollingsworth:2020kjg,
      author         = "Hollingsworth, Jacob and Whiteson, Daniel",
      title          = "{Resonance Searches with Machine Learned Likelihood Ratios}",
      year           = "2020",
      eprint         = "2002.04699",
      archivePrefix  = "arXiv",
      primaryClass   = "hep-ph",
      SLACcitation   = "%%CITATION = ARXIV:2002.04699;%%"
}

@article{Andreassen:2020gtw,
    author = "Andreassen, Anders and Hsu, Shih-Chieh and Nachman, Benjamin and Suaysom, Natchanon and Suresh, Adi",
    title = "{Parameter Estimation using Neural Networks in the Presence of Detector Effects}",
    eprint = "2010.03569",
      doi="10.1103/PhysRevD.103.036001",
    archivePrefix = "arXiv",
    primaryClass = "hep-ph",
    journal="Phys. Rev. D",
    volume="103",
    pages="036001",
    year = "2021"
}

@article{Coogan:2020yux,
    author = "Coogan, Adam and Karchev, Konstantin and Weniger, Christoph",
    title = "{Targeted Likelihood-Free Inference of Dark Matter Substructure in Strongly-Lensed Galaxies}",
    journal = "{34th Conference on Neural Information Processing Systems}",
    eprint = "2010.07032",
    archivePrefix = "arXiv",
    primaryClass = "astro-ph.CO",
    month = "10",
    year = "2020"
}

@article{Flesher:2020kuy,
    author = "Flesher, Forrest and Fraser, Katherine and Hutchison, Charles and Ostdiek, Bryan and Schwartz, Matthew D.",
    title = "{Parameter Inference from Event Ensembles and the Top-Quark Mass}",
    eprint = "2011.04666",
      journal="JHEP",
      volume="09",
      pages="058",
      doi="10.1007/JHEP09(2021)058",
    archivePrefix = "arXiv",
    primaryClass = "hep-ph",
    month = "11",
    year = "2020"
}

@article{Bieringer:2020tnw,
    author = {Bieringer, Sebastian and Butter, Anja and Heimel, Theo and H\"oche, Stefan and K\"othe, Ullrich and Plehn, Tilman and Radev, Stefan T.},
    title = "{Measuring QCD Splittings with Invertible Networks}",
    eprint = "2012.09873",
      journal="SciPost Phys.",
      volume="10",
      pages="126",
      doi="10.21468/SciPostPhys.10.6.126",
    archivePrefix = "arXiv",
    primaryClass = "hep-ph",
    reportNumber = "FERMILAB-PUB-20-665-T",
    month = "12",
    year = "2020"
}

@article{Nachman:2021yvi,
    author = "Nachman, Benjamin and Thaler, Jesse",
    title = "{E Pluribus Unum Ex Machina: Learning from Many Collider Events at Once}",
    eprint = "2101.07263",
      journal="Phys.Rev.D",
      volume="103",
      pages="116013",
      doi="10.1103/PhysRevD.103.116013",
    archivePrefix = "arXiv",
    primaryClass = "physics.data-an",
    reportNumber = "MIT-CTP 5271",
    month = "1",
    year = "2021"
}

@article{NEURIPS2020_a878dbeb,
 author = "Shirobokov, Sergey and Belavin, Vladislav and Kagan, Michael and Ustyuzhanin, Andrei and Baydin, Atilim Gunes",
 booktitle = "{Advances in Neural Information Processing Systems}",
 editor = "H. Larochelle and M. Ranzato and R. Hadsell and M. F. Balcan and H. Lin",
 pages = "14650--14662",
 publisher = "Curran Associates, Inc.",
 title = "{Black-Box Optimization with Local Generative Surrogates}",
 volume = "33",
 eprint = "2002.04632",
 archivePrefix = "arXiv",
 primaryClass = "cs.LG",
 month = "Feb",
 year = "2020"
}

@article{Mishra-Sharma:2021oxe,
    author = "Mishra-Sharma, Siddharth and Cranmer, Kyle",
    title = "{A neural simulation-based inference approach for characterizing the Galactic Center $\gamma$-ray excess}",
    eprint = "2110.06931",
      journal="Phys.Rev.D",
      volume="105",
      pages="063017",
      doi="10.1103/PhysRevD.105.063017",
    archivePrefix = "arXiv",
    primaryClass = "astro-ph.HE",
    reportNumber = "MIT-CTP/5337",
    month = "10",
    year = "2021"
}

@article{Arganda:2022qzy,
    author = "Arganda, Ernesto and Marcano, Xabier and Lozano, V\'\i{}ctor Mart\'\i{}n and Medina, Anibal D. and Perez, Andres D. and Szewc, Manuel and Szynkman, Alejandro",
    title = "{A method for approximating optimal statistical significances with machine-learned likelihoods}",
    eprint = "2205.05952",
      journal="Eur.Phys.J.C",
      volume="82",
      pages="993",
      doi="10.1140/epjc/s10052-022-10944-3",
    archivePrefix = "arXiv",
    primaryClass = "hep-ph",
    reportNumber = "IFT-UAM/CSIC-22-52",
    month = "5",
    year = "2022"
}

@article{Arganda:2022zbs,
    author = "Arganda, Ernesto and Perez, Andres D. and de los Rios, Martin and Sand\'a Seoane, Rosa Mar\'\i{}a",
    title = "{Machine-Learned Exclusion Limits without Binning}",
    eprint = "2211.04806",
    journal="Eur.Phys.J.C",
    volume="83",
    pages="1158",
    doi="10.1140/epjc/s10052-023-12314-z",
    archivePrefix = "arXiv",
    primaryClass = "hep-ph",
    reportNumber = "IFT-UAM/CSIC-22-134",
    month = "11",
    year = "2022"
}

@article{Butter:2022vkj,
    author = "Butter, Anja and Heimel, Theo and Martini, Till and Peitzsch, Sascha and Plehn, Tilman",
    title = "{Two Invertible Networks for the Matrix Element Method}",
    eprint = "2210.00019",
      journal="SciPost Phys.",
      volume="15",
      pages="094",
      doi="10.21468/SciPostPhys.15.3.094",
    archivePrefix = "arXiv",
    primaryClass = "hep-ph",
    month = "9",
    year = "2022"
}

@article{Neubauer:2022gbu,
    author = "Neubauer, Mark and Feickert, Matthew and Katare, Mihir and Roy, Avik",
    title = "{Deep Learning for the Matrix Element Method}",
    eprint = "2211.11910",
    archivePrefix = "arXiv",
    primaryClass = "hep-ex",
    doi = "10.22323/1.414.0246",
    journal = "PoS",
    volume = "ICHEP2022",
    pages = "246",
    year = "2022"
}

@article{Rizvi:2023mws,
    author = "Rizvi, Shahzar and Pettee, Mariel and Nachman, Benjamin",
    title = "{Learning Likelihood Ratios with Neural Network Classifiers}",
    eprint = "2305.10500",
    journal="JHEP",
    volume="02",
    pages="136",
    doi="10.1007/JHEP02(2024)136",
    archivePrefix = "arXiv",
    primaryClass = "hep-ph",
    month = "5",
    year = "2023"
}

@article{Heinrich:2023bmt,
    author = "Heinrich, Lukas and Mishra-Sharma, Siddharth and Pollard, Chris and Windischhofer, Philipp",
    title = "{Hierarchical Neural Simulation-Based Inference Over Event Ensembles}",
    eprint = "2306.12584",
    archivePrefix = "arXiv",
    primaryClass = "stat.ML",
    reportNumber = "MIT-CTP/5576",
    month = "6",
    year = "2023"
}

@article{Morandini:2023pwj,
    author = "Morandini, Alessandro and Ferber, Torben and Kahlhoefer, Felix",
    title = "{Reconstructing axion-like particles from beam dumps with simulation-based inference}",
    eprint = "2308.01353",
    journal="Eur.Phys.J.C",
    volume="84",
    pages="200",
    doi="10.1140/epjc/s10052-024-12557-4",
    archivePrefix = "arXiv",
    primaryClass = "hep-ph",
    reportNumber = "TTP23-031, P3H-23-054",
    month = "8",
    year = "2023"
}

@article{Heimel:2023mvw,
    author = "Heimel, Theo and Huetsch, Nathan and Winterhalder, Ramon and Plehn, Tilman and Butter, Anja",
    title = "{Precision-Machine Learning for the Matrix Element Method}",
    eprint = "2310.07752",
    journal="SciPost Phys.",
    volume="17",
    pages="129",
    doi="10.21468/SciPostPhys.17.5.129",
    archivePrefix = "arXiv",
    primaryClass = "hep-ph",
    reportNumber = "IRMP-CP3-23-55",
    month = "10",
    year = "2023"
}

@article{Chai:2024zyl,
    author = "Chai, Shengdu and Gu, Jiayin and Li, Lingfeng",
    title = "{From Optimal Observables to Machine Learning: an Effective-Field-Theory Analysis of $e^+e^- \to W^+W^-$ at Future Lepton Colliders}",
    eprint = "2401.02474",
    journal="JHEP",
    volume="05",
    pages="292",
    doi="10.1007/JHEP05(2024)292",
    archivePrefix = "arXiv",
    primaryClass = "hep-ph",
    month = "1",
    year = "2024"
}

@article{Alvarez:2024owq,
    author = "Alvarez, Ezequiel and Da Rold, Leandro and Szewc, Manuel and Szynkman, Alejandro and Tanco, Santiago A. and Tarutina, Tatiana",
    title = "{Improvement and generalization of ABCD method with Bayesian inference}",
    eprint = "2402.08001",
    journal="SciPost Phys.Core",
    volume="7",
    pages="043",
    doi="10.21468/SciPostPhysCore.7.3.043",
    archivePrefix = "arXiv",
    primaryClass = "hep-ph",
    month = "2",
    year = "2024"
}

@article{Diaz:2024yfu,
    author = "Diaz, Mauricio A. and Cerro, Giorgio and Dasmahapatra, Srinandan and Moretti, Stefano",
    title = "{Bayesian Active Search on Parameter Space: a 95 GeV Spin-0 Resonance in the ($B-L$)SSM}",
    eprint = "2404.18653",
    archivePrefix = "arXiv",
    primaryClass = "hep-ph",
    month = "4",
    year = "2024"
}

@article{JETSCAPE:2024cqe,
    author = "{JETSCAPE Collaboration}",
    title = "{Bayesian Inference analysis of jet quenching using inclusive jet and hadron suppression measurements}",
    eprint = "2408.08247",
    archivePrefix = "arXiv",
    primaryClass = "hep-ph",
    month = "8",
    year = "2024"
}

@article{Maitre:2024hzp,
    author = "Ma\^\i{}tre, Daniel and Ngairangbam, Vishal S. and Spannowsky, Michael",
    title = "{Optimal Equivariant Architectures from the Symmetries of Matrix-Element Likelihoods}",
    eprint = "2410.18553",
    archivePrefix = "arXiv",
    primaryClass = "hep-ph",
    reportNumber = "IPPP/24/69",
    month = "10",
    year = "2024"
}

@article{Heimel:2024drk,
    author = "Heimel, Theo and Plehn, Tilman and Schmal, Nikita",
    title = "{Profile Likelihoods on ML-Steroids}",
    eprint = "2411.00942",
    archivePrefix = "arXiv",
    primaryClass = "hep-ph",
    month = "11",
    year = "2024"
}

@article{Blance:2019ibf,
      author         = "Blance, Andrew and Spannowsky, Michael and Waite, Philip",
      title          = "{Adversarially-trained autoencoders for robust unsupervised new physics searches}",
      journal        = "JHEP",
      volume         = "10",
      year           = "2019",
      pages          = "047",
      doi            = "10.1007/JHEP10(2019)047",
      eprint         = "1905.10384",
      archivePrefix  = "arXiv",
      primaryClass   = "hep-ph",
      reportNumber   = "IPPP/19/41",
      SLACcitation   = "%%CITATION = ARXIV:1905.10384;%%"
}

@article{Englert:2018cfo,
      author         = "Englert, Christoph and Galler, Peter and Harris, Philip
                        and Spannowsky, Michael",
      title          = "{Machine Learning Uncertainties with Adversarial Neural Networks}",
      journal        = "Eur. Phys. J.",
      volume         = "C79",
      year           = "2019",
      number         = "1",
      pages          = "4",
      doi            = "10.1140/epjc/s10052-018-6511-8",
      eprint         = "1807.08763",
      archivePrefix  = "arXiv",
      primaryClass   = "hep-ph",
      reportNumber   = "IPPP/18/61",
      SLACcitation   = "%%CITATION = ARXIV:1807.08763;%%"
}

@article{Dolen:2016kst,
      author         = "Dolen, James and Harris, Philip and Marzani, Simone and
                        Rappoccio, Salvatore and Tran, Nhan",
      title          = "{Thinking outside the ROCs: Designing Decorrelated Taggers (DDT) for jet substructure}",
      journal        = "JHEP",
      volume         = "05",
      year           = "2016",
      pages          = "156",
      doi            = "10.1007/JHEP05(2016)156",
      eprint         = "1603.00027",
      archivePrefix  = "arXiv",
      primaryClass   = "hep-ph",
      reportNumber   = "FERMILAB-PUB-16-046-PPD",
      SLACcitation   = "%%CITATION = ARXIV:1603.00027;%%"
}

@article{Moult:2017okx,
    author = "Moult, Ian and Nachman, Benjamin and Neill, Duff",
    title = "{Convolved Substructure: Analytically Decorrelating Jet Substructure Observables}",
    eprint = "1710.06859",
    archivePrefix = "arXiv",
    primaryClass = "hep-ph",
    doi = "10.1007/JHEP05(2018)002",
    journal = "JHEP",
    volume = "05",
    pages = "002",
    year = "2018"
}

@article{Stevens:2013dya,
 author         = "Stevens, Justin and Williams, Mike",
 title          = "{uBoost: A boosting method for producing uniform selection efficiencies from multivariate classifiers}",
 journal        = "JINST",
 volume         = "8",
 year           = "2013",
 pages          = "P12013",
 doi            = "10.1088/1748-0221/8/12/P12013",
 eprint         = "1305.7248",
 archivePrefix  = "arXiv",
 primaryClass   = "nucl-ex",
 SLACcitation   = "%%CITATION = ARXIV:1305.7248;%%"
}

@article{Shimmin:2017mfk,
 author         = "Shimmin, Chase and Sadowski, Peter and Baldi, Pierre and
 Weik, Edison and Whiteson, Daniel and Goul, Edward and
 Søgaard, Andreas",
 title          = "{Decorrelated Jet Substructure Tagging using Adversarial Neural Networks}",
 year           = "2017",
 eprint         = "1703.03507",
      doi="10.1103/PhysRevD.96.074034",
 archivePrefix  = "arXiv",
 primaryClass   = "hep-ex",
}

@article{Bradshaw:2019ipy,
      author         = "Bradshaw, Layne and Mishra, Rashmish K. and Mitridate,
                        Andrea and Ostdiek, Bryan",
      title          = "{Mass Agnostic Jet Taggers}",
      year           = "2019",
      eprint         = "1908.08959",
      doi="10.21468/SciPostPhys.8.1.011",
      archivePrefix  = "arXiv",
      primaryClass   = "hep-ph",
      SLACcitation   = "%%CITATION = ARXIV:1908.08959;%%"
}

@article{ATLAS:2018ibz,
    author = "ATLAS",
    title = "{Performance of mass-decorrelated jet substructure observables for hadronic two-body decay tagging in ATLAS}",
    reportNumber = "ATL-PHYS-PUB-2018-014",
    year = "2018"
}

@article{DiscoFever,
      author         = "Kasieczka, Gregor and Shih, David",
      title          = "{DisCo Fever: Robust Networks Through Distance Correlation}",
      year           = "2020",
      eprint         = "2001.05310",
      doi="10.1103/PhysRevLett.125.122001",
      archivePrefix  = "arXiv",
      primaryClass   = "hep-ph",
      SLACcitation   = "%%CITATION = ARXIV:2001.05310;%%"
}

@article{Wunsch:2019qbo,
      author         = "Wunsch, Stefan and J\'{o}rger, Simon and Wolf, Roger and
                        Quast, Gunter",
      title          = "{Reducing the dependence of the neural network function to systematic uncertainties in the input space}",
      year           = "2019",
      eprint         = "1907.11674",
      doi="10.1007/s41781-020-00037-9",
      archivePrefix  = "arXiv",
      primaryClass   = "physics.data-an",
      SLACcitation   = "%%CITATION = ARXIV:1907.11674;%%"
}

@article{Rogozhnikov:2014zea,
 author         = "Rogozhnikov, Alex and Bukva, Aleksandar and Gligorov, V.
 V. and Ustyuzhanin, Andrey and Williams, Mike",
 title          = "{New approaches for boosting to uniformity}",
 journal        = "JINST",
 volume         = "10",
 year           = "2015",
 number         = "03",
 pages          = "T03002",
 doi            = "10.1088/1748-0221/10/03/T03002",
 eprint         = "1410.4140",
 archivePrefix  = "arXiv",
 primaryClass   = "hep-ex",
 SLACcitation   = "%%CITATION = ARXIV:1410.4140;%%"
}

@article{10.1088/2632-2153/ab9023,
author="{CMS Collaboration}",
title="{A deep neural network to search for new long-lived particles decaying to jets}",
journal={Machine Learning: Science and Technology},
eprint="1912.12238",
doi="10.1088/2632-2153/ab9023",
year={2020}
}

@article{clavijo2020adversarial,
    title="{Adversarial domain adaptation to reduce sample bias of a high energy physics classifier}",
    author={Jose M. Clavijo and Paul Glaysher and Judith M. Katzy},
    year={2020},
    eprint="2005.00568",
      journal="Mach.Learn.Sci.Tech.",
      volume="3",
      pages="015014",
      doi="10.1088/2632-2153/ac3dde",
    archivePrefix={arXiv},
    primaryClass={stat.ML}
}

@article{Kasieczka:2020pil,
    author = "Kasieczka, Gregor and Nachman, Benjamin and Schwartz, Matthew D. and Shih, David",
    title = "{ABCDisCo: Automating the ABCD Method with Machine Learning}",
    eprint = "2007.14400",
      doi="10.1103/PhysRevD.103.035021",
    archivePrefix = "arXiv",
    primaryClass = "hep-ph",
    month = "7",
    year = "2020"
}

@article{Kitouni:2020xgb,
    author = "Kitouni, Ouail and Nachman, Benjamin and Weisser, Constantin and Williams, Mike",
    title = "{Enhancing searches for resonances with machine learning and moment decomposition}",
    eprint = "2010.09745",
      journal="JHEP",
      volume="04",
      pages="070",
      doi="10.1007/JHEP04(2021)070",
    archivePrefix = "arXiv",
    primaryClass = "hep-ph",
    month = "10",
    year = "2020"
}

@article{Estrade:2019gzk,
    author = "Estrade, Victor and Germain, C\'ecile and Guyon, Isabelle and Rousseau, David",
    title = "{Systematic aware learning - A case study in High Energy Physics}",
    doi = "10.1051/epjconf/201921406024",
    journal = "EPJ Web Conf.",
    volume = "214",
    pages = "06024",
    year = "2019"
}

@article{Ghosh:2021hrh,
    author = "Ghosh, Aishik and Nachman, Benjamin",
    title = "{A Cautionary Tale of Decorrelating Theory Uncertainties}",
    eprint = "2109.08159",
      journal="Eur.Phys.J.C",
      volume="82",
      pages="46",
      doi="10.1140/epjc/s10052-022-10012-w",
    archivePrefix = "arXiv",
    primaryClass = "hep-ph",
    month = "9",
    year = "2021"
}

@article{Nachman:2019dol,
    author = "Nachman, Benjamin",
    archivePrefix = "arXiv",
    eprint = "1909.03081",
      doi="10.21468/SciPostPhys.8.6.090",
    month = "9",
    primaryClass = "hep-ph",
    title = "{A guide for deploying Deep Learning in LHC searches: How to achieve optimality and account for uncertainty}",
    year = "2019"
}

@article{Kasieczka:2020vlh,
    author = "Kasieczka, Gregor and Luchmann, Michel and Otterpohl, Florian and Plehn, Tilman",
    archivePrefix = "arXiv",
    eprint = "2003.11099",
      doi="10.21468/SciPostPhys.9.6.089",
    month = "3",
    primaryClass = "hep-ph",
    title = "{Per-Object Systematics using Deep-Learned Calibration}",
    year = "2020"
}

@article{Bollweg:2019skg,
    author = "Bollweg, Sven and Haußmann, Manuel and Kasieczka, Gregor and Luchmann, Michel and Plehn, Tilman and Thompson, Jennifer",
    archivePrefix = "arXiv",
    doi = "10.21468/SciPostPhys.8.1.006",
    eprint = "1904.10004",
    journal = "SciPost Phys.",
    number = "1",
    pages = "006",
    primaryClass = "hep-ph",
    title = "{Deep-Learning Jets with Uncertainties and More}",
    volume = "8",
    year = "2020"
}

@article{Araz:2021wqm,
    author = "Araz, Jack Y. and Spannowsky, Michael",
    title = "{Combine and Conquer: Event Reconstruction with Bayesian Ensemble Neural Networks}",
    eprint = "2102.01078",
    archivePrefix = "arXiv",
    primaryClass = "hep-ph",
    reportNumber = "IPPP/20/74",
    doi = "10.1007/JHEP04(2021)296",
    journal = "JHEP",
    volume = "04",
    pages = "296",
    year = "2021"
}

@article{Bellagente:2021yyh,
    author = "Bellagente, Marco and Hau\ss{}mann, Manuel and Luchmann, Michel and Plehn, Tilman",
    title = "{Understanding Event-Generation Networks via Uncertainties}",
    eprint = "2104.04543",
      journal="SciPost Phys.",
      volume="13",
      pages="003",
      doi="10.21468/SciPostPhys.13.1.003",
    archivePrefix = "arXiv",
    primaryClass = "hep-ph",
    month = "4",
    year = "2021"
}

@article{Wunsch:2020iuh,
    author = "Wunsch, Stefan and Jörger, Simon and Wolf, Roger and Quast, Günter",
    archivePrefix = "arXiv",
    eprint = "2003.07186",
      doi="10.1007/s41781-020-00049-5",
    month = "3",
    primaryClass = "physics.data-an",
    title = "{Optimal statistical inference in the presence of systematic uncertainties using neural network optimization based on binned Poisson likelihoods with nuisance parameters}",
    year = "2020"
}

@article{Elwood:2020pik,
    author = {Elwood, A. and Kr\"ucker, D. and Shchedrolosiev, M.},
    title = "{Direct optimization of the discovery significance in machine learning for new physics searches in particle colliders}",
    doi = "10.1088/1742-6596/1525/1/012110",
    journal = "J. Phys. Conf. Ser.",
    volume = "1525",
    pages = "012110",
    year = "2020"
}

@article{Xia:2018kgd,
      author         = "Xia, Li-Gang",
      title          = "{QBDT, a new boosting decision tree method with systematical uncertainties into training for High Energy Physics}",
      journal        = "Nucl. Instrum. Meth.",
      volume         = "A930",
      year           = "2019",
      pages          = "15-26",
      doi            = "10.1016/j.nima.2019.03.088",
      eprint         = "1810.08387",
      archivePrefix  = "arXiv",
      primaryClass   = "physics.data-an",
      SLACcitation   = "%%CITATION = ARXIV:1810.08387;%%"
}

@article{deCastro:2018mgh,
    author = "De Castro, Pablo and Dorigo, Tommaso",
    archivePrefix = "arXiv",
    doi = "10.1016/j.cpc.2019.06.007",
    eprint = "1806.04743",
    journal = "Comput. Phys. Commun.",
    pages = "170--179",
    primaryClass = "stat.ML",
    title = "{INFERNO: Inference-Aware Neural Optimisation}",
    volume = "244",
    year = "2019"
}

@article{Charnock_2018,
    author = "Charnock, Tom and Lavaux, Guilhem and Wandelt, Benjamin D.",
    title = "{Automatic physical inference with information maximizing neural networks}",
    eprint = "1802.03537",
    archivePrefix = "arXiv",
    primaryClass = "astro-ph.IM",
    doi = "10.1103/PhysRevD.97.083004",
    journal = "Phys. Rev. D",
    volume = "97",
    number = "8",
    pages = "083004",
    year = "2018"
}

@article{Alsing:2019dvb,
    author = "Alsing, Justin and Wandelt, Benjamin",
    title = "{Nuisance hardened data compression for fast likelihood-free inference}",
    eprint = "1903.01473",
    archivePrefix = "arXiv",
    primaryClass = "astro-ph.CO",
    doi = "10.1093/mnras/stz1900",
    journal = "Mon. Not. Roy. Astron. Soc.",
    volume = "488",
    number = "4",
    pages = "5093--5103",
    year = "2019"
}

@inproceedings{Simpson:2022suz,
    author = "Simpson, Nathan and Heinrich, Lukas",
    title = "{neos: End-to-End-Optimised Summary Statistics for High Energy Physics}",
    booktitle = "{20th International Workshop on Advanced Computing and Analysis Techniques in Physics Research}: {AI Decoded - Towards Sustainable, Diverse, Performant and Effective Scientific Computing}",
    eprint = "2203.05570",
    archivePrefix = "arXiv",
    primaryClass = "physics.data-an",
    doi = "10.48550/arXiv.2203.05570",
    month = "3",
    year = "2022"
}

@article{Feichtinger:2021uff,
    author = "Feichtinger, P. and others",
    title = "{Punzi-loss: a non-differentiable metric approximation for sensitivity optimisation in the search for new particles}",
    eprint = "2110.00810",
    archivePrefix = "arXiv",
    primaryClass = "hep-ex",
    doi = "10.1140/epjc/s10052-022-10070-0",
    journal = "Eur. Phys. J. C",
    volume = "82",
    number = "2",
    pages = "121",
    year = "2022"
}

@article{Layer:2023lwi,
    author = "Layer, Lukas and Dorigo, Tommaso and Strong, Giles",
    title = "{Application of Inferno to a Top Pair Cross Section Measurement with CMS Open Data}",
    eprint = "2301.10358",
    archivePrefix = "arXiv",
    primaryClass = "hep-ex",
    month = "1",
    year = "2023"
}

@article{Kong:2022rnd,
    author = "Kong, Kyoungchul and Matchev, Konstantin T. and Mrenna, Stephen and Shyamsundar, Prasanth",
    title = "{New Machine Learning Techniques for Simulation-Based Inference: InferoStatic Nets, Kernel Score Estimation, and Kernel Likelihood Ratio Estimation}",
    eprint = "2210.01680",
    archivePrefix = "arXiv",
    primaryClass = "stat.ML",
    reportNumber = "FERMILAB-PUB-22-741-SCD",
    month = "10",
    year = "2022"
}

@article{scikit-learn,
 title={Scikit-learn: Machine Learning in {P}ython},
 author={Pedregosa, F. and Varoquaux, G. and Gramfort, A. and Michel, V.
         and Thirion, B. and Grisel, O. and Blondel, M. and Prettenhofer, P.
         and Weiss, R. and Dubourg, V. and Vanderplas, J. and Passos, A. and
         Cournapeau, D. and Brucher, M. and Perrot, M. and Duchesnay, E.},
 journal={Journal of Machine Learning Research},
 volume={12},
 pages={2825--2830},
 year={2011}
}

@article{Barrue:2023ysk,
    author = "Barru\'e, Ricardo and Conde-Mu\'\i{}\~no, Patricia and Dao, Valerio and Santos, Rui",
    title = "{Simulation-based inference in the search for CP violation in leptonic WH production}",
    eprint = "2308.02882",
    archivePrefix = "arXiv",
    primaryClass = "hep-ph",
    doi = "10.1007/JHEP04(2024)014",
    journal = "JHEP",
    volume = "04",
    pages = "014",
    year = "2024"
}

@article{Bahl:2024meb,
    author = "Bahl, Henning and Bres\'o, Victor and De Crescenzo, Giovanni and Plehn, Tilman",
    title = "{Advancing Tools for Simulation-Based Inference}",
    eprint = "2410.07315",
    archivePrefix = "arXiv",
    primaryClass = "hep-ph",
    month = "10",
    year = "2024"
}

@article{CMS:2014wdm,
    author = "Chatrchyan, Serguei and others",
    collaboration = "CMS",
    title = "{Evidence for the 125 GeV Higgs boson decaying to a pair of $\tau$ leptons}",
    eprint = "1401.5041",
    archivePrefix = "arXiv",
    primaryClass = "hep-ex",
    reportNumber = "CMS-HIG-13-004, CERN-PH-EP-2014-001",
    doi = "10.1007/JHEP05(2014)104",
    journal = "JHEP",
    volume = "05",
    pages = "104",
    year = "2014"
}

@article{CMS:2017zyp,
    author = "Sirunyan, Albert M and others",
    collaboration = "CMS",
    title = "{Observation of the Higgs boson decay to a pair of $\tau$ leptons with the CMS detector}",
    eprint = "1708.00373",
    archivePrefix = "arXiv",
    primaryClass = "hep-ex",
    reportNumber = "CMS-HIG-16-043, CERN-EP-2017-181",
    doi = "10.1016/j.physletb.2018.02.004",
    journal = "Phys. Lett. B",
    volume = "779",
    pages = "283--316",
    year = "2018"
}

@article{CMS:2022kdi,
    author = "Tumasyan, Armen and others",
    collaboration = "CMS",
    title = "{Measurements of Higgs boson production in the decay channel with a pair of $\tau $ leptons in proton\textendash{}proton collisions at $\sqrt{s}=13$ TeV}",
    eprint = "2204.12957",
    archivePrefix = "arXiv",
    primaryClass = "hep-ex",
    reportNumber = "CMS-HIG-19-010, CERN-EP-2022-027",
    doi = "10.1140/epjc/s10052-023-11452-8",
    journal = "Eur. Phys. J. C",
    volume = "83",
    number = "7",
    pages = "562",
    year = "2023"
}

@article{CMS:2021gxc,
    author = "Tumasyan, Armen and others",
    collaboration = "CMS",
    title = "{Measurement of the inclusive and differential Higgs boson production cross sections in the decay mode to a pair of $\tau$ leptons in pp collisions at $\sqrt{s} = $ 13 TeV}",
    eprint = "2107.11486",
    archivePrefix = "arXiv",
    primaryClass = "hep-ex",
    reportNumber = "CMS-HIG-20-015, CERN-EP-2021-134",
    doi = "10.1103/PhysRevLett.128.081805",
    journal = "Phys. Rev. Lett.",
    volume = "128",
    number = "8",
    pages = "081805",
    year = "2022"
}

@article{CMS:2024jbe,
    author = "Hayrapetyan, Aram and others",
    collaboration = "CMS",
    title = "{Measurement of the production cross section of a Higgs boson with large transverse momentum in its decays to a pair of \ensuremath{\tau} leptons in proton-proton collisions at s=13TeV}",
    eprint = "2403.20201",
    archivePrefix = "arXiv",
    primaryClass = "hep-ex",
    reportNumber = "CMS-HIG-21-017, CERN-EP-2023-300",
    doi = "10.1016/j.physletb.2024.138964",
    journal = "Phys. Lett. B",
    volume = "857",
    pages = "138964",
    year = "2024"
}

@article{ATLAS:2015xst,
    author = "Aad, Georges and others",
    collaboration = "ATLAS",
    title = "{Evidence for the Higgs-boson Yukawa coupling to tau leptons with the ATLAS detector}",
    eprint = "1501.04943",
    archivePrefix = "arXiv",
    primaryClass = "hep-ex",
    reportNumber = "CERN-PH-EP-2014-262",
    doi = "10.1007/JHEP04(2015)117",
    journal = "JHEP",
    volume = "04",
    pages = "117",
    year = "2015"
}

@article{ATLAS:2018ynr,
    author = "Aaboud, Morad and others",
    collaboration = "ATLAS",
    title = "{Cross-section measurements of the Higgs boson decaying into a pair of $\tau$-leptons in proton-proton collisions at $\sqrt{s}=13$ TeV with the ATLAS detector}",
    eprint = "1811.08856",
    archivePrefix = "arXiv",
    primaryClass = "hep-ex",
    reportNumber = "CERN-EP-2018-232",
    doi = "10.1103/PhysRevD.99.072001",
    journal = "Phys. Rev. D",
    volume = "99",
    pages = "072001",
    year = "2019"
}

@article{ATLAS:2022yrq,
    author = "Aad, Georges and others",
    collaboration = "ATLAS",
    title = "{Measurements of Higgs boson production cross-sections in the~$H\to\tau^{+}\tau^{-}$ decay channel in pp collisions at $ \sqrt{s} $ = 13 TeV with the ATLAS detector}",
    eprint = "2201.08269",
    archivePrefix = "arXiv",
    primaryClass = "hep-ex",
    reportNumber = "CERN-EP-2021-217",
    doi = "10.1007/JHEP08(2022)175",
    journal = "JHEP",
    volume = "08",
    pages = "175",
    year = "2022"
}

@article{ATLAS:2024wfv,
    author = "Aad, Georges and others",
    collaboration = "ATLAS",
    title = "{Differential cross-section measurements of Higgs boson production in the H \textrightarrow{} \ensuremath{\tau}$^{+}$\ensuremath{\tau}$^{-}$ decay channel in pp collisions at $ \sqrt{s} $ = 13 TeV with the ATLAS detector}",
    eprint = "2407.16320",
    archivePrefix = "arXiv",
    primaryClass = "hep-ex",
    reportNumber = "CERN-EP-2024-198",
    doi = "10.1007/JHEP03(2025)010",
    journal = "JHEP",
    volume = "03",
    pages = "010",
    year = "2025"
}

@article{ATLAS:2024ynn,
    author = "Aad, Georges and others",
    collaboration = "ATLAS",
    title = "{An implementation of neural simulation-based inference for parameter estimation in ATLAS}",
    eprint = "2412.01600",
    archivePrefix = "arXiv",
    primaryClass = "hep-ex",
    reportNumber = "CERN-EP-2024-305",
    month = "12",
    year = "2024"
}

@article{Bhimji:2024bcd,
    author = "Bhimji, Wahid and others",
    title = "{FAIR Universe HiggsML Uncertainty Challenge Competition}",
    eprint = "2410.02867",
    archivePrefix = "arXiv",
    primaryClass = "hep-ph",
    month = "10",
    year = "2024"
}

@article{ATLAS:2011tau,
    author = "ATLAS, CMS, LHC Higgs Combination Group",
    title = "{Procedure for the LHC Higgs boson search combination in Summer 2011}",
    reportNumber = "CMS-NOTE-2011-005, ATL-PHYS-PUB-2011-011, ATL-PHYS-PUB-2011-11",
    month = "8",
    year = "2011"
}

@article{Chatterjee:2024pbp,
    author = {Chatterjee, Suman and Cruz, Sergio S\'anchez and Sch\"ofbeck, Robert and Schwarz, Dennis},
    title = "{Rotation-equivariant graph neural network for learning hadronic SMEFT effects}",
    eprint = "2401.10323",
    archivePrefix = "arXiv",
    primaryClass = "hep-ph",
    doi = "10.1103/PhysRevD.109.076012",
    journal = "Phys. Rev. D",
    volume = "109",
    pages = "076012",
    year = "2024"
}

@article{dAgnolo:2021aun,
    author = "d'Agnolo, Raffaele Tito and Grosso, Gaia and Pierini, Maurizio and Wulzer, Andrea and Zanetti, Marco",
    title = "{Learning new physics from an imperfect machine}",
    eprint = "2111.13633",
    archivePrefix = "arXiv",
    primaryClass = "hep-ph",
    doi = "10.1140/epjc/s10052-022-10226-y",
    journal = "Eur. Phys. J. C",
    volume = "82",
    pages = "275",
    year = "2022"
}

@inproceedings{Thais:2022iok,
    author = "Thais, Savannah and Calafiura, Paolo and Chachamis, Grigorios and DeZoort, Gage and Duarte, Javier and Ganguly, Sanmay and Kagan, Michael and Murnane, Daniel and Neubauer, Mark S. and Terao, Kazuhiro",
    title = "{Graph Neural Networks in Particle Physics: Implementations, Innovations, and Challenges}",
    booktitle = "{2022 Snowmass Summer Study}",
    eprint = "2203.12852",
    archivePrefix = "arXiv",
    primaryClass = "hep-ex",
    month = "3",
    year = "2022"
}

@article{Grosso:2023scl,
    author = "Grosso, Gaia and Letizia, Marco and Pierini, Maurizio and Wulzer, Andrea",
    title = "{Goodness of fit by Neyman-Pearson testing}",
    eprint = "2305.14137",
    archivePrefix = "arXiv",
    primaryClass = "hep-ph",
    doi = "10.21468/SciPostPhys.16.5.123",
    journal = "SciPost Phys.",
    volume = "16",
    number = "5",
    pages = "123",
    year = "2024"
}

@article{Letizia:2022xbe,
    author = "Letizia, Marco and Losapio, Gianvito and Rando, Marco and Grosso, Gaia and Wulzer, Andrea and Pierini, Maurizio and Zanetti, Marco and Rosasco, Lorenzo",
    title = "{Learning new physics efficiently with nonparametric methods}",
    eprint = "2204.02317",
    archivePrefix = "arXiv",
    primaryClass = "hep-ph",
    doi = "10.1140/epjc/s10052-022-10830-y",
    journal = "Eur. Phys. J. C",
    volume = "82",
    number = "10",
    pages = "879",
    year = "2022"
}

@article{Benevedes:2025nzr,
    author = "Benevedes, Sean and Thaler, Jesse",
    title = "{Frequentist Uncertainties on Neural Density Ratios with wifi Ensembles}",
    eprint = "2506.00113",
    archivePrefix = "arXiv",
    primaryClass = "hep-ph",
    reportNumber = "MIT-CTP/5874",
    month = "5",
    year = "2025"
}

@article{Cranmer:2015bka,
    author = "Cranmer, Kyle and Pavez, Juan and Louppe, Gilles",
    title = "Approximating Likelihood Ratios with Calibrated Discriminative  Classifiers",
    eprint = "1506.02169",
    archivePrefix = "arXiv",
    primaryClass = "stat.AP",
    year = "2015"
}

@article{Cowan:2010js,
    author = "Cowan, Glen and Cranmer, Kyle and Gross, Eilam and Vitells, Ofer",
    title = "{Asymptotic formulae for likelihood-based tests of new physics}",
    eprint = "1007.1727",
    archivePrefix = "arXiv",
    primaryClass = "physics.data-an",
    doi = "10.1140/epjc/s10052-011-1554-0",
    journal = "Eur. Phys. J. C",
    volume = "71",
    pages = "1554",
    year = "2011",
    note = "[Erratum: Eur.Phys.J.C 73, 2501 (2013)]"
}

@article{Brehmer:2018kdj,
    author = "Brehmer, Johann and Cranmer, Kyle and Louppe, Gilles and Pavez, Juan",
    title = "Constraining Effective Field Theories with Machine Learning",
    eprint = "1805.00013",
    archivePrefix = "arXiv",
    primaryClass = "hep-ph",
    doi = "10.1103/PhysRevLett.121.111801",
    journal = "Phys. Rev. Lett.",
    volume = "121",
    pages = "111801",
    year = "2018"
}

@article{Brehmer:2018eca,
    author = "Brehmer, Johann and Cranmer, Kyle and Louppe, Gilles and Pavez, Juan",
    title = "A Guide to Constraining Effective Field Theories with Machine Learning",
    eprint = "1805.00020",
    archivePrefix = "arXiv",
    primaryClass = "hep-ph",
    doi = "10.1103/PhysRevD.98.052004",
    journal = "Phys. Rev. D",
    volume = "98",
    pages = "052004",
    year = "2018"
}

@article{Brehmer:2018hga,
    author = "Brehmer, Johann and Louppe, Gilles and Pavez, Juan and Cranmer, Kyle",
    title = "Mining gold from implicit models to improve likelihood-free inference",
    eprint = "1805.12244",
    archivePrefix = "arXiv",
    primaryClass = "stat.ML",
    doi = "10.1073/pnas.1915980117",
    journal = "Proc. Nat. Acad. Sci.",
    volume = "117",
    pages = "5242",
    year = "2020"
}

@article{Brehmer:2019xox,
    author = "Brehmer, Johann and Kling, Felix and Espejo, Irina and Cranmer, Kyle",
    title = "{MadMiner}: Machine learning-based inference for particle physics",
    eprint = "1907.10621",
    archivePrefix = "arXiv",
    primaryClass = "hep-ph",
    doi = "10.1007/s41781-020-0035-2",
    journal = "Comput. Softw. Big Sci.",
    volume = "4",
    pages = "3",
    year = "2020"
}

@article{Chen:2020mev,
    author = "Chen, Siyu and Glioti, Alfredo and Panico, Giuliano and Wulzer, Andrea",
    title = "Parametrized classifiers for optimal {EFT} sensitivity",
    eprint = "2007.10356",
    archivePrefix = "arXiv",
    primaryClass = "hep-ph",
    doi = "10.1007/JHEP05(2021)247",
    journal = "JHEP",
    volume = "05",
    pages = "247",
    year = "2021"
}

@article{Chatterjee:2021nms,
    author = {Chatterjee, Suman and Frohner, Nikolaus and Lechner, Lukas and Sch\"ofbeck, Robert and Schwarz, Dennis},
    title = "Tree boosting for learning {EFT} parameters",
    eprint = "2107.10859",
    archivePrefix = "arXiv",
    primaryClass = "hep-ph",
    doi = "10.1016/j.cpc.2022.108385",
    journal = "Comput. Phys. Commun.",
    volume = "277",
    pages = "108385",
    year = "2022"
}

@article{Chen:2023ind,
    author = "Chen, Siyu and Glioti, Alfredo and Panico, Giuliano and Wulzer, Andrea",
    title = "{Boosting likelihood learning with event reweighting}",
    eprint = "2308.05704",
    archivePrefix = "arXiv",
    primaryClass = "hep-ph",
    doi = "10.1007/JHEP03(2024)117",
    journal = "JHEP",
    volume = "03",
    pages = "117",
    year = "2024"
}

@article{deFavereau:2013fsa,
    author = "de Favereau, J. and Delaere, C. and Demin, P. and Giammanco, A. and Lema\^\i{}tre, V. and Mertens, A. and Selvaggi, M.",
    collaboration = "DELPHES 3",
    title = "{DELPHES} 3, A modular framework for fast simulation of a generic collider experiment",
    eprint = "1307.6346",
    archivePrefix = "arXiv",
    primaryClass = "hep-ex",
    doi = "10.1007/JHEP02(2014)057",
    journal = "JHEP",
    volume = "02",
    pages = "057",
    year = "2014"
}

@article{Sjostrand:2014zea,
    author = {Sj\"ostrand, Torbj\"orn and Ask, Stefan and Christiansen, Jesper R. and Corke, Richard and Desai, Nishita and Ilten, Philip and Mrenna, Stephen and Prestel, Stefan and Rasmussen, Christine O. and Skands, Peter Z.},
    title = "An introduction to {PYTHIA} 8.2",
    eprint = "1410.3012",
    archivePrefix = "arXiv",
    primaryClass = "hep-ph",
    reportNumber = "LU-TP-14-36, MCNET-14-22, CERN-PH-TH-2014-190, FERMILAB-PUB-14-316-CD, DESY-14-178, SLAC-PUB-16122",
    doi = "10.1016/j.cpc.2015.01.024",
    journal = "Comput. Phys. Commun.",
    volume = "191",
    pages = "159",
    year = "2015"
}

@article{CMS:2024onh,
    author = "Hayrapetyan, Aram and others",
    collaboration = "CMS",
    title = "{The CMS statistical analysis and combination tool: COMBINE}",
    eprint = "2404.06614",
    archivePrefix = "arXiv",
    primaryClass = "physics.data-an",
    reportNumber = "CMS-CAT-23-001, CERN-EP-2024-078",
    month = "4",
    year = "2024"
}

@inproceedings{Cranmer:2014lly,
    author = "Cranmer, Kyle",
    title = "{Practical Statistics for the LHC}",
    booktitle = "{2011 European School of High-Energy Physics}",
    eprint = "1503.07622",
    archivePrefix = "arXiv",
    primaryClass = "physics.data-an",
    doi = "10.5170/CERN-2014-003.267",
    pages = "267",
    year = "2014"
}

@article{Benato:2025rgo,
    author = {Benato, Lisa and Giordano, Cristina and Krause, Claudius and Li, Ang and Sch{\"o}fbeck, Robert and Schwarz, Dennis and Shooshtari, Maryam and Wang, Daohan},
    title = "{Unbinned inclusive cross-section measurements with machine-learned systematic uncertainties}",
    eprint = "2505.05544",
    archivePrefix = "arXiv",
    primaryClass = "hep-ph",
    reportNumber = "HEPHY-ML-25-01",
    month = "5",
    year = "2025"
}

@article{Ghosh:2025fma,
    author = "Ghosh, Aishik and Griese, Maximilian and Haisch, Ulrich and Park, Tae Hyoun",
    title = "{Neural simulation-based inference of the Higgs trilinear self-coupling via off-shell Higgs production}",
    eprint = "2507.02032",
    archivePrefix = "arXiv",
    primaryClass = "hep-ph",
    reportNumber = "MPP-2025-123",
    month = "7",
    year = "2025"
}

@article{Farrell:2022lfd,
    author = "Farrell, Delaney and Baldi, Pierre and Ott, Jordan and Ghosh, Aishik and Steiner, Andrew W. and Kavitkar, Atharva and Lindblom, Lee and Whiteson, Daniel and Weber, Fridolin",
    title = "{Deducing neutron star equation of state parameters directly from telescope spectra with uncertainty-aware machine learning}",
    eprint = "2209.02817",
    archivePrefix = "arXiv",
    primaryClass = "astro-ph.HE",
    doi = "10.1088/1475-7516/2023/02/016",
    journal = "JCAP",
    volume = "02",
    pages = "016",
    year = "2023"
}

@article{Brandes:2024vhw,
    author = "Brandes, Len and Modi, Chirag and Ghosh, Aishik and Farrell, Delaney and Lindblom, Lee and Heinrich, Lukas and Steiner, Andrew W. and Weber, Fridolin and Whiteson, Daniel",
    title = "{Neural simulation-based inference of the neutron star equation of state directly from telescope spectra}",
    eprint = "2403.00287",
    archivePrefix = "arXiv",
    primaryClass = "astro-ph.HE",
    doi = "10.1088/1475-7516/2024/09/009",
    journal = "JCAP",
    volume = "09",
    pages = "009",
    year = "2024"
}

@misc{SAGE,
  title        = {{SAGE Code repository}},
  author       = {He, Minxuan and Krause, Claudius and Wang, Daohan},
  year         = 2025,
  howpublished         = {\url{https://github.com/Dorhand/SAGE}}
}

@article{Elsharkawy:2025six,
    author = "Elsharkawy, Ibrahim and Kahn, Yonatan",
    title = "{Contrastive Normalizing Flows for Uncertainty-Aware Parameter Estimation}",
    eprint = "2505.08709",
    archivePrefix = "arXiv",
    primaryClass = "physics.data-an",
    month = "5",
    year = "2025"
}

@article{Ethier:2021bye,
    author = "Ethier, Jacob J. and Magni, Giacomo and Maltoni, Fabio and Mantani, Luca and Nocera, Emanuele R. and Rojo, Juan and Slade, Emma and Vryonidou, Eleni and Zhang, Cen",
    collaboration = "SMEFiT",
    title = "{Combined SMEFT interpretation of Higgs, diboson, and top quark data from the LHC}",
    eprint = "2105.00006",
    archivePrefix = "arXiv",
    primaryClass = "hep-ph",
    reportNumber = "OUTP-20-05P, Nikhef-2020-020, CP3-21-12, MCNET-21-07,
  MAN/HEP/2021/004",
    doi = "10.1007/JHEP11(2021)089",
    journal = "JHEP",
    volume = "11",
    pages = "089",
    year = "2021"
}

@article{CMS:2022dwd,
    author = "Tumasyan, Armen and others",
    collaboration = "CMS",
    title = "{A portrait of the Higgs boson by the CMS experiment ten years after the discovery.}",
    eprint = "2207.00043",
    archivePrefix = "arXiv",
    primaryClass = "hep-ex",
    reportNumber = "CMS-HIG-22-001, CERN-EP-2022-039",
    doi = "10.1038/s41586-022-04892-x",
    journal = "Nature",
    volume = "607",
    number = "7917",
    pages = "60--68",
    year = "2022",
    note = "[Erratum: Nature 623, (2023)]"
}

@article{ATLAS:2022vkf,
    author = "Aad, Georges and others",
    collaboration = "ATLAS",
    title = "{A detailed map of Higgs boson interactions by the ATLAS experiment ten years after the discovery}",
    eprint = "2207.00092",
    archivePrefix = "arXiv",
    primaryClass = "hep-ex",
    reportNumber = "CERN-EP-2022-057",
    doi = "10.1038/s41586-022-04893-w",
    journal = "Nature",
    volume = "607",
    number = "7917",
    pages = "52--59",
    year = "2022",
    note = "[Erratum: Nature 612, E24 (2022)]"
}

@techreport{Cranmer:2012sba,
    author      = "Cranmer, Kyle and Lewis, George and Moneta, Lorenzo and Shibata, Akira and Verkerke, Wouter",
    title       = "{HistFactory: A tool for creating statistical models for use with RooFit and RooStats}",
    institution = "CERN",
    number      = "CERN-OPEN-2012-016",
    year        = "2012",
    doi         = "10.17181/CERN-OPEN-2012-016"
}

@article{Bahl:2024gyt,
    author = "Bahl, Henning and Elmer, Nina and Favaro, Luigi and Hau{\ss}mann, Manuel and Plehn, Tilman and Winterhalder, Ramon",
    title = "{Accurate Surrogate Amplitudes with Calibrated Uncertainties}",
    eprint = "2412.12069",
    archivePrefix = "arXiv",
    primaryClass = "hep-ph",
    month = "12",
    year = "2024"
}

@article{ATLAS:2024rpl,
    author = "Aad, Georges and others",
    collaboration = "ATLAS",
    title = "{Precision calibration of calorimeter signals in the ATLAS experiment using an uncertainty-aware neural network}",
    eprint = "2412.04370",
    archivePrefix = "arXiv",
    primaryClass = "hep-ex",
    reportNumber = "CERN-EP-2024-317",
    month = "12",
    year = "2024"
}
\end{document}